\documentclass[a4paper,fleqn,10pt]{article}
\pdfoutput=1

\usepackage{authblk}

\usepackage{todonotes}
\presetkeys{todonotes}{inline}{}
\usepackage{amsmath}
\usepackage{amssymb}
\usepackage{commath}
\usepackage{cite}
\usepackage{array}
\usepackage{calc}
\usepackage{longtable}
\usepackage{multirow}
\usepackage{graphicx}
\graphicspath{{figures/}}
\usepackage{xspace}
\usepackage{
  pgf,
  tikz}
\usetikzlibrary{
  patterns,
  shapes.multipart,
  arrows,
  trees,
  scopes,
  decorations.pathreplacing,
  decorations.pathmorphing,
  decorations.markings,
  decorations.text,
  positioning,
  calc
}
\usepackage{mciteplus}
\usepackage{booktabs}
\usepackage[a4paper,pdfborder={0 0 0}]{hyperref}
\usepackage[format=hang,labelfont=bf,hypcap=true]{caption}
\usepackage{sectsty}
\usepackage{relsize}
\usepackage[frozencache=true]{minted}
\usepackage{xcolor}
\definecolor{listingsbg}{rgb}{0.95,0.95,0.95}
\usepackage{siunitx}
\usepackage[symbol]{footmisc}

\allsectionsfont{\sffamily}
\subsubsectionfont{\mdseries\itshape\large}

\usepackage{tocloft}

\let\spreprint\empty
\newcommand{\preprint}[1]{\def\spreprint{\protect#1}}
\let\sinstitute\empty

\makeatletter
\renewcommand{\maketitle}{\begingroup
  \null\thispagestyle{empty}%
    \ifx\spreprint\empty
      \vskip 5ex
    \else
      \flushright\large\spreprint\vskip 2ex
    \fi
    \vskip 5ex
    \flushleft
      {\sffamily\bfseries\huge\@title}\vskip 2ex
      \@author\vskip 2ex
      \ifx\sinstitute\empty
      \else
        {\small\sinstitute}
      \fi
    \vskip 5ex
  \endgroup
}
\makeatother
\renewenvironment{abstract}{\begin{center}
  {\large\sffamily\bfseries Abstract: }
  \begin{minipage}[t]{0.75\textwidth}
}{\end{minipage}\end{center}\vskip 10ex}

\numberwithin{equation}{section}

\newcommand{\order}{\ensuremath{\mathcal{O}}}

\newcommand{\mbb}[1]{\mathbb{#1}}
\newcommand{\mr}[1]{\mathrm{#1}}

\newcommand{\bea}{\begin{align}}
\newcommand{\eea}{\end{align}}
\newcommand{\beq}{\begin{equation}}
\newcommand{\eeq}{\end{equation}}

\newcommand{\bs}{\begin{split}}
\newcommand{\es}{\end{split}}

\newcommand{\bi}{\begin{itemize}}
\newcommand{\ei}{\end{itemize}}
\newcommand{\bc}{\begin{center}}
\newcommand{\ec}{\end{center}}
\newcommand{\bac}{\begin{array}{c}}
\newcommand{\bacc}{\begin{array}{cc}}
\newcommand{\ea}{\end{array}}

\def\spa#1.#2{\langle#1\,#2\rangle}
\def\spb#1.#2{[#1\,#2]}

\newcommand{\qt}{\ensuremath{q_\mathrm{T}}\xspace}
\newcommand{\qT}{\qt}
\newcommand{\qTcut}{\ensuremath{q_{\mathrm{T},\text{cut}}}\xspace}

\newcommand{\epluseminus}{\ensuremath{e^+ e^-}\xspace}
\newcommand{\ee}{\epluseminus\xspace}

\newcommand{\EWvirt}{\ensuremath{\text{EW}_\text{virt}}}
\newcommand{\EWsud}{\ensuremath{\text{EW}_\text{sud}}}

\newcommand{\QCDpEW}{\ensuremath{\text{QCD}+\text{EW}}\xspace}
\newcommand{\QCDtEW}{\ensuremath{\text{QCD}\times\text{EW}}\xspace}
\newcommand{\LO}{\ensuremath{\text{LO}}\xspace}
\newcommand{\NLO}{\ensuremath{\text{NLO}}\xspace}
\newcommand{\NNLO}{\ensuremath{\text{N$^2$LO}}\xspace}

\newcommand{\NLL}{\ensuremath{\text{NLL}}\xspace}

\newcommand{\NNLLNNLO}{\ensuremath{\text{N$^2$LL+N$^2$LO}}\xspace}

\newcommand{\NNLLpNNLO}{\ensuremath{\text{N$^2$LL$'$+N$^2$LO}}\xspace}
\newcommand{\NNNLLpNNLO}{\ensuremath{\text{N$^3$LL$'$+N$^2$LO}}\xspace}

\newcommand{\NNLP}{\ensuremath{\text{N$^2$LP}}\xspace}
\newcommand{\SCET}{\ensuremath{\text{SCET}}\xspace}

\newcommand{\sla}[1]{\ensuremath{{#1\kern-0.45em/}}}
\newcommand{\Bbar}{\ensuremath{\overline{\mr{B}}}}

\newcommand{\SecRef}[1]{Sec.~\ref{#1}\xspace}

\newcommand\lep{L\scalebox{0.8}{EP}\xspace}

\newcommand\LEP{\lep}

\newcommand\HERA{H\scalebox{0.8}{ERA}\xspace}
\newcommand\EIC{E\scalebox{0.8}{IC}\xspace}

\newcommand\LHC{L\protect\scalebox{0.8}{HC}\xspace}

\newcommand\hera{H\scalebox{0.8}{ERA}\xspace}
\newcommand\hone{H1\xspace}
\newcommand\zeus{Z\scalebox{0.8}{EUS}\xspace}

\newcommand{\MCatNLO}{M\protect\scalebox{0.8}{C}@N\protect\scalebox{0.8}{LO}\xspace}
\newcommand{\SMCatNLO}{S-M\protect\scalebox{0.8}{C}@N\protect\scalebox{0.8}{LO}\xspace}
\newcommand{\Powheg}{P\protect\scalebox{0.8}{OWHEG}\xspace}

\newcommand{\UNLOPS}{UN\scalebox{0.8}{LO}P\scalebox{0.8}{S}\xspace}
\newcommand{\UNNLOPS}{UN$^2$\scalebox{0.8}{LO}P\scalebox{0.8}{S}\xspace}

\newcommand{\MEPSatNLO}{M\scalebox{0.8}{E}P\scalebox{0.8}{S}@N\protect\scalebox{0.8}{LO}\xspace}
\newcommand{\MEPSatLOOP}{M\scalebox{0.8}{E}P\scalebox{0.8}{S}@L\protect\scalebox{0.8}{OOP}$^2$\xspace}

\newcommand{\Vegas}{V\protect\scalebox{0.8}{EGAS}\xspace}

\newcommand{\UFO}{U\scalebox{0.8}{FO}\xspace}
\newcommand{\Herwig}{H\protect\scalebox{0.8}{ERWIG}\xspace}

\newcommand{\Pythia}{P\protect\scalebox{0.8}{YTHIA}\xspace}

\newcommand{\Vincia}{V\protect\scalebox{0.8}{INCIA}\xspace}

\newcommand{\aMCatNLO}{aM\protect\scalebox{0.8}{C}@N\protect\scalebox{0.8}{LO}\xspace}

\newcommand{\Caesar}{C\protect\scalebox{0.8}{AESAR}\xspace}

\newcommand{\MadLoop}{M\protect\scalebox{0.8}{AD}L\protect\scalebox{0.8}{OOP}\xspace}
\newcommand{\OpenLoops}{O\protect\scalebox{0.8}{PEN}L\protect\scalebox{0.8}{OOPS}\xspace}
\newcommand{\Recola}{R\protect\scalebox{0.8}{ECOLA}\xspace}

\newcommand{\MCFM}{M\protect\scalebox{0.8}{C}F\protect\scalebox{0.8}{M}\xspace}

\newcommand{\Sherpa}{S\protect\scalebox{0.8}{HERPA}\xspace}
\newcommand{\Comix}{C\protect\scalebox{0.8}{OMIX}\xspace}

\newcommand{\Amegic}{A\protect\scalebox{0.8}{MEGIC}\xspace}
\newcommand{\CSS}{CSS\protect\scalebox{0.8}{HOWER}\xspace}
\newcommand{\Dire}{D\protect\scalebox{0.8}{IRE}\xspace}
\newcommand{\Alaric}{A\protect\scalebox{0.8}{LARIC}\xspace}

\newcommand{\Photons}{P\protect\scalebox{0.8}{HOTONS}\xspace}
\newcommand{\PhotonSplitter}{P\protect\scalebox{0.8}{HOTON}S\protect\scalebox{0.8}{PLITTER}\xspace}
\newcommand{\YFS}{Y\protect\scalebox{0.8}{FS}\xspace}
\newcommand{\Rivet}{R\protect\scalebox{0.8}{IVET}\xspace}

\newcommand{\HepMC}{H\protect\scalebox{0.8}{EP}MC\xspace}

\newcommand{\EvtGen}{E\protect\scalebox{0.8}{VT}G\protect\scalebox{0.8}{EN}\xspace}

\newcommand{\Tauola}{T\protect\scalebox{0.8}{AUOLA}\xspace}

\newcommand{\LHAPDF}{L\protect\scalebox{0.8}{HAPDF}\xspace}

\newcommand{\MCgrid}{MC\protect\scalebox{0.8}{GRID}\xspace}
\newcommand{\fastNLO}{\protect\scalebox{0.8}{FAST}NLO\xspace}
\newcommand{\APPLgrid}{APPL\protect\scalebox{0.8}{GRID}\xspace}
\hypersetup{
  pdfauthor={
Enrico Bothmann,
Lois Flower,
Christian G\"utschow,
Stefan H{\"o}che,
Mareen Hoppe,
Joshua Isaacson,
Max Knobbe,
Frank Krauss,
Peter Meinzinger,
Davide Napoletano,
Alan Price,
Daniel Reichelt,
Marek Sch\"onherr,
Steffen Schumann,
Frank Siegert
  },
  pdftitle={Event generation with Sherpa 3}
}

\preprint{IPPP/24/67\\LTH-1385\\FERMILAB-PUB-24-0748-T\\ZU-TH 51/24\\CERN-TH-2024-171\\MCNET-24-17}

\author[1]{Enrico~Bothmann}

\author[2,3]{Lois~Flower}

\author[4,5]{Christian~G\"utschow}

\author[6]{Stefan~H{\"o}che}

\author[7]{Mareen~Hoppe}

\author[6]{Joshua~Isaacson}

\author[1,6]{Max~Knobbe}

\author[2]{Frank~Krauss}

\author[2,8]{Peter~Meinzinger}

\author[9]{Davide~Napoletano}

\author[10]{Alan~Price}

\author[2,11]{Daniel~Reichelt}

\author[2]{Marek~Sch\"onherr}

\author[1]{Steffen~Schumann}

\author[7]{Frank~Siegert}

\affil[1]{Institut f\"ur Theoretische Physik, Georg-August-Universit\"at G\"ottingen, Friedrich-Hund-Platz 1, 37077 G\"ottingen, Germany}
\affil[2]{Institute for Particle Physics Phenomenology, Durham University, Durham DH1 3LE, UK}
\affil[3]{Department of Mathematical Sciences, University of Liverpool, Liverpool L69 3BX, UK}
\affil[4]{Department of Physics \& Astronomy, University College London, Gower~Street, WC1E~6BT, London, UK}
\affil[5]{Centre for Advanced Research Computing, University College London, Gower Street, London, WC1E~6BT, UK}
\affil[6]{Theoretical Physics Division, Fermi National Accelerator Laboratory, P.O. Box 500, Batavia, IL 60510, USA}
\affil[7]{Institut f\"ur Kern- und Teilchenphysik, Technische Universit\"at Dresden, 01062 Dresden, Germany}
\affil[8]{Physik-Institut, Universität Zürich, Winterthurerstrasse 190, CH-8057 Zürich, Switzerland}
\affil[9]{Universit\`a degli Studi di Milano-Bicocca \& INFN, Piazza della Scienza 3, Milano 20126, Italy}
\affil[10]{Jagiellonian University,ul.\ prof.\ Stanis\l{}awa \L{}ojasiewicza 11, 30-348 Krak\'{o}w, Poland}
\affil[11]{CERN, Theoretical Physics Department, CH-1211 Geneva 23, Switzerland}

\title{Event generation with \Sherpa~3}

\begin{document}
\maketitle

\begin{abstract}
  \Sherpa is a general-purpose Monte Carlo event generator for the
  simulation of particle collisions in high-energy collider experiments.
  We summarise new developments, essential features, and ongoing
  improvements within the \Sherpa 3 release series.
  Physics improvements include higher-order electroweak corrections,
  simulations of photoproduction and hard diffraction at NLO QCD,
  heavy-flavour matching in NLO multijet merging,
  spin-polarised cross section calculations,
  and a new model of colour reconnections. In addition, the modelling of
  hadronisation, the underlying event and QED effects in both production
  and decay has been improved, and the overall event generation efficiency
  has been enhanced.
\end{abstract}

\newpage

\renewcommand{\baselinestretch}{0.95}\normalsize
\tableofcontents
\renewcommand{\baselinestretch}{1.0}\normalsize

\section{Introduction}
\label{sec:intro}
\Sherpa is a multi-purpose Monte Carlo event generator for simulations
in high-energy particle physics, mainly in the context of collider
experiments. Event generators such as \Sherpa play a unique role
in the analysis of experimental data, the design, construction,
and improvement of ongoing and future measurements, and the
refinement and validation of theoretical ideas and their application
to phenomenology~\cite{Buckley:2011ms,Campbell:2022qmc}.
The three large event generator projects \Herwig~\cite{Bellm:2015jjp,
  Bewick:2023tfi}, \Pythia~\cite{Sjostrand:2014zea,Bierlich:2022pfr}
and \Sherpa are central to the success of present and future
collider experiments and contribute to the continued development
of the field of particle physics, including in particular the analysis
of data from recent and upcoming runs of the Large Hadron Collider
(\LHC). They are also crucial for the preparation of future
experiments~\cite{EuropeanStrategyforParticlePhysicsPreparatoryGroup:2019qin,
  Butler:2023glv}.

\begin{figure}[h]
  \centerline{\includegraphics[width=0.5\textwidth]{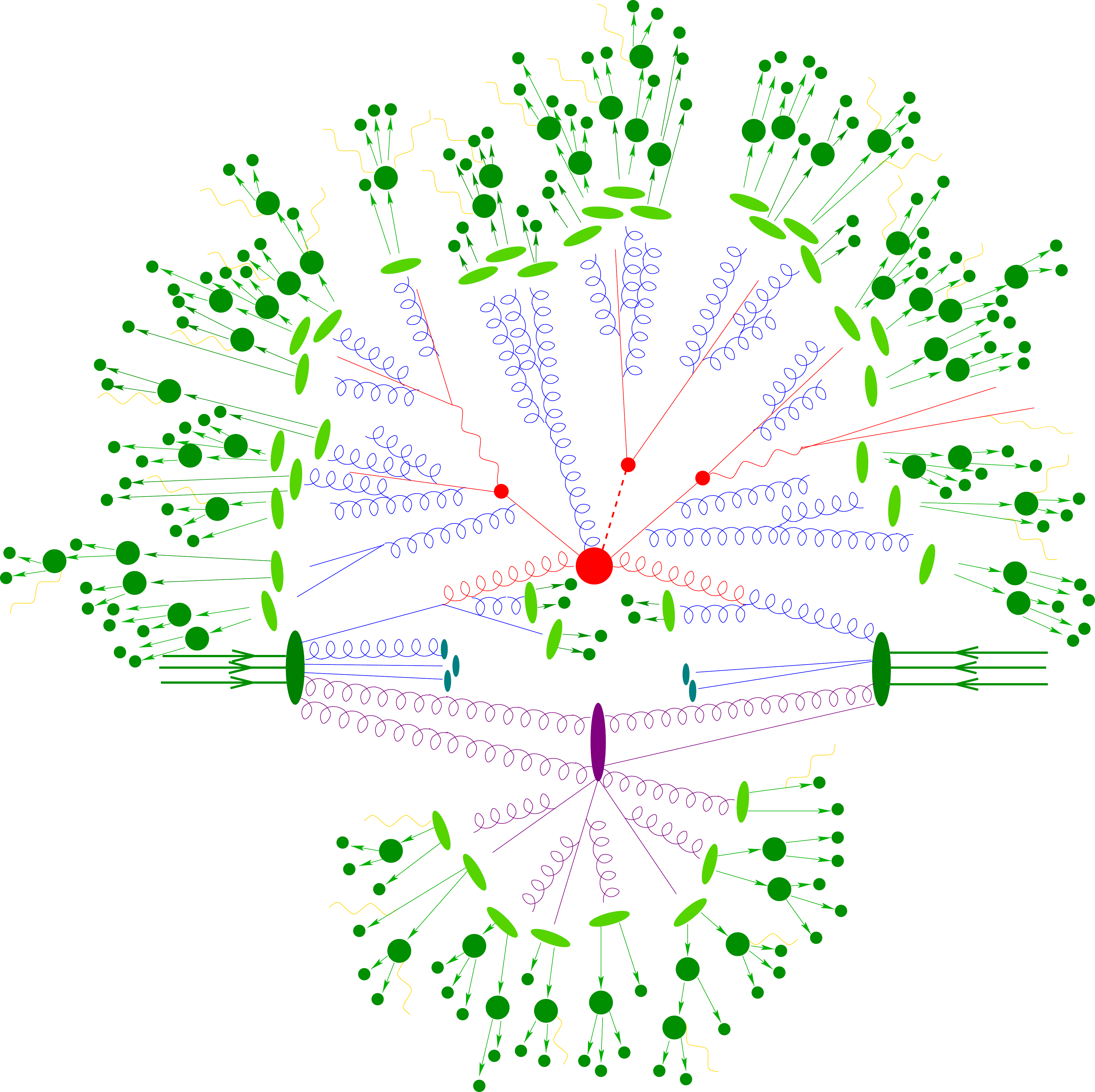}}
  \caption{Event structure of a typical \LHC collision. See the text for details.\label{fig:event}}
\end{figure}

The current version of \Sherpa builds on a series of previous
developments~\cite{Gleisberg:2003xi,Gleisberg:2008ta,Bothmann:2019yzt},
which successively broadened the range of physics effects that could
be simulated with the generator. These effects span the full range of distance 
scales encountered in high-energy particle collisions, across multiple 
parts of a collision event. A general overview of the event structure
typically found in \LHC collisions is sketched in Fig.~\ref{fig:event}.
The simulation of such events relies on the factorisation of perturbative
QCD effects described by the hard collision (central red blob) from the
QCD evolution (blue, tree-like structure), and the transition to the
non-perturbative regime through the hadronisation process (light green blobs).
Additional aspects are non-factorisable QCD corrections (purple blob and
lines), the effects of QED and electroweak radiative corrections
(yellow lines), and the decays of hadronic resonances (dark green blobs).
The \Sherpa framework is highly modular, in that the details of the physics models
implementing these effects are separated from the interfaces that let
different parts of the code interact with each other. This allows us to
systematically improve or replace the implemented physics models,
and to develop and add new models without deprecation of the old ones.
The modular structure has proved particularly useful for the steady improvement
of the formal perturbative accuracy achievable in \Sherpa, for example
the inclusion of higher-order electroweak corrections in different
approximations, and the construction of new parton showers with increased
logarithmic accuracy. Other recent examples include the addition of new
models for non-perturbative physics, such as a dedicated module for the
description of colour-reconnection effects.

In this manuscript we discuss some of these developments and recent additions
in detail, focusing on the ones that significantly enhance the physics
capabilities of \Sherpa. They will be presented in \SecRef{sec:sherpa}
and include the following major new features:
\begin{itemize}
\item the computation of electroweak corrections at full NLO accuracy
  and in various approximations,
\item higher-order QED corrections in production processes
  and from photon splittings in decays,
\item spin-polarised cross section calculations,
\item treatment of collider setups with resolved photons
  and other non-trivial beam spectra,
\item heavy-flavour matching in multijet merging,
\item improvements to the multiple interactions and hadronisation modelling,
\item new models for beam remnants and colour reconnections, and
\item technical developments leading to higher event generation efficiency.
\end{itemize}
In addition to its main use case for the simulation of exclusive final states
at colliders, \Sherpa also serves as a development platform for spin-off
projects.  For example, recent efforts enabled the support of automated or
semi-automated resummation tools and event generation for neutrino physics
within the \Sherpa framework.
In addition, we aim to address the physics simulation and computational needs
of the \LHC community and of future collider experiments not only by better
physics modelling, but also by systematically enhancing the performance
of the code through algorithmic improvements. These include
AI/ML techniques, and developments for new hardware platforms,
and will be described in \SecRef{sec:sherpa:pipeline}.
%

All updates and new versions of \Sherpa, including an extensive and
up-to-date manual describing the technical details and
options for running the code, are available online at
\begin{center}
  \url{https://sherpa-team.gitlab.io}.
\end{center}
The software download (see App.\ \ref{app:install}) includes a pdf
version of the relevant manual.

\section{The physics model of the \Sherpa Monte Carlo event generator}
\label{sec:sherpa}

\begin{figure}[h]
  \centerline{\includegraphics[width=0.85\textwidth]{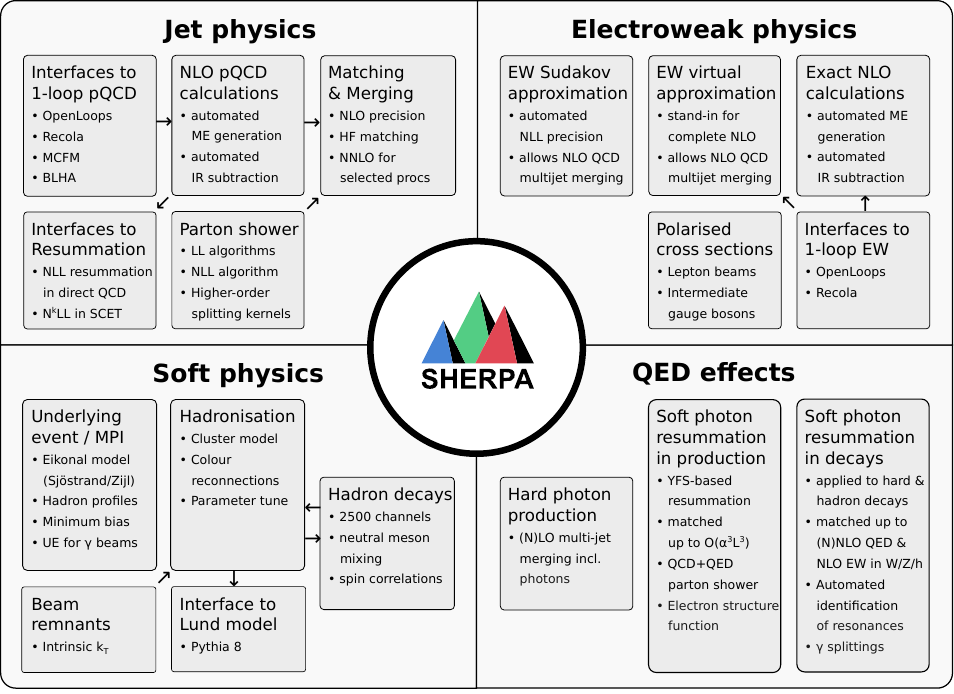}}
  \caption{Schematic overview of the physics effects simulated by \Sherpa,
    as well as major characteristics of their implementation
    in the event generator.\label{fig:physics_overview}
  }
\end{figure}
A schematic overview of the \Sherpa 3 event generation framework
is shown in Fig.~\ref{fig:physics_overview}. \Sherpa itself is
the centerpiece that coordinates the computation of QED, electroweak
and QCD effects leading to the emergence of the many-body final state
in a scattering experiment. We discuss these computations in detail
in the following subsections thereby putting particular emphasis on
newly added features in \Sherpa 3.

\subsection{The initial state}
\label{sec:sherpa:is}

\begin{figure}
  \centering
  \includegraphics[width=0.65\textwidth]{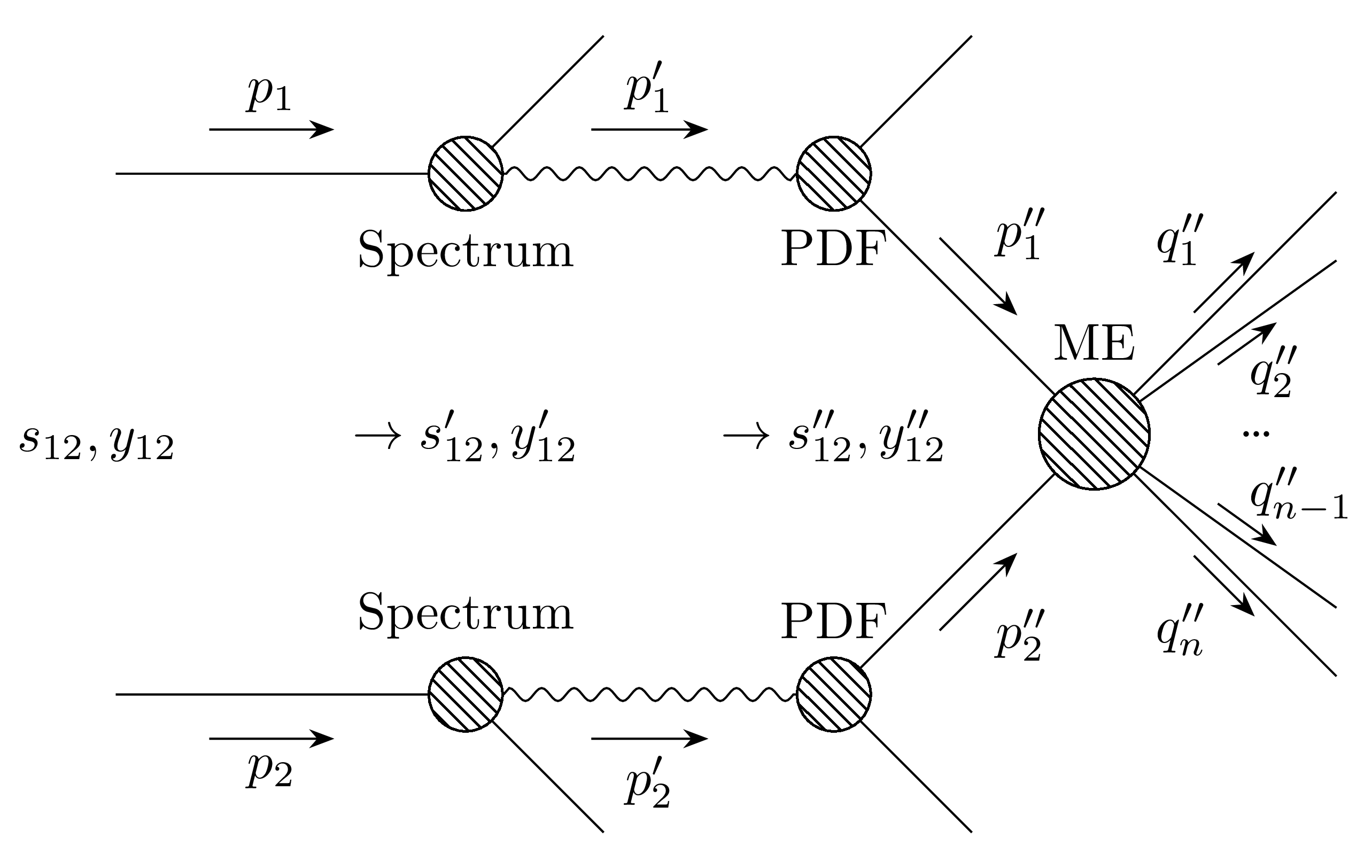}
  \caption{
    Incident-beam setup for \protect\Sherpa simulations.
    Each beam particle may (but does not have to) produce
    a secondary beam of particles parametrised by a beam spectrum.
    The secondary beam particles may (but do not have to) be resolved
    through a structure function or PDF. The emerging ``partons''
    with momenta $p_1''$ and $p_2''$ then undergo a hard collision
    which is described by a matrix element (ME) and produces $n$
    final-state particles.
    \label{fig:is:phase-space}
  }
\end{figure}

\Sherpa is capable of simulating scattering events from a wide range
of incident beam particles. This includes situations where an initial
composite or elementary beam particle initiates a secondary beam particle
according to a given momentum spectrum.  The secondary beam particles
themselves may be either composite or elementary. A prominent example is
resolved vs.\ unresolved photons, where the former fluctuate into a hadronic
structure with a parton distribution function (PDF) while the latter remain
point-like.  Accordingly, the sampling of the initial state comprises
a two-stage procedure: The (optional)
beam-spectrum sampling and the (optional) beam-substructure modelling through
PDFs. Figure~\ref{fig:is:phase-space} illustrates the most
general initial-state setup supported by \Sherpa.

\subsubsection{Beam spectra}
\label{sec:sherpa:is:spectra}

Depending on the experimental situation, the particles which define 
the beam may produce other particles, and these secondary particles 
may produce the collision event instead. In these cases, the secondary 
beam particles follow a momentum distribution (a beam spectrum) 
which is either perturbatively calculable or can be parametrised.
Such spectra are of prime importance
to describe, for example, the interactions of photons from charged 
particle beams.
\Sherpa supports two methods to obtain photon beams: either via
laser backscattering, where incident leptons convert to photons
through Compton scattering, or via the Equivalent Photon Approximation
(EPA).  In the EPA the beam particles serve as quasi-classical sources
of photons~\cite{vonWeizsacker:1934sx,Williams:1934ad,Budnev:1974de}.
In \Sherpa, the above spectra are implemented for electron/positron, 
(anti-)proton and ion beams.
In the case of electrons, there are two versions, the original formula
quoted in~\cite{Budnev:1974de}, and the improved version derived
in~\cite{Frixione:1993yw}.
Other available spectra are the pomeron and reggeon fluxes as used in
parametrisations of Diffractive
PDFs~\cite{Collins:1997sr,Ingelman:1984ns,Newman:2013ada} to
calculate diffractive jet production,
see Sec.~\ref{sec:sherpa:mm:photoproduction_harddiffraction}.

\subsubsection{Parton densities and structure functions}
\label{sec:sherpa:is:pdfs}

To resolve the substructure of (composite) incoming particles,
theoretical calculations make use of dedicated PDFs based on collinear
factorisation. \Sherpa provides access to all commonly used sets either
through an interface to \LHAPDF~\cite{Buckley:2014ana}, or through
a dedicated interface which is specific to the required PDF.

For proton beams \Sherpa defaults to the use of \LHAPDF and
the \texttt{PDF4LHC21\_40\_pdfas} PDF set~\cite{PDF4LHCWorkingGroup:2022cjn}.
See App.\ \ref{app:install} for instructions on enabling \LHAPDF support.
If \LHAPDF support is not enabled, dedicated interfaces to the NNPDF 3.1 PDF
set~\cite{NNPDF:2017mvq} and the CT14 sets~\cite{Dulat:2015mca} are
also provided.
When using the \LHAPDF interface, the value and the perturbative order
for the running of $\alpha_s$ in \Sherpa are automatically set to the values
used in the PDF fit.
At the expense of possible inconsistencies, the user can choose to
override this behaviour by explicitly defining the value of
$\alpha_s(m_\mathrm{Z}^2)$ and the loop-order of its evolution.
For the simulation of Multi-Parton Interactions (MPI, see
Sec.\ \ref{sec:sherpa:mi}), the PDFs can be selected independently.
This treatment is motivated by the fact that the MPI tunes
are highly sensitive to the PDFs and describe an effect beyond
the collinear factorisation theorems underpinning the
perturbative QCD calculations. While a different PDF
potentially introduces small inconsistencies in the description
of the parton content of a given proton, these mismatches
lie entirely within the inherent uncertainties of the MPI model.
It is recommended to use \LHAPDF version 6.4.0 or later with \Sherpa,
as this version includes very significant performance improvements
that are relevant for typical event generation use cases,
benchmarked using standard \Sherpa setups~\cite{Bothmann:2022thx}.

In case of incoming photons, either as a monochromatic beam
or produced through one of the beam spectra described above,
a range of PDF sets is supported, namely
\texttt{GRV}~\cite{Gluck:1991ee,Gluck:1991jc},
\texttt{GRS}~\cite{Gluck:1999ub},
\texttt{SAL}~\cite{Slominski:2005bw},
\texttt{CJK}~\cite{Cornet:2002iy,Cornet:2004nb},
and \texttt{SaS}~\cite{Schuler:1995fk,Schuler:1996fc}, with
the \texttt{SaS1M} set of the \texttt{SaS} family the default.
For incident pomerons, which are only available as the product of a beam
spectrum, the \hone Diffractive PDF fit has been
interfaced~\cite{H1:2006zyl}.
Similary, for incident reggeons we default to the
\texttt{GRVPI0} PDF fit in \LHAPDF.

Finally, \Sherpa provides an analytical QED structure function
\cite{Kuraev:1985hb} for incoming lepton beams (electrons, positrons and 
(anti-)muons). It encodes the leading logarithmic (LL)
corrections arising from collinear photon emissions, resummed
using the DGLAP evolution equations.
The resulting universal factors are matched to higher-order
corrections, leading to some additional terms described
in~\cite{Bardin:1993bh,Beenakker:1994vn,Montagna:1994qu,Berends:1994pv}.
For details on the implementation see \cite{Krauss:2022ajk}.

\subsection{The hard scattering}
\label{sec:sherpa:me}

The actual simulation of individual events starts from a partonic
hard-scattering configuration, where the momenta of all initial
and final state particles are distributed according to the
corresponding transition matrix element.
These matrix elements are stochastically sampled to determine
the total inclusive production cross section as well as
arbitrary differential distributions of final state particles.
To address the large number of interesting scattering
processes at the LHC and other past and future colliders, \Sherpa's
matrix element generators are built with a high degree of
automation. 

\subsubsection[LO accuracy]
              {Hard scatterings at \texorpdfstring{\LO}{LO} accuracy}
\label{sec:sherpa:me:lo}

\Sherpa includes two in-house automated tree-level
matrix element generators, \Amegic \cite{Krauss:2001iv}
and \Comix \cite{Gleisberg:2008fv}.
They are capable of generating scattering matrix elements
for any process within the Standard Model and a number of
frequently used extensions like the Higgs Effective Field
Theory (HEFT)~\cite{Ellis:1975ap,
  Wilczek:1977zn,Shifman:1979eb,Ellis:1979jy}, and are only
limited by the available computing resources.
Additionally, \Comix supports most models formulated using the
\UFO standard~\cite{Degrande:2011ua,Darme:2023jdn},
see Sec.\ \ref{sec:sherpa:me:ufo}.

When using either generator, it simultaneously generates 
suitable phase-space parametrisations 
using a combination of: inverse transform methods on propagator virtualities
and polar angles~\cite{Byckling:1969sx}, the multi-channel method described,
e.g., in~\cite{Kleiss:1994qy,Berends:1994xn},
and \Vegas optimisation routines \cite{Lepage:1977sw,Ohl:1998jn}.
Where appropriate, this also includes sampling of the colour and helicity spaces.
This procedure allows for an efficient integration of multi-particle
final states both in the bulk of the phase space
and in intricate corners.

Finally, \Sherpa also allows users to compute scattering cross sections
for loop-induced processes, whose lowest-order contribution is mediated
by one-loop diagrams.
These calculations are facilitated by an interface to external
one-loop providers, for details see Sec.\ \ref{sec:sherpa:me:nlo}.
The phase-space parametrisations are obtained in a semi-automated
fashion by using a tree-level proxy process which contains
similar propagator and spin structures~\cite{Cascioli:2013gfa,
  Goncalves:2015mfa,Jones:2017giv,Bothmann:2021led}.

\subsubsection[NLO accuracy]
              {Hard scatterings at \texorpdfstring{\NLO}{NLO} accuracy}
\label{sec:sherpa:me:nlo}

Hard-scattering cross sections at \NLO accuracy, comprising
the inclusion of QCD, electroweak (EW), as well as mixed QCD-EW corrections,
are computed by combining the Born-level expressions that constitute the \LO
expression and its real and virtual \NLO corrections.  When using a
Monte Carlo integration framework, the calculation must be performed in
four space-time dimensions, necessitating a subtraction formalism
to render all integrands finite~\cite{Frixione:1995ms,Catani:1996vz}.
In \Sherpa, the Catani--Seymour subtraction
formalism~\cite{Catani:1996vz,Catani:2002hc} is used to construct the
corresponding infrared subtraction terms.
To assemble the tree-level expressions for Born and real-emission
corrections, \Sherpa relies on its matrix element generators
\Amegic and \Comix, which also provide the corresponding phase-space
parametrisations as described above. 
The infrared subtraction automatically identifies both
QCD~\cite{Gleisberg:2007md} and QED~\cite{Schonherr:2017qcj} divergences
and constructs the relevant counterterms.
Processes with simultaneous QCD and QED divergences can
be handled by this procedure in \Sherpa as well.
External photons can be treated both as resolved and
unresolved partons~\cite{Schonherr:2017qcj,Kallweit:2017khh,
  Chiesa:2017gqx,Greiner:2017mft,Reyer:2019obz}.

To compute the UV-renormalised one-loop corrections, \Sherpa includes a small
library of purpose-built renormalised one-loop matrix elements and provides
a number of interfaces to one-loop providers (OLPs), namely
\OpenLoops~\cite{Cascioli:2011va,Buccioni:2019sur},
\Recola~\cite{Actis:2012qn,Denner:2017wsf,Biedermann:2017yoi},
or \MadLoop~\cite{Hirschi:2011pa}.
A recent addition is the interface to \MCFM~\cite{Campbell:2021vlt}.
\MCFM's fast analytic one-loop matrix elements can provide significant
overall event generation speed-ups, particularly when combined
with \Sherpa's pilot run strategy~\cite{Bothmann:2022thx}.
Details on how to enable \Sherpa's OLP interfaces are given
in App.~\ref{app:install}.

\begin{figure}[t!]
  \centering
  \includegraphics[width=0.47\textwidth]{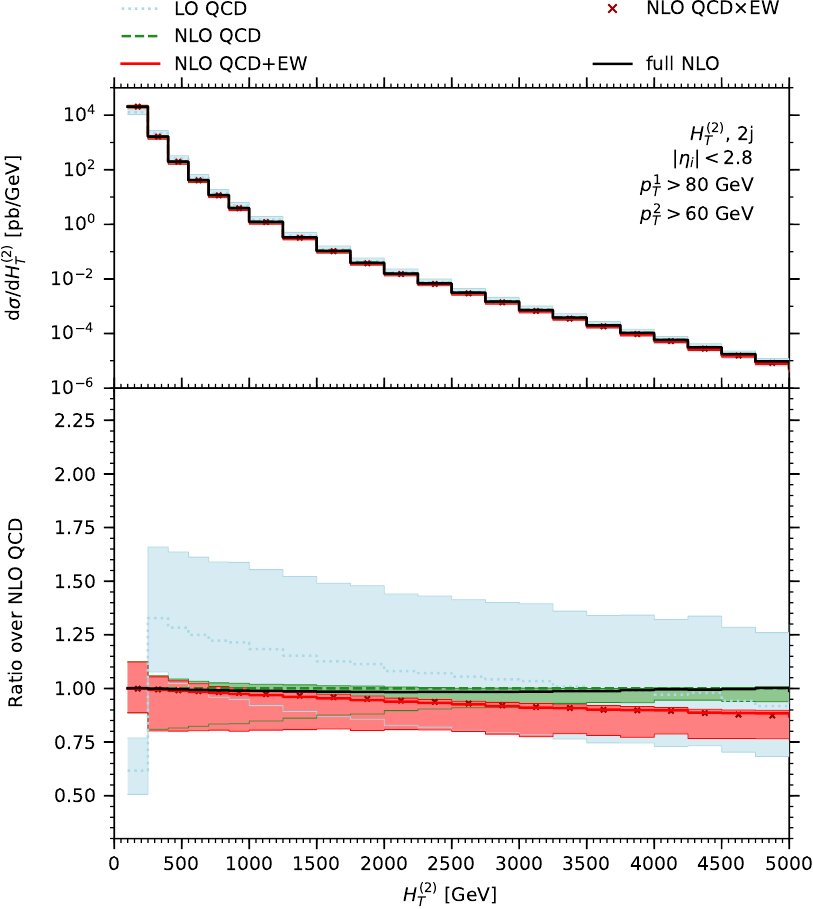}
  \hfill
  \includegraphics[width=0.47\textwidth]{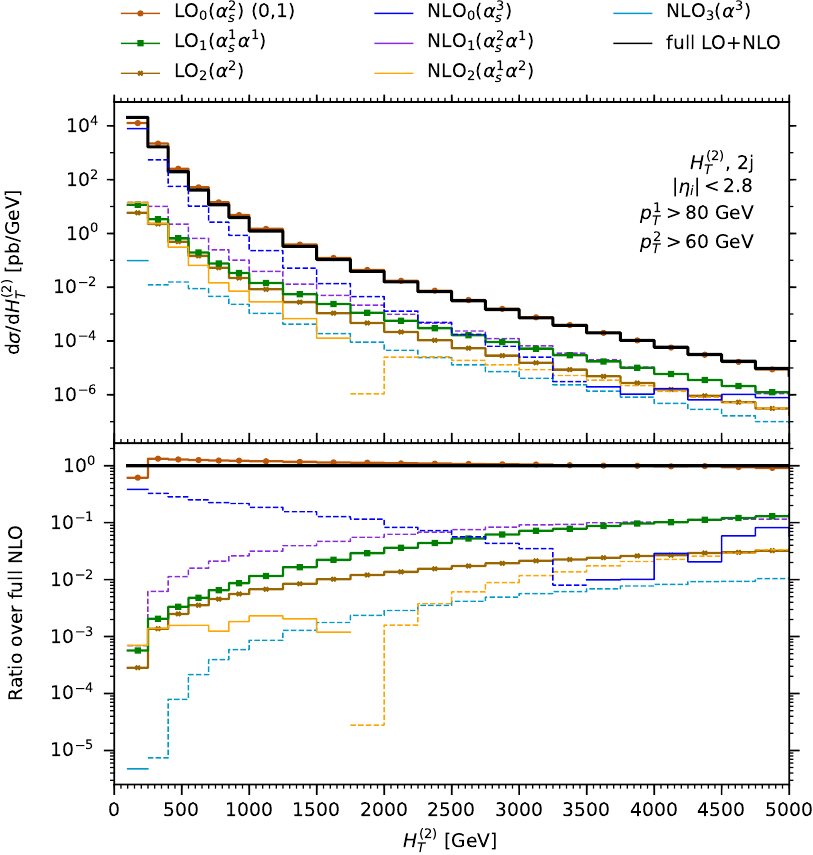}
  \caption{
    The full \NLO corrections to the scalar sum of leading and
    sub-leading jet transverse momenta, $H_{\mr{T}}^{(2)}$ in
    inclusive dijet production at the LHC.
    The left panel shows the LO, NLO QCD, NLO \QCDpEW, NLO \QCDtEW,
    and full NLO results, including their scale uncertainties.
    The right panel shows the decomposition of the full NLO computation.
    The results were computed in the setup and the conventions
    of~\cite{Reyer:2019obz}.
    Dashed lines denote the absolute of an otherwise negative
    contribution.
    \label{fig:sherpa:me:nlo:dijet:HT}
  }
\end{figure}

Figure~\ref{fig:sherpa:me:nlo:dijet:HT} shows an example application
of the above machinery to calculate the complete NLO predictions to
inclusive dijet production at the LHC in the setup of~\cite{Reyer:2019obz}.
This process, despite its limited number of external legs, is comprised
of a multitude of subprocesses contributing at various different coupling
orders in perturbation theory.
As a consequence, the \NLO corrections contain both QCD and QED
divergences which must be addressed by a suitable subtraction procedure.

\subsubsection[NNLO accuracy]{Hard scatterings at \texorpdfstring{\NNLO}{NNLO} accuracy}
\label{sec:sherpa:me:nnlo}
A few selected calculations of high phenomenological importance have been
implemented in \Sherpa at \NNLO precision.
In particular, Drell--Yan lepton-pair production~\cite{Hoche:2014uhw},
Higgs-boson production~\cite{Hoche:2014dla}
and Deep-Inelastic Scattering~\cite{Hoche:2018gti} simulations
can be carried out fully differentially at the parton level,
making use of \qT slicing~\cite{Catani:2007vq,Catani:2009sm}
or Projection-to-Born~\cite{Cacciari:2015jma} techniques.
The \qT slicing method separates the \NNLO corrections
into a contribution with $\qT<\qTcut$, the so-called zero-\qT bin,
and the remaining spectrum with $\qT>\qTcut$. The first contribution is integrated out
analytically using approximate expressions that become exact in the
limit $\qTcut\to 0$. The second comprises an NLO calculation
of the Born-plus-one-jet process and is computed using the tools
introduced in the previous section.
The Projection-to-Born method, on the other hand, introduces an arbitrary
but infrared-safe  and unambiguous mapping from the real and double-real
phase spaces onto the Born phase space and evaluates all components
of the \NNLO calculation using this mapping.
Consequently, all singularities cancel locally in the Born
phase space, leaving a finite result.
The mismatch introduced by projecting the real and double-real phase-space
contributions is corrected through dedicated lower-order calculations
at \NLO and \LO, respectively, using the methods discussed in the previous
sections.

\subsubsection{Decays of unstable particles}
\label{sec:sherpa:me:dec}

Within \Sherpa, there are various options to simulate the decays
of massive unstable particles produced in the hard scattering process.
A full off-shell treatment in the matrix element yields the
most complete calculation, but might not always be feasible,
either because of the inefficient generation of unweighted events due the
high final-state complexity or because the user is interested in an
inclusive simulation of multiple decay final states.
For such cases, \Sherpa provides a module to simulate decays in an
automatic way~\cite{Hoche:2014kca}.

In its automated treatment of massive unstable particle decays, \Sherpa employs
an improved narrow-width approximation, where the hard scattering
of Secs.\ \ref{sec:sherpa:me:lo}--\ref{sec:sherpa:me:nnlo}
constitutes the production process, while its decays are calculated
at LO accuracy. Spin correlations are taken into account using the algorithm
described in~\cite{Collins:1987cp,Knowles:1987cu,Knowles:1988vs,Richardson:2001df},
and off-shell effects are modelled by \emph{a posteriori} adjusting the resonance
kinematics according to its Breit--Wigner distribution.
QCD and QED radiative corrections can be effected through interfaces
to \Sherpa's parton shower and soft-photon resummation, see
Secs.\ \ref{sec:sherpa:showers} and \ref{sec:sherpa:yfs}, respectively.
Unless specified by the user,
event-by-event decay channels are selected according to their branching ratios,
determined from the automatically generated decay matrix elements and
decay widths using tree-level expressions.
Their generation is model-specific and, in addition to the Standard 
Model, is applicable to beyond the Standard Model theories 
using the \UFO format (see Sec.~\ref{sec:sherpa:me:ufo} for details).
This enables a decay simulation that is fully
consistent with the production process.

The decay framework is used, for example, for Standard Model processes
involving top quarks, massive vector bosons, or the Higgs boson.
Decays can be added on top of a wide variety of simulations of
the hard scattering, calculated both at LO and NLO, and both
in fixed-order or in parton-shower matched/merged simulations.

\subsubsection[Polarised lepton beams]
              {Cross sections for polarised lepton beams}
\label{sec:dec:beampol}

The ability to polarise the incoming beams is a defining feature of
various proposed future lepton-collider
experiments~\cite{ILC:2013jhg,CLICdp:2018cto}.
Such setups can be used to considerably enhance signal rates
while also suppressing unwanted background processes, because the
cross sections of scattering processes within and beyond the Standard Model
often depend on the helicities of incoming particles.
By varying the polarisations of the incoming beams the properties
of the produced final-state particles such as their chiral couplings
and quantum numbers can be probed~\cite{Moortgat-Pick:2005jsx}.

In \Sherpa, it is possible to simulate events with longitudinally-polarised
beams by reweighting the helicity amplitudes with the corresponding
fractional polarisation, $P_{e^\pm}$, using the \Amegic matrix element
generator, which is a common approach in event generators~\cite{Kilian:2007gr,Alwall:2014hca}.
This can then be combined with a soft-photon resummation in the YFS formalism,
as described in \SecRef{sec:sherpa:yfs:ee}, to further improve the accuracy of
the simulation. Since YFS resummation is based on the soft-photon limit,
the matching of higher-order corrections with beam polarisation can be
achieved in a straightforward fashion. This will allow \Sherpa to provide NLO EW
predictions for polarised collider experiments in the future.
In Tab.~\ref{table:cross_sections} we illustrate the
effect of beam polarisation on various relevant Higgs-boson production
processes. It can be observed that due to the chiral nature of the
weak-current interaction significant enhancements or suppressions
depending on the chosen initial-state polarisations can be realised.
The YFS corrections included here are purely at the resummation level
and are not matched to any higher-order corrections.

\begin{table}[t!]
	\centering
	\begin{tabular}{c | c c | c c | c c}
		$(P_{e^-}, P_{e^+})$ & \multicolumn{2}{c|}{$HZ$}
		                     & \multicolumn{2}{c|}{$H\nu_e\bar{\nu}_e$}
		                     & \multicolumn{2}{c }{$He^+e^-$} \\
		\hline
		\hline
    \rule{0pt}{3ex}
                 & $\sigma^{\text{LO}}/\sigma^{\text{LO}}_\text{(0, 0)}$ & $\sigma^{\text{YFS}}/\sigma^{\text{YFS}}_\text{(0, 0)}$     
                 & $\sigma^{\text{LO}}/\sigma^{\text{LO}}_\text{(0, 0)}$ & $\sigma^{\text{YFS}}/\sigma^{\text{YFS}}_\text{(0, 0)}$
                 & $\sigma^{\text{LO}}/\sigma^{\text{LO}}_\text{(0, 0)}$ & $\sigma^{\text{YFS}}/\sigma^{\text{YFS}}_\text{(0, 0)}$      \\
		\hline
    (-0.8, 0)    & 1.17 & 1.17   & 1.70 & 1.68   & 1.18 & 1.19   \\
    (-0.8, 0.3)  & 1.47 & 1.47   & 2.20 & 2.18   & 1.49 & 1.50   \\
    (-0.8, -0.3) & 0.87 & 0.87   & 1.20 & 1.19   & 0.87 & 0.88   \\
    (0.8, 0.0)   & 0.83 & 0.83   & 0.30 & 0.32   & 0.82 & 0.83   \\
    (0.8, 0.3)   & 0.65 & 0.65   & 0.32 & 0.34   & 0.65 & 0.65   \\
    (0.8, -0.3)  & 1.01 & 1.00   & 0.28 & 0.31   & 0.99 & 1.00   \\
		\hline
	\end{tabular}
	\caption{
     Normalised cross sections for various $e^+e^- \rightarrow H + X$ production processes 
     evaluated with different initial-state polarisation 
     combinations at $\sqrt{s} = 380\,\text{GeV}$, 
     both at LO and including YFS corrections.
      \label{table:cross_sections}
    }
\end{table}

\subsubsection[Polarised intermediate gauge bosons]
              {Cross sections for polarised intermediate gauge bosons}
\label{sec:dec:pols}

The ability to predict cross sections for polarised vector boson
production is of great interest, as they probe the structure
of the electroweak interaction. With \Sherpa 3 it is now possible to
compute cross sections for polarised intermediate vector bosons
in the $s$-channel~\cite{Hoppe:2023uux}. The efficiency of 
the implementation is guaranteed by the simultaneous computation
of all polarisation combinations.
Each combination is added as an additional event weight
to the unpolarised sample, using the techniques described in~\cite{Bothmann:2022pwf}.
The spin-correlated narrow-width approximation~\cite{Hoche:2014kca}
is used to compute the various contributions (see
\SecRef{sec:sherpa:me:dec} for details). The different polarisation 
components are based on the complete helicity-dependent amplitude,
such that interferences between different polarisations
are also accessible on an event-by-event basis.
Within a single generator run, multiple reference frames can be 
studied.

The calculation of polarised cross sections is not
limited to LO and can also be performed at approximate NLO QCD
accuracy, referred to as nLO QCD.
They can be matched to \Sherpa's parton shower via the \MCatNLO
method and be included in multijet merging, see \SecRef{sec:sherpa:mm}.
Within the nLO QCD calculation, polarisation fractions are
calculated depending on the event type in the \MCatNLO
formalism.
For $\mbb{H}$- and resolved $\mbb{S}$-events, the corresponding
amplitude information is constructed using the complete real
emission corrections.
Hence, the exact polarisation fractions, up to NLO QCD, are
used for both soft and hard emissions.
For unresolved $\mbb{S}$-events, however, the amplitude information
is based on the Born expression, i.e.\ all corrections
stemming from virtual and ultra-soft and/or collinear emission
are neglected.
As the number of events in this category is generally exceedingly
small in typical LHC setups, and in any case this construction is only used
to determine the polarisation fractions in the otherwise fully
NLO QCD-accurate unpolarised sample, the error introduced in
this way is expected to be small.

\begin{figure}[t]
	\centering
	\includegraphics[width=0.425\textwidth]{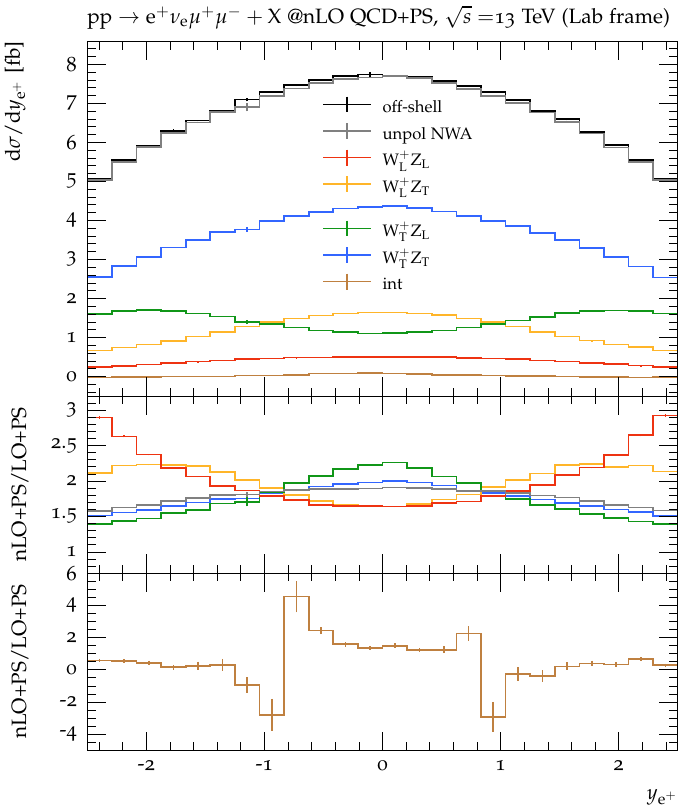}\hfill
	\includegraphics[width=0.55\textwidth,clip,trim=0 10mm 0 0]{plots/Polarisation_VB/WZ_merging_NLOtoLO_COM.pdf}
	\caption{Left: Double-polarised distributions of the positron
          rapidity $y_{e^+}$ in inclusive $\text{W}^+$Z production at nLO QCD+PS
          (polarised distribution)/ NLO QCD+PS (unpol, full) accuracy;
          polarisation states are defined in the laboratory (Lab).
          $K$-factors (bottom panel) are the ratio of n(N)LO QCD+PS
          over LO+PS cross sections.
          Right: Integrated polarisation fractions for inclusive
          $\text{W}^+$Z boson production at the LHC (13~TeV) for LO+1j
          simulations matched to parton shower using different merging
          scales. For comparison, also polarisation fractions at nLO QCD+PS
          and LO+PS, as well as at fixed NLO QCD taken
          from~\cite{Denner:2020eck} are shown. The polarisation is
          defined in the $\text{W}^+$Z boson centre-of-mass frame.
          (Figures adapted from~\cite{Hoppe:2023uux}).}
	\label{fig:pols:merging}
\end{figure}

Fig.~\ref{fig:pols:merging} (left) shows the nLO QCD+PS contribution to
polarised inclusive $\text{W}^+$Z production for the lepton
rapidity in the laboratory frame, as an example.
It illustrates the importance of including higher order QCD effects
in polarisation templates, since they can be very
large and non-trivial. Comparisons with complete NLO QCD fixed-order
calculations~\cite{Denner:2020eck}
confirm that our nLO QCD approximation can reproduce all main
contributions of the full calculation, as these are strongly
dominated by real corrections.
Hence, also  multijet-merged calculations are able to describe
the bulk of the NLO QCD effects, if small merging scales are
used, as demonstrated on the right hand side of Fig.~\ref{fig:pols:merging}.

\subsubsection{Physics within the Standard Model -- instantons}
\label{sec:sherpa:me:instantons}

In addition to standard perturbative scattering amplitudes,
discussed in Secs.\ \ref{sec:sherpa:me:lo}--\ref{sec:sherpa:me:nnlo},
the Standard Model contains other manifestly non-perturbative
solutions like the QCD instanton~\cite{Kuzmin:1985mm}, which emerge
as a consequence of the non-trivial structure of the
Yang--Mills vacuum~\cite{Fukugita:1986hr,tHooft:1976rip}.
While the QCD instanton violates $\text{B}+\text{L}$ symmetries, 
it conserves $\text{B}-\text{L}$ as well as chirality.
Despite otherwise large inclusive production rates, the cross section
falls rapidly with increasing instanton mass, and, as a consequence,
the existence of instantons has not been confirmed experimentally yet.

\Sherpa comprises an implementation of QCD-instanton-mediated
multiparton production processes~\cite{Khoze:2019jta},
including important quantum
corrections due to initial- and final-state gluon interactions.
The $\hat{s}$-dependent production cross sections, where $\hat{s}$
is the partonic c.m. energy squared, is taken from an interpolation
table included in the runcard. This table also provides results for
different scale choices and allows, to some extent, to vary these
scales to obtain some idea about related uncertainties.  The outgoing
quarks, anti-quarks, and a Poisson-distributed number of
gluons populate the phase space isotropically in the instanton
rest frame.

\subsubsection{Physics models beyond the Standard Model --
  \texorpdfstring{\UFO}{UFO}}
\label{sec:sherpa:me:ufo}

\Sherpa provides a versatile framework for the simulation of new
physics signals~\cite{Christensen:2009jx,Hoche:2014kca}, through
built-in models (Higgs Effective Field Theory~\cite{Ellis:1975ap,
  Wilczek:1977zn,Shifman:1979eb,Ellis:1979jy},
the Minimal Supersymmetric Standard Model~\cite{Martin:1997ns} with inputs in the
SLHA format~\cite{Skands:2003cj}, and various anomalous Triple
and Quartic Gauge Couplings Models~\cite{Hagiwara:1986vm,Belyaev:1998ih,
  Eboli:2000ad,Eboli:2003nq,Eboli:2006wa}) or, more generally through a
\UFO~\cite{Degrande:2011ua} interface to \Comix.
The latter has been used in a variety of analyses within the context
of the Standard Model Effective Field Theory (SMEFT), see e.g.\
\cite{Biekotter:2020flu,Biekotter:2021int,Banerjee:2024eyo}.
For details of how to enable \Sherpa's \UFO support during installation,
see App.~\ref{app:install}.

Recently, the \UFO format was updated to address and standardise
several extensions that have been implemented since the first version was
proposed~\cite{Darme:2023jdn}, thus ensuring
portability and compatibility between generators.  The updated \UFO
opens new possibilities and options, such as customised
propagators~\cite{Christensen:2013aua}, the inclusion of particle decay
information~\cite{Alwall:2014bza}, and the renormalisation
group running of model parameters~\cite{Aoude:2022aro}, all of which
are expected in future versions of \Sherpa.
In addition, it allows the inclusion of form factors associated with specific
Lorentz structures in the vertices. These are enabled in \Sherpa but
currently need to be manually implemented; we expect this process to be
semi-automatic in future releases.


\subsection{Parton showers}
\label{sec:sherpa:showers}

The role of parton showers~\cite{Webber:1983if,Bengtsson:1986gz,
  Bengtsson:1986et,Marchesini:1987cf}
and dipole showers~\cite{Gustafson:1987rq,Andersson:1989ki,Lonnblad:1992tz}
is to link the particles involved in
the hard-scattering process, as well as in possible secondary scatterings,
to an ensemble of comparably low-energetic QCD quanta that undergo hadronisation.
\Sherpa provides two built-in parton-shower algorithms, \CSS and
\Alaric. While the former is the current default, the latter is the
development platform towards higher formal accuracy, including \NLL
and ultimately NLO QCD precision.
In addition, the legacy algorithm \Dire is also still part of the code base.

\subsubsection[LL accuracy]{LL accuracy -- \texorpdfstring{\CSS}{CSShower} and
               \texorpdfstring{\Dire}{Dire}}
\label{sec:sherpa:showers:css}

\Sherpa's default parton shower, called \CSS~\cite{Schumann:2007mg}, is
based on Catani--Seymour dipole factorisation of NLO matrix elements
for massless and massive partons~\cite{Catani:1996vz,Catani:2002hc},
as first suggested in~\cite{Nagy:2006kb}.
\CSS simulates
emissions by splitting pseudo-dipoles involving an emitter and
a spectator, each being either an initial or final-state particle,
giving four types of dipole configuration. The shower evolution variable
governing the sequence of emissions corresponds to the relative
transverse momentum between emitter and emitted particle. For 
initial-state splittings, this is the transverse momentum with 
respect to the beam.
Details on the available options for the momentum mapping
and evolution variable are described in~\cite{Hoeche:2014lxa}.  
The \CSS fully respects mass effects in the kinematics, allowing one
to consistently perform four- and five-flavour scheme
calculations~\cite{Krauss:2016orf,Krauss:2017wmx}, see also
\SecRef{sec:sherpa:mm:fusing}. The \CSS implementation provides all
necessary functionalities and extensions for combining the shower
with LO, NLO, and NNLO matrix elements, see \SecRef{sec:sherpa:mm}.
It also contains an implementation
of QED splitting functions~\cite{Hoeche:2009xc},
which, when active, co-evolve with their QCD counterparts.
A fully-fledged QED parton shower, which includes all dipoles, 
the full charge correlators and the correct collinear limit for 
photon splittings \cite{Schonherr:2017qcj}, will be included in
a future release.
The recommended option for including QED radiation
in \Sherpa, however, is the YFS soft-photon resummation,
described in section \ref{sec:sherpa:yfs}.

The \Dire parton shower~\cite{Hoche:2015sya} is an alternative
QCD evolution model within \Sherpa, and served as a test bed for
various systematic improvements.
\Dire hosts the first implementation of fully
exclusive triple-collinear and double-soft splitting functions, which are
needed for any NLL accurate parton shower \cite{Hoche:2017iem,
  Dulat:2018vuy,Gellersen:2021eci}.
It has been shown~\cite{Dasgupta:2018nvj} that the kinematic
mappings in \Dire are not \NLL safe, therefore the model has been deprecated
and is no longer actively supported. We note, however, that a solution
to the known NLL violation in \Dire at the level of the second emission
was proposed in the context of the fully differential two-loop soft
corrections~\cite{Dulat:2018vuy}. This has inspired the development
of the novel \Alaric parton-shower model, described in the following section.

\subsubsection[NLL accuracy]{NLL accuracy -- \texorpdfstring{\Alaric}{Alaric}}
\label{sec:sherpa:showers:alaric}

In addition to \Sherpa's default dipole-like parton shower
described above, a new method for
QCD evolution is implemented in the \Alaric module
\cite{Herren:2022jej}\footnote{
  Note that \Alaric is included as of release \Sherpa-3.1.
}.
It has been constructed to address the shortcomings of the
\CSS and \Dire wrt.\ their formal resummation accuracy pointed
out in~\cite{Dasgupta:2018nvj,Dasgupta:2020fwr}
and has been shown to be NLL accurate \cite{Herren:2022jej}.
The basic algorithm has since been extended to account for massive-quark
effects \cite{Assi:2023rbu}, as well as multijet merging. Further studies 
have assessed the impact of
certain uncertainties at sub-leading power that arise from
different kinematics parametrisations \cite{Hoche:2024dee}.
A unique aspect of the \Alaric method is the non-trivial dependence
of splitting functions on the azimuthal emission angle, even when spin
correlations are not included.  This allows simulation of the complete
one-loop soft radiation pattern without the need for angular ordering.
Since it is well known that kinematic edge effects play an important role
in the effective description of data by parton showers \cite{Hoeche:2017jsi},
\Alaric allows the variation of key components such as the recoil system,
evolution variable and splitting parameters in order to probe remaining
ambiguities beyond NLL accuracy. \Alaric is the prospective default
parton shower of future \Sherpa releases, once crucial features including NLO
matching and MEPS@NLO merging have been provided in full generality.
The corresponding integrated splitting functions were presented
for the massless and massive case in~\cite{Herren:2022jej}
and~\cite{Assi:2023rbu}, respectively.
In addition, \Alaric will be equipped with higher-order corrections
to the splitting functions in a fully differential form, using the
methods of~\cite{Hoche:2017iem,Dulat:2018vuy,Gellersen:2021eci}.

\begin{figure}[t!]
  \includegraphics[width=.49\textwidth]{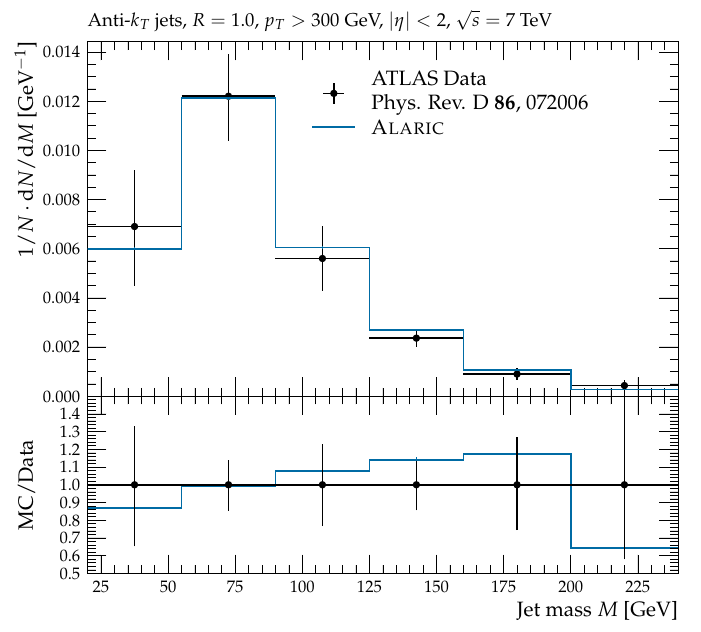}
  \includegraphics[width=.49\textwidth]{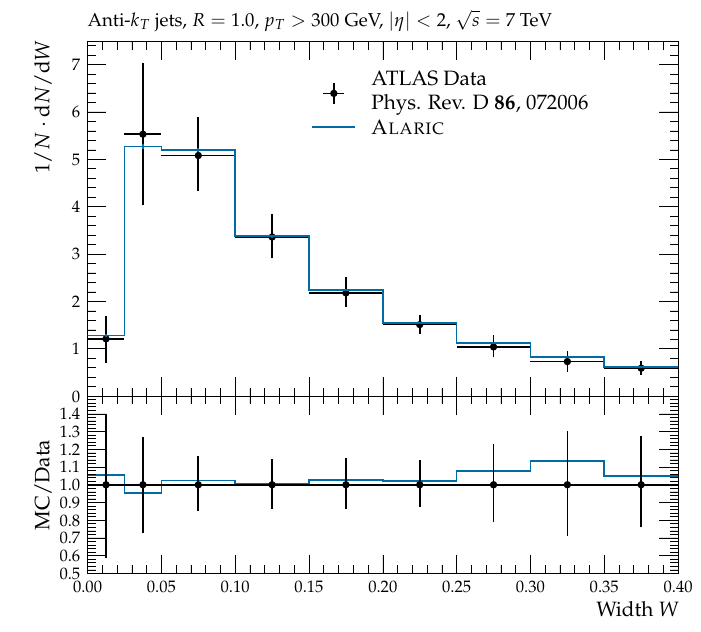}
  \caption{
    Jet mass and jet width measured by the ATLAS collaboration
    \cite{ATLAS:2012nnf} at $\sqrt{s}=7~\text{TeV}$, compared
    to predictions from \Alaric using the setup of \cite{Hoche:2024dee}.
    \label{fig:sherpa:showers:alaric:jets}
  }
\end{figure}

We illustrate the quality of the predictions achieved by the \Alaric method in
Fig.~\ref{fig:sherpa:showers:alaric:jets}, where we compare to
jet shape data measured by
ATLAS at $\sqrt{s}=7~\text{TeV}$ \cite{ATLAS:2012nnf}.
We show the jet mass and jet width, measured on high transverse
momentum jets $p_\mathrm{T}>300~\text{GeV}$,
clustered with the anti-$k_t$ algorithm with $R=1$, in the central rapidity
region $|\eta|<2$. We observe an excellent agreement, within the
uncertainty of the experimental data.
A more comprehensive overview of LHC phenomenology with \Alaric has been
presented in~\cite{Hoche:2024dee}.

\subsection{Matching and multijet merging}
\label{sec:sherpa:mm}

Going beyond simulations at the lowest order
in perturbation theory in an event generator,
whether through the inclusion of either higher-order loop
corrections or multiple-emission exact matrix elements,
inevitably introduces overlap in the perturbative description
of a scattering process between the hard matrix element
and the parton-shower evolution.
In this section we review the different options
to address this problem.

\subsubsection{NLO matching methods}
\label{sec:sherpa:mm:mcnlo}

To combine the higher-order calculations of
Sec.~\ref{sec:sherpa:me:nlo} with the parton showers of
Sec.~\ref{sec:sherpa:showers}, a number of techniques
are available in the literature.
While the most commonly used are known as
\MCatNLO \cite{Frixione:2002ik}
and \Powheg \cite{Nason:2004rx,Frixione:2007vw},
\textsc{KrkNLO} \cite{Jadach:2015mza},
\UNLOPS \cite{Lonnblad:2012ix},
and the multiplicative-accumulative matching of \cite{Nason:2021xke}
provide alternative formalisms.
In \Sherpa, the \SMCatNLO matching technique
\cite{Hoeche:2011fd,Hoeche:2012fm,Hoeche:2013mua},
an extension of the \MCatNLO method, is used.
The algorithm is implemented in complete generality, both for
massless and for massive processes.
It is the only publicly available
implementation of a matching procedure that includes the complete
colour and spin information of the matrix elements at the single-emission level
for arbitrary hard processes.
The matching has been compared against
other, publicly available implementations of \MCatNLO and \Powheg
in a number of community studies, which found the expected level of
agreement in the physics modelling between the various simulation tools
\cite{Bellm:2019yyh,Buckley:2021gfw}.
\Sherpa's implementation of the matching procedure has also been tested
in simulations of up to $W+3$~jets~\cite{Hoche:2012tae} and
$H+3$~jets~\cite{Andersen:2016qtm,ATLAS:2018hxb}
at NLO precision.
Approximate NLO EW corrections can be included for almost every process,
see Sec.\ \ref{sec:sherpa:ew}.

In some carefully validated cases it can be beneficial to disable
the full colour- and full spin-correlation treatments of the \SMCatNLO
technique and fall back to the leading-colour spin-averaged
approximations of the standard \MCatNLO matching method.
Combined with additional modifications that do not reduce the formal
accuracy, this has been shown to reduce not only the negative weight
fraction of inclusive event samples by a factor two~\cite{Danziger:2021xvr},
but also the average CPU resources required per unweighted event
by a similar factor~\cite{Bothmann:2022thx}. In a typical
$Z$+jets or $t\bar{t}$+jets NLO multijet merged calculation,
this translates into a significant reduction
of the computing time required
to further process the resulting event samples,
e.g.\ for detector simulations.

For selected processes, \Sherpa provides a matching
of \NNLO matrix elements to its default shower
using the \UNNLOPS method~\cite{Lonnblad:2012ix}.
Such a matching is particularly useful to cross-check the quality of
\Sherpa's multijet merged simulations.
Due to a large number of negative weights, however, we do not recommend
the \UNNLOPS technique to be used directly in experimental simulation
campaigns.

\subsubsection{Photoproduction and hard diffraction at NLO}
\label{sec:sherpa:mm:photoproduction_harddiffraction}

\begin{figure}
  \centering
  \includegraphics[width=.45\textwidth]{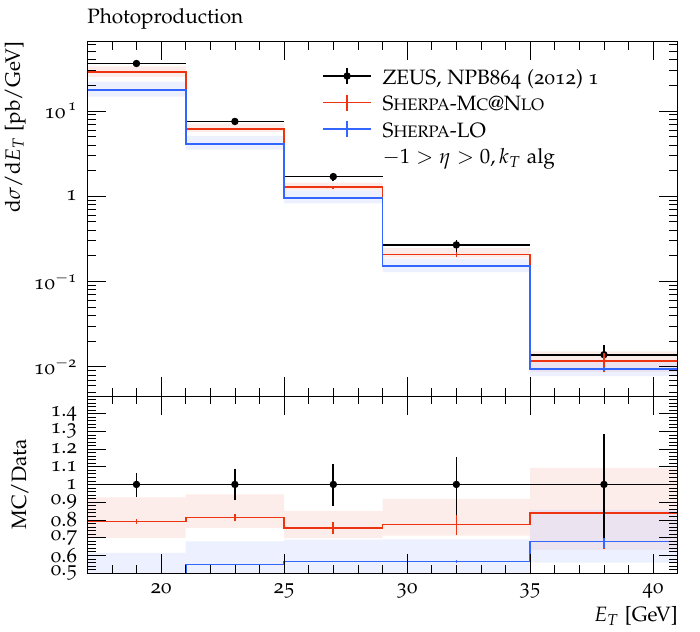}
  \includegraphics[width=.45\textwidth]{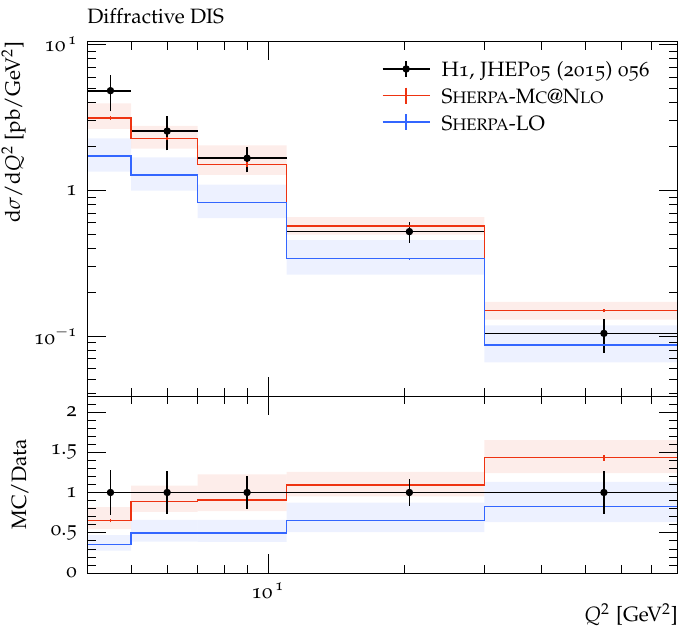}
  \caption{
    Left: Distribution of inclusive jet transverse energy for $k_T$-clustered jets
    in the pseudo-rapidity bin $-1<\eta<0$ in photoproduction,
    comparing \Sherpa \MCatNLO results with \zeus Run 2
    data~\cite{ZEUS:2012pcn}.
    Right: Distribution of photon virtuality $Q^2$ in diffractive DIS,
    comparing \Sherpa leading order (LO) and \MCatNLO results
    with \hone data~\cite{H1:2015okx}.
    \label{fig:sherpa:mm:photoproduction}
  }
\end{figure}

The NLO matching methods of the previous section have recently been
applied to photoproduction and diffractive jet production.
The photoproduction regime is characterised by beam-spectrum
photons with a small virtuality, and
gives significant contributions to the total cross section in
lepton--lepton~\cite{L3:2004ehh,OPAL:1996iet,OPAL:2007jeb} and
lepton--hadron~\cite{ZEUS:1997fwn,ZEUS:2012pcn,H1:2006rre} collisions,
and has also been studied in ion--ion collisions in the context of
Ultra Peripheral Collisions~\cite{ATLAS:2017fur,STAR:2004bzo,STAR:2019wlg,STAR:2009giy}.
In contrast to regular NLO-matched calculations, photoproduction
features a second convolution with a beam spectrum for the incident
photons, see Sec.~\ref{sec:sherpa:is:spectra}.
In particular, while the flux of quasi-real photons is typically
computed  in the Equivalent Photon Approximation, these quasi-real
photons are then resolved by means of parton-in-photon
PDFs~\cite{Schuler:1996en,Schuler:1995fk}.
These PDFs encode both non-perturbative contributions arising from
the photons' mixing with neutral vector mesons (vector meson dominance),
and perturbative contributions by means of $\gamma\to q\bar{q}$ splittings.
Both these effects have to be combined with the ``direct" interaction,
where the photon remains intact.
To match this computation at NLO, the varying beam energies as well as the QED
and QCD divergences have to be taken into account. The latter can be handled by
leveraging the combined automated QED+QCD subtraction, while for the former, the
momentum fractions that appear in the matching algorithm must be computed with
respect to the photon momentum given by the phase-space point. The implementation in
\Sherpa has been validated against data from \LEP and
\HERA experiments~\cite{Hoeche:2023gme}. An example is illustrated in
the left plot in Fig.~\ref{fig:sherpa:mm:photoproduction}, where predictions at
\MCatNLO accuracy for jet production at \HERA are compared to data from the \zeus
experiment~\cite{ZEUS:2012pcn}. We observe large corrections when comparing to LO,
which can be associated with the phase space being filled up by the real correction.
First matched NLO predictions for the \EIC have
been presented in~\cite{Meinzinger:2023xuf}.

Hard-diffractive events are defined by the beam proton
undergoing an elastic scattering or dissociation into a low-mass excitation.
Diffraction contributed about 10\% to the total cross section at
\hera~\cite{Newman:2013ada}, and will be studied at the \EIC as well~\cite{AbdulKhalek:2021gbh}.
Diffractive jet production, including both diffractive DIS and diffractive
photoproduction, has been implemented in \Sherpa and validated against \hone and
\zeus data~\cite{Krauss:2024dzj} by implementing an interface to the \hone DPDF
fit and the corresponding flux~\cite{H1:2006zyl}. The matching procedure
is the same as for photoproduction, and we show a comparison to \hone
data~\cite{H1:2015okx} for diffractive DIS in the right plot in
Fig.~\ref{fig:sherpa:mm:photoproduction}. Again, large corrections can be seen
with respect to LO, associated with filled-up phase space. These methods have been
used for predictions of diffraction at the \EIC, and are the first
fully-differential hadron-level calculations of hard diffraction at
matched NLO accuracy~\cite{Krauss:2024dzj}. Figure~\ref{fig:sherpa:mm:photoproduction}
shows a significant improvement compared to a leading-order prediction. Both the photoproduction
and hard-diffraction implementations can also be applied at hadron colliders.

\subsubsection{Multijet merging procedures}
\label{sec:sherpa:mm:mets}
\begin{table}[h!]
  \centering
  \begin{tabular}{|l|c|c|c|}
    \hline
    Process & Highest additional & References & Comments \\
            & jet mult.\ at NLO  &            & \\\hline\hline
    $e^+e^-\to\text{hadrons}$ & $4$
    & \cite{Gehrmann:2012yg,Baberuxki:2019ifp} & \\
    $e^+e^-\to e^+e^-jj$ & --
    & \cite{Hoeche:2023gme} & in photoproduction limit \\\hline
    $ep\to e+\text{jets}$ & $3$
    & \cite{Hoche:2018gti,Knobbe:2023ehi} & in DIS limit \\
    $ep\to ejj$ & --
    & \cite{Hoeche:2023gme} & in photoproduction limit \\
    $ep\to epjj$ & --
    & \cite{Krauss:2024dzj} & diffractive photoproduction/DIS \\\hline
    $pp\to\text{jets}$ & $3$
    & \cite{Baron:2020xoi} & \\\hline
    $pp\to V+\text{jets}$ & $3$
    & \cite{Hoeche:2012yf,Hoche:2012tae} & \\
    $pp\to\gamma+\text{jets}$ & $2$
    & \cite{ATLAS:2019iaa} & \\
    $pp\to H+\text{jets}$ & $3$
    & \cite{Hoeche:2014lxa,Andersen:2016qtm,ATLAS:2018hxb,
             Buckley:2021gfw}
    & ggF in HEFT, incl.\ finite $m_t$, $m_b$ \\\hline
    $pp\to Vjj$ & --
    & \cite{Lindert:2022ejn}
    & in VBF topologies \\
    $pp\to Hjj$ & --
    & \cite{Buckley:2021gfw}
    & in VBF topologies \\\hline
    $pp\to VV+\text{jets}$ & $1$
    & \cite{Cascioli:2013gfa,Brauer:2020kfv,Bothmann:2021led} & \\
    $pp\to V\gamma+\text{jets}$ & $1$
    & \cite{Krause:2017nxq} & \\
    $pp\to\gamma\gamma+\text{jet}$ & $1$
    & \cite{Siegert:2016bre,ATLAS:2021mbt} & \\
    $pp\to VH+\text{jets}$ & $1$
    & \cite{Hoeche:2014rya,Goncalves:2016bkl,Goncalves:2015mfa} & \\
    $pp\to HH$ & --
    & \cite{Jones:2017giv}
    & full loop-induced, incl.\ finite $m_t$ \\\hline
    $pp\to VVjj$ & --
    & \cite{Denner:2024ufg}
    & $t$- (VBS), $s$-ch.\;(semilep.\ $VVV$) \\\hline
    $pp\to VVV+\text{jets}$ & $1$
    & \cite{Hoeche:2014rya} & \\
    $pp\to VV\gamma$ & --
    & \cite{ATLAS:2023zkw} & \\
    $pp\to V\gamma\gamma$ & --
    & \cite{ATLAS:2022wmu,ATLAS:2023avk} & \\
    $pp\to\gamma\gamma\gamma+\text{jets}$ & $1$
    & \cite{ATLAS:2021lsc} & \\\hline
    $pp\to\gamma\gamma\gamma\gamma$ & --
    & \cite{ATLAS:2021lsc} & \\\hline
    $pp\to tj$ & --
    & \cite{Bothmann:2017jfv}
    & $t$- and $s$-channel \\
    $pp\to tW$ & --
    & \cite{Bothmann:2017jfv}
    & using diagram removal (DR) \\
    $pp\to t\bar{t}+\text{jets}$ & $2$
    & \cite{Hoeche:2014qda,Gutschow:2018tuk} & \\
    $pp\to t\bar{t}V+\text{jets}$ & $1$
    & \cite{ATLAS:2020esn} & \\
    $pp\to t\bar{t}\gamma+\text{jets}$ & $1$
    & \cite{ATLAS:2020esn} & \\
    $pp\to t\bar{t}b\bar{b}$ & --
    &\cite{Cascioli:2013era}
    & full $m_b$ dependence \\
    $pp\to t\bar{t}t\bar{t}$ & --
    & \cite{ATLAS:2020esn} & \\\hline\hline
    $pp\to V+\text{HF}$ & $2$
    & \cite{Hoche:2019ncc}
    & in fusing scheme, see Sec. \ref{sec:sherpa:mm:fusing} \\
    $pp\to t\bar{t}+\text{HF}$ & $2$
    &\cite{Ferencz:2024pay}
    & in fusing scheme, see Sec. \ref{sec:sherpa:mm:fusing} \\\hline
  \end{tabular}
  \caption{
    Usage of \Sherpa's matching and merging capabilities in
    the literature.
    $V$ generically denotes the off-shell production of a
    $W$ or $Z$ boson, decaying leptonically.
    Maximal jet multiplicities at NLO largely depend on the
    hardware available, the stated multiplicities correspond
    to the largest one that was used in the cited references
    and not a limitation in principle.
    In almost all cases additional multiplicities were merged
    on top of the quoted NLO multiplicities using the techniques
    of \cite{Hoche:2010kg,Hoeche:2012yf,Hoeche:2014rya}.
    \label{tab:sherpa:mm:procs}
  }
\end{table}

One of the strengths of the physics modelling with \Sherpa is the
control over both the matrix-element calculation and the parton-shower
simulation in one single framework.
This facilitates the implemention of techniques to systematically improve
the simulation of jet production. Such methods include
multijet merging at leading-order~\cite{Catani:2001cc,
  Lonnblad:2001iq,Krauss:2002up,Alwall:2007fs,Hoeche:2009rj}
and at next-to-leading order in QCD~\cite{Lonnblad:2011xx,
  Hoeche:2012yf,Gehrmann:2012yg,Frederix:2012ps,
  Lonnblad:2012ix,Platzer:2012bs}.
\Sherpa implements the leading-order merging
methods described in~\cite{Hoeche:2009rj,Lonnblad:2001iq}, and
the next-to-leading order techniques from~\cite{Hoche:2010kg,
 Hoeche:2012yf,Gehrmann:2012yg,Hoeche:2014rya}.
They incorporate leading-order or next-to-leading
order calculations with sufficiently separated parton-level jets into
parton-shower predictions, while maintaining both the logarithmic
accuracy of the parton shower resummation and the fixed-order accuracy
of the hard matrix elements.

\Sherpa's implementation of the matching and merging procedures at
\SMCatNLO and \MEPSatNLO precision has been studied in great detail
for a large number of processes, see Tab.\ \ref{tab:sherpa:mm:procs}
for an overview.
In many cases, additional jet multiplicities are merged at LO accuracy
on top of the highest multiplicity at NLO,
and approximate EW corrections can be included for almost every
process, see Sec.\ \ref{sec:sherpa:ew} and
Fig.\ \ref{fig:sherpa:ew:ewsud:ewapprox}.
The availability of such corrections is only limited by computational
resources and has been accomplished, for example, for
$W+\{\le 9\}\,\text{jets}$~\cite{Hoche:2019flt},
$H+\{\le 7\}\,\text{jets}$~\cite{Bothmann:2023ozs} and -- in
combination with \Vincia\ -- for VBF Higgs production
${}+\{\le 4\}\,\text{jets}$~\cite{Hoche:2021mkv}.
It has also been used to compute observables in neutral-current DIS
at HERA with up to 5 jets in the final state~\cite{Carli:2009cg}.
Further, the multijet merging technology has been extended to
loop-induced processes at LO accuracy, \MEPSatLOOP, in
\cite{Cascioli:2013gfa,Goncalves:2016bkl,
  Goncalves:2015mfa,Bothmann:2021led}, which is of particular
relevance for diboson processes at the \LHC.

\subsubsection[Heavy-flavour matching]{Combining four- and five-flavour calculations: fusing}
\label{sec:sherpa:mm:fusing}

The merging method described in Sec.~\ref{sec:sherpa:mm:mets}
is a well established algorithm to describe multijet observables with
NLO accuracy within an inclusive calculation.
Originally, the algorithm was limited to the case of massless
quarks in the hard scattering matrix element.
Although this is a useful approximation in cases where the
quark mass is small compared to the typical scales of the observables
in question, quark masses often play an important role.
Using a scheme with five active quark flavours (5FS) does not
allow the use of fixed-order matrix elements to correct the parton-shower
resummation in the regions of collinear $g\to b\bar{b}$ splittings.
Conversely a scheme with 4 active flavours, (4FS) and massive $b$-quarks
can provide consistent fixed-order predictions in this region, but lacks
the resummation of $b$-jet production at high energies.

A four-flavour scheme simulation of processes involving $b$-quarks creates
additional complications.
Experimental analyses involving heavy flavour final states rely on
precise simulations involving light jets since these may fake
$b$-quark-initiated jets in detectors.
Consequently, both the 4FS and 5FS simulations have to be used
simultaneously and their overlap needs to be removed.
To overcome this problem, the ``fusing'' approach~\cite{Hoche:2019ncc}
has been developed.
It rigorously incorporates matrix elements with massive $b$-quarks
into the existing merged predictions while keeping all resummation
features and avoiding double counting.
In this approach, the hardest heavy-flavour emissions stem from 4FS
matrix-element calculations supplemented by Sudakov form factors
(``direct'' component), whereas softer $b$-quarks and light jets
are still produced by the 5FS multijet matrix elements and the
parton shower (``fragmentation'' component).

Similar to the FONLL method~\cite{Forte:2016sja}, we need appropriate
counter-terms to treat the 4FS matrix elements within a prediction using
5FS PDFs and a 5FS running $\alpha_s$.
In fusing calculations with \Sherpa, these are provided as event weights, such as to make
the massless multijet event generation usable both inclusively or as a
fragmentation component.
Applications of \Sherpa's fusing implementation have been published
for $Z$ + heavy flavour~\cite{Hoche:2019ncc} and
$t\bar{t}$ + heavy flavour~\cite{Ferencz:2024pay} final states.

\subsection{Approximate electroweak corrections}
\label{sec:sherpa:ew}

Full NLO EW calculations are available with \Sherpa
for fixed-order calculations only.
Nonetheless, approximate higher-order electroweak corrections
can be included in particle-level event generation, including
parton showering and hadronisation, using the EW virtual (\EWvirt)
scheme, or alternatively through EW Sudakov
(\EWsud) logarithms.
Formally, both correction schemes evaluate the same
logarithms, up to NLL, that dominate the electroweak
corrections in the high-energy regime, but they differ
in the inclusion of finite terms.
Both schemes offer the possibility to exponentiate
the corrections, resulting in an approximate resummation
that estimates the electroweak corrections beyond
$\order(\alpha)$ \cite{Denner:2024yut}.

In practical terms, the two schemes differ in their availability
and computational overhead.
To fully benefit from their respective advantages, samples with
\EWvirt\ and \EWsud\ corrections can be combined \emph{a posteriori}~\cite{Bothmann:2021led}.
The effect of the corrections is usually given via alternative event weights
in the output event sample, see Sec.~\ref{sec:otf},
which allows to compare the corrected predictions with the
baseline (QCD only) result. In the following, we discuss
the details of the two approximation schemes.

\subsubsection{EW virtual approximation}
\label{sec:sherpa:ew:ewvirt}

The electroweak virtual approximation (\EWvirt), introduced in
\cite{Kallweit:2015dum,Gutschow:2018tuk},
calculates approximate electroweak and subleading
mixed QCD-EW corrections which can be incorporated in \MCatNLO-matched
simulations, including \MEPSatNLO multijet merged ones.
It supplements the \MCatNLO
$\Bbar$-function with an EW correction built by using
exact NLO EW renormalised virtual corrections as well as
approximated NLO EW real-emission corrections integrated
over their real-emission phase space, either in an additive,
multiplicative, or exponentiated manner
\cite{Brauer:2020kfv,Bothmann:2021led}.
The correct electroweak input-parameter and renormalisation scheme
dependence is preserved by
construction~\cite{Gutschow:2018tuk,Bothmann:2021led}.

This approximation reproduces the exact NLO EW corrections
in regions with large momentum transfers that are dominated
by virtual weak-boson exchanges and renormalisation corrections.
The integrated real-photon emission part of the electroweak
correction, of particular importance for leptons in the final state,
can reliably be recovered by including a soft-photon resummation
\cite{Gutschow:2020cug}, see Sec. \ref{sec:sherpa:yfs}.

\subsubsection{EW Sudakov approximation}
\label{sec:sherpa:ew:ewsud}

The electroweak Sudakov approximation (\EWsud) comprises
the leading-logarithmic corrections
induced by EW higher orders 
in the strict high-energy limit.
At one-loop, these have been derived
by Denner and Pozzorini~\cite{Denner:2000jv,Denner:2001gw},
and implemented in a fully automated and process independent way
for the first time
in \Sherpa~\cite{Bothmann:2020sxm,Bothmann:2021led}.
Similar implementations are also available in
\aMCatNLO~\cite{Pagani:2021vyk,Pagani:2023wgc} and
\OpenLoops~\cite{Lindert:2023fcu}.
In the high energy limit, the leading higher-order
corrections factorise and can be computed by taking
ratios of tree-level diagrams, which, in turn, can be evaluated
using \Sherpa's internal ME generator \Comix.
The strength of this approach is twofold: not only does it
reproduce the leading and next-to-leading behaviour of higher-order
EW corrections, but it also allows the user to combine this prediction
with the existing QCD
technology, such as parton showering and multijet merging.
For details on how to enable \Sherpa's EW Sudakov calculations, see App.~\ref{app:install}.

In practical terms, which energy range corresponds to the high-energy regime depends on the process and
the observable, and can be controlled by the user.
Intermediate resonances, such as
$Z\to\ell^+\ell^-$ within $pp\to\ell^+\ell^-+\;\text{jets}$
processes,
formally spoil the high-energy limit, as they are associated
with moderate scales of the order of the resonant particle's mass.
The implementation in \Sherpa disentangles resonant (associated with
scales of the order of the resonant mass) and non-resonant (potentially
associated with resonance-independent large scales) topologies
using the algorithm described in Sec.\ \ref{sec:sherpa:yfs:res}.
If a resonant decay has been identified and clustered,
EW Sudakov corrections are computed for the clustered process.

In addition, various subleading contributions can be included to
extend the range of validity of the approximation.
To be precise, we allow for both the inclusion of logarithms of ratios
of invariants which are not formally large, as well as the inclusion
of purely imaginary phases appearing when considering $2\to n$ processes
with $n>2$. These two types of terms were shown to be non-negligible
in some cases \cite{Pagani:2021vyk}.
While logarithms of ratios of intermediate invariants
are not strictly controlled by the \EWsud\ approximation,
they can be used as either a way to estimate the uncertainty of the
approximation or as a way to extend it to lower energies.
It is thus advised to include them when comparing to the full NLO EW
corrections. On the other hand, the purely imaginary phases should always
be included, which is the default.

\begin{figure}[t!]
  \begin{center}
    \includegraphics[width=0.32\textwidth]{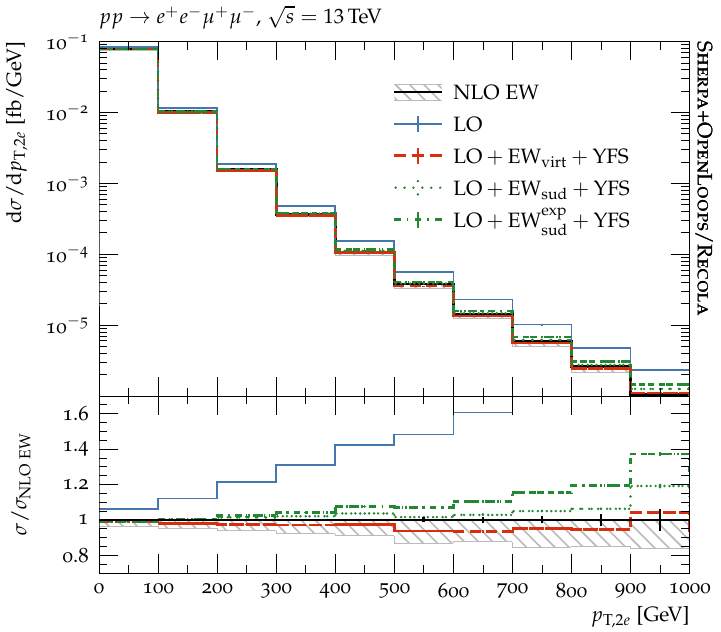}
    \includegraphics[width=0.32\textwidth]{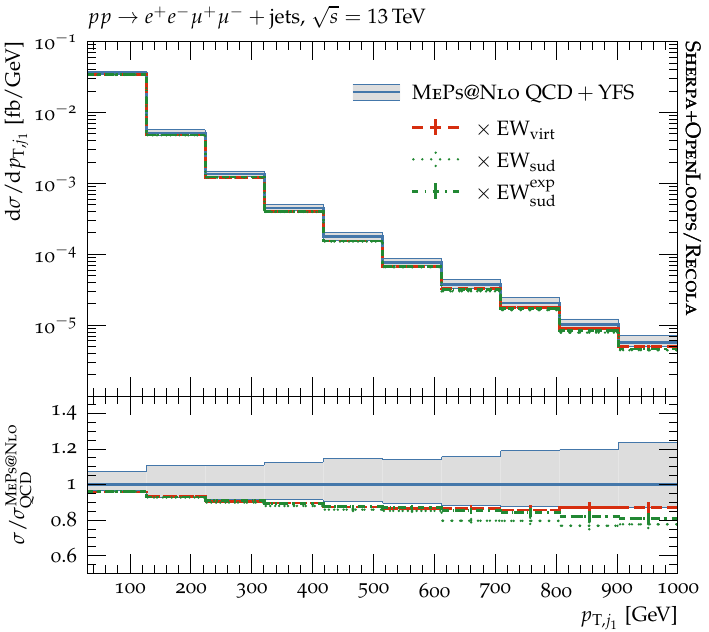}
    \includegraphics[width=0.32\textwidth]{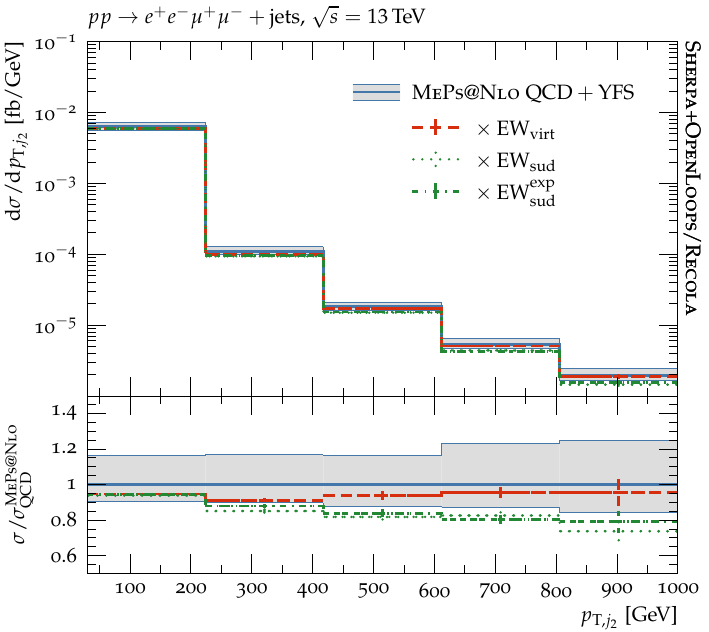}
  \end{center}
  \caption{
    Effects of higher-order EW corrections and their approximations for
    four-lepton production in the setup of \cite{Bothmann:2021led}.
    Left: 
    We show a comparison between exact NLO EW corrections, the \EWvirt\
    and the \EWsud\ approximations at fixed order in QCD.
    Both approximations reproduce the EW corrections in the high-energy
    limit well.
    Centre and right:
    We show a \MEPSatNLO QCD multijet-merged calculation including either
    the \EWvirt\ or \EWsud\ approximations for higher-order EW corrections.
    It is important to note that, as seen in the right panel, as the
    implementation of both approximations differs in the treatment of
    the lower-multiplicity \MCatNLO $\boldsymbol{\mathbb{H}}$- and
    higher-multiplicity LO-events, the two results may differ when these
    event types have a sizable contribution.
    \label{fig:sherpa:ew:ewsud:ewapprox}
  }
\end{figure}

In the left panel of Fig.~\ref{fig:sherpa:ew:ewsud:ewapprox} we
compare the fixed-order exact NLO EW to both the \EWvirt\ and \EWsud\
approximations, including final-state QED radiation corrections in
the YFS soft-photon resummation of Sec.~\ref{sec:sherpa:yfs:dec}
in the setup described in \cite{Bothmann:2021led}.
We find excellent agreement that extends well beyond the strict
high-energy limit.
The na\"ively exponentiated Sudakov logarithms can be used to estimate
the size of the $\order(\alpha^2)$ corrections.
Unlike the fixed-order NLO EW calculation, both approximations allow
their direct incorporation in the QCD parton-shower-matched and
multijet-merged machinery of \Sherpa.
We show their impact in the centre and right panels of
Fig.\ \ref{fig:sherpa:ew:ewsud:ewapprox}, and find that both
approximations agree well with each other, indicating a robust
prediction.
The right panel, however, also exhibits a limitation of the
computationally intensive \EWvirt\ approximation -- it is often not
available for the higher multiplicities.
Here, the \EWsud\ approximation, being entirely based on tree-level
diagrams, can
provide support to all relevant multiplicities and calculate
the corresponding EW corrections.

\subsection{QED radiative corrections}
\label{sec:sherpa:yfs}

\Sherpa calculates higher-order QED corrections using the
soft-photon resummation of Yennie, Frautschi and Suura (YFS)
\cite{Yennie:1961ad}.
The YFS formalism uses the universal structure of real and
virtual soft-photon radiation, constructing an all-orders
approximation that retains all relevant mass effects.
Two implementations exist within the \Sherpa framework,
with \Photons \cite{Schonherr:2008av},
detailed in Sec.\ \ref{sec:sherpa:yfs:dec},
focusing on higher-order corrections to particle decays,
while \YFS \cite{Krauss:2022ajk}, see Sec.\ \ref{sec:sherpa:yfs:ee},
implements corrections to incident leptons as well.

\subsubsection{Soft-photon resummation for particle decays}
\label{sec:sherpa:yfs:dec}

QED final-state radiation is implemented in the \Photons
\cite{Schonherr:2008av} module of \Sherpa, based on the
YFS soft-photon resummation algorithm \cite{Yennie:1961ad}.
To improve its accuracy away from the soft-photon limit,
i.e.\ for hard photon radiation, universal spin-dependent
hard collinear emission corrections are applied by default.
For dedicated decays, $\tau\to\ell\nu_\ell\nu_\tau$ and
some hadron decays \cite{Schonherr:2008av,Alioli:2016fum,
  Bernlochner:2010fc}, exact NLO QED corrections are available.
NLO QED $+$ NLO EW and NNLO QED $+$ NLO EW corrections
are implemented for $W\to\ell\nu$ as well as $Z\to\ell\ell$
and $h\to\ell\ell$ \cite{Krauss:2018djz}, respectively,
where the highest precision is needed.
Care has to be taken, however, when final states contain
multiple competing resonances.
This is the topic of Sec.\ \ref{sec:sherpa:yfs:res}.

In order to not interfere with the QCD parton showering,
YFS soft-photon resummed higher-order QED corrections are
only applied to decay processes that do not involve coloured
particles.
An alternative prescription using a collinear factorisation
picture exists in the form of a QED parton shower \cite{Hoeche:2009xc}
in the \CSS, see Sec.\ \ref{sec:sherpa:showers:css}.
While this method allows for co-evolving QED and QCD splitting
functions, this co-evolution needs only to be considered if
QED emissions off quarks are relevant.
In addition, currently it lacks both the soft-photon
coherence inherent in the YFS soft-photon resummation
and the dedicated higher-order corrections.

\subsubsection{Resonance identification}
\label{sec:sherpa:yfs:res}

Complex final states often contain (multiple) internal
resonances.
Thus, additional care is required when effecting
higher-order QED corrections in order to preserve
these structures.
To this end, \Sherpa employs a universal resonance
identification algorithm~\cite{Kallweit:2017khh}.
First, all possible resonances occurring in the chosen
model are identified by scanning the final state of
a scattering process for possible recombinations
into resonant states.
Second, all such combinations are ordered in increasing
distance from the nominal on-shell resonance in units of
its width, using
$\Delta=|m_\text{kin}^\text{inv}-m_\text{res}|/\Gamma_\text{res}$.
Starting with the recombination with the smallest $\Delta$,
resonances are identified as present in the current
configuration, and recombination candidates with
$\Delta>\Delta_\text{thr}$ are classified as non-resonant.
Identified resonances are treated separately,
ensuring that no momentum is transferred outside a
resonant-decay system through the application of
higher-order QED corrections.
Finally, all non-resonantly produced final states are
corrected using the universal YFS soft-photon resummation
together with the universal hard-collinear corrections.

\subsubsection{Photon-splitting corrections}
\label{sec:sherpa:yfs:split}

\begin{figure}
  \centering
  \includegraphics[width=0.47\textwidth]{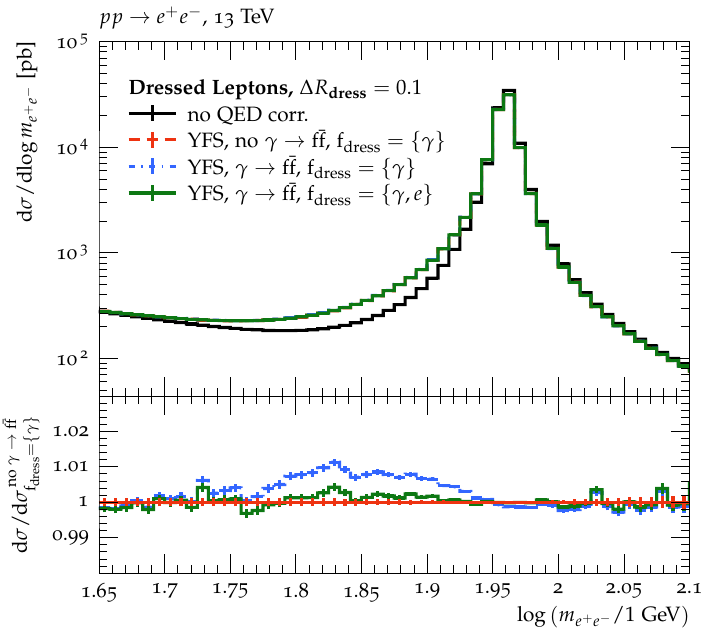}
  \hfill
  \includegraphics[width=0.47\textwidth]{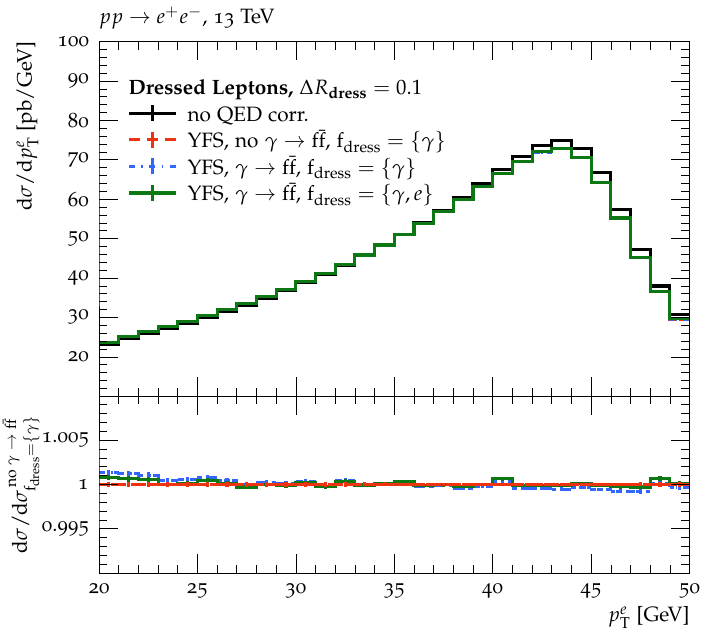}
  \caption{
    The dressed dilepton invariant mass $m_{e^+e^-}$ (left)
    and electron transverse momentum $p_{\mathrm{T}}^{e}$ (right)
    in Drell--Yan production as described at leading order (black),
    by the YFS soft-photon
    resummation at YFS+NLO QED+NLO EW accuracy without photon-splitting
    corrections (red) or additionally resolving the
    photons further into pairs of charged particles (blue and green),
    including mitigating effects of adequately improved dressing strategies.
    For details see \cite{Flower:2022iew}.
    \label{fig:yfs:photonsplittings}
  }
\end{figure}

A feature of higher-order QED effects absent
in the YFS soft-photon resummation are photon-splitting
corrections.
To account for these effects, \Sherpa's \Photons module
has been extended with \PhotonSplitter \cite{Flower:2022iew}.
Despite being of relative $\order(\alpha^2)$ and only
enhanced by single collinear logarithms compared
to the Born configuration, these corrections can
play an appreciable role when hard primary photons
split into pairs of light charged particles.
The possibility for a photon 
to be replaced by an electron or muon pair also has
important consequences for lepton dressing. 
This module therefore 
allows photons to split into electrons, muons and/or light charged
hadrons (the relevant QCD degrees of freedom at this energy) in a
parton-shower-like collinear evolution, starting from the
primary photon ensemble generated by YFS.
Figure \ref{fig:yfs:photonsplittings} shows the impact of 
photon-splitting corrections on Drell--Yan electron-pair production
at the LHC, including their mitigation using a modified lepton
dressing procedure.
For details see \cite{Flower:2022iew}.

\subsubsection{Soft-photon resummation for \texorpdfstring
 {$\text{e}^+\text{e}^-$}{ee} colliders}
\label{sec:sherpa:yfs:ee}

At lepton--lepton colliders, an important source of uncertainty
which must be included
is the modelling of photon emissions in the initial state. Such
emissions can spoil the perturbative expansion as they lead to
potentially large logarithms, which arise from the emission of soft
and/or collinear photons. To ensure the stability of theory
predictions, and to reduce the overall uncertainty, these
logarithms must be resummed. \Sherpa currently supports two
different approaches to the treatment of QED ISR. The first approach
uses the electron structure function, which is a solution
of the DGLAP evolution equations~\cite{Altarelli:1977zs,
  Gribov:1972ri,Lipatov:1974qm,Dokshitzer:1977sg}
using LO initial conditions \cite{Skrzypek:1990qs},
see Sec.\ \ref{sec:sherpa:is:pdfs}.
This analytic approach can be combined with a
traditional parton shower, extended to QED~\cite{Hoeche:2009xc}, to
generate exclusive kinematic distributions for the collinear photons.
In the second approach, we use the YFS
theorem~\cite{Yennie:1961ad} to resum the emission of soft photons
to all orders in $\alpha$. In this method, the photon emissions are
considered in a fully differential form where the photons are
explicitly created and the treatment of their phase space is exact.

\begin{figure}[t!]
  \centering
  \includegraphics[width=0.47\textwidth]{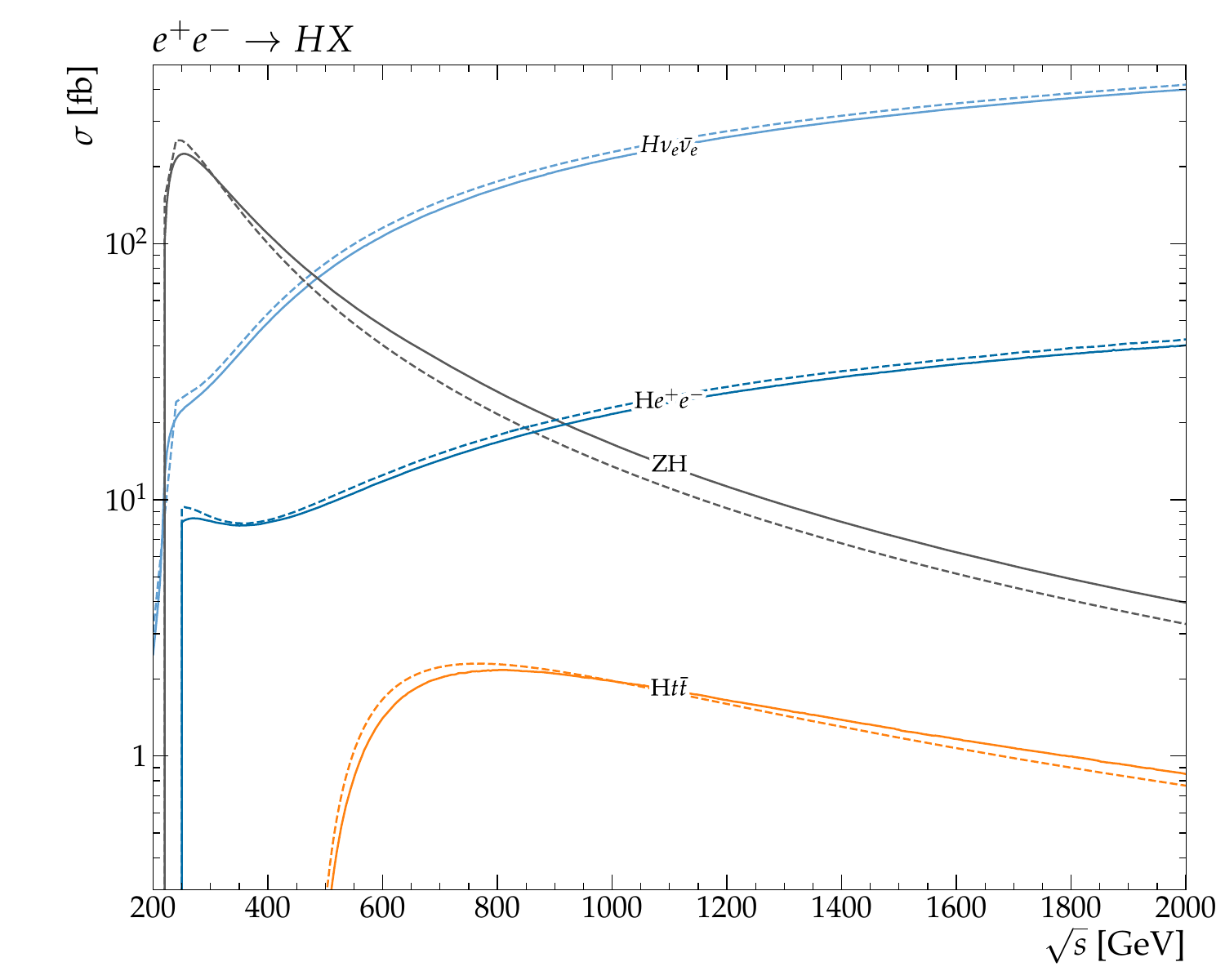}
  \caption{Total cross section for $\ee \rightarrow HX$ at the Born level (dashed) and with ISR (solid) corrections included.}
  \label{fig:yfs:xs}
\end{figure}

The YFS approach was originally implemented in process-specific Monte Carlo
tools~\cite{Jadach:2001mp,Jadach:1999vf,Jadach:1991by,Jadach:2022mbe} that
were predominantly used for the LEP physics programme and were
crucial for its electroweak precision measurements~\cite{ALEPH:2005ab}.
Despite these tools still being available and being further developed,
having a process-independent Monte Carlo event generator
based on the YFS formalism is highly desirable.
Hence, while such a YFS-based implementation for QED FSR
for arbitrary final states was available for some time,
see Sec.\ \ref{sec:sherpa:yfs:dec}, this framework has been
extended to include QED ISR for initial-state leptons in
a process-independent fashion in \cite{Krauss:2022ajk}.
Therein, corrections related to hard collinear photon emissions
are available in a leading-logarithmic
formulation up to $\order(\alpha^3L^3)$.
An automated calculation of the complete fixed-order corrections
at full NLO EW accuracy is envisioned for future \Sherpa versions.
As an example application, we present a number of Higgs production
cross sections in Fig.~\ref{fig:yfs:xs} as a function of the
collider centre-of-mass energy.


\subsection{Multi-parton interactions}
\label{sec:sherpa:mi}

Multi-parton interactions (MPIs) have long been established as an important
physics model for collider event simulation which ensures that particle
production, and its scaling behaviour with the hadronic centre-of-mass energy,
are correctly described~\cite{Sjostrand:1987su}.
The \Sherpa model for this effect builds on the original Sj{\"o}strand--van~Zijl
approach~\cite{Sjostrand:1987su}. While newer versions of \Pythia integrate
the secondary scatterings into the initial-state parton
evolution~\cite{Sjostrand:2004pf,Sjostrand:2004ef}, and add final-state
parton showering as well as hadronic rescattering effects~\cite{Corke:2009tk},
\Sherpa treats the scatterings as independent, apart from momentum-conserving
and colour reconnection effects. The perturbatively computable (regularised)
parton-level cross section is normalised to the non-diffractive hadron-level
cross section and exponentiated in an expression similar to a Sudakov factor.
This expression is used to generate a sequence of
secondary interactions, which
individually undergo parton-shower evolution in the initial and final states.
The production of the secondary interactions is integrated into the multijet
merging algorithm used to describe the hard scattering~\cite{Alekhin:2005dx}.

As a new feature in version 3, \Sherpa is now capable of modelling
multiple scattering effects in processes with resolved photons. It
is also now possible to veto additional scatters between beam particles,
which is useful in measurements of large rapidity gaps and diffractive
jet production~\cite{Krauss:2024dzj}, for example.
In this way, survival probabilities can be computed
as the probability for no further scatters to occur,
akin to their estimation in \cite{Kaidalov:2001iz,Kaidalov:2003xf,Kaidalov:2009fp}.

\subsection{Hadronisation}
\label{sec:sherpa:had}

The transition from the region where QCD partons are asymptotically free
to the regime where they are bound into hadrons is the traditional
domain of Monte Carlo event generators. This region cannot be described
using first-principles calculations due to the complications of the
transition from the perturbative to the non-perturbative regime
of the theory. A number of models, which are rooted in a few theoretically
calculable quantities and experimental observations, are therefore
employed. The ones used in \Sherpa are described in this section.

\subsubsection{Beam remnant handling}
\label{sec:sherpa:had:beamremnant}

In collisions involving hadronic initial states, one or more partons
are typically extracted from the incoming beam particles. This is modelled
via the perturbatively described hard scattering and the perturbatively
modelled, but softer, multi-parton interactions, see
Secs.~\ref{sec:sherpa:me} and \ref{sec:sherpa:mi}.
Both types of calculations are dressed with the parton
showers of Sec.~\ref{sec:sherpa:showers} which ultimately terminate at low
transverse-momentum scales of around 1 GeV.

The breakup of incoming hadrons (and other beam particles with
substructure), and the formation of the beam
remnants, begins with a list of shower initiators from the above
perturbative descriptions.
The breakup of incoming particles is guided by flavour compensation,
colour compensation,
longitudinal momentum distribution according to the PDFs,
and transverse momentum distribution according to a 
polynomially-suppressed Gaussian.
In addition, the valence structure of the beam particle is
respected; in particular, baryons are considered to constitute
a valence quark-diquark pair.
Details on this model can be found in
App.~\ref{sec:sherpa:had:beamremnant_app}.

\subsubsection{Colour reconnections}
\label{sec:sherpa:had:colrec}

The partons produced by the hard scattering and multi-parton interactions,
their
subsequent parton showers, and the break-ups of the beam remnants all turn into so-called
primary hadrons. These primary hadrons have their colours assigned in the large-$N_c$ limit.
This means that effectively, at this stage, every colour in the parton
ensemble has exactly one anti-colour and vice versa.  The difference between
the $N_c\to\infty$ approximation and the actual value $N_c=3$, and the
existence of potential soft non-perturbative gluon interactions (so-called
``gluers''~\cite{Dokshitzer:1991wu}), suggests that 
this model can be improved by a rearrangement of the original 
colour assignments of the partons.
In particular,
the non-perturbative nature of the long-range strong interactions introduces
significant liberty in the modelling of such colour reconnections (CRs).

The model in \Sherpa incorporates various ideas from earlier literature~\cite{
  Sjostrand:1987su,Skands:2007zg,Sjostrand:1993hi,Khoze:1994fu,Khoze:1999up,
  Lonnblad:2023stc,Sjostrand:2004pf,Webber:1997iw,Gieseke:2012ft,
  Gieseke:2017clv,Gieseke:2018gff,Bellm:2019wrh,Platzer:2022jny}.  Assuming
that $N$ different colours (matched by exactly $N$ anti-colours) emerge from
the scatters, showers, and beam remnants, the model checks $N^2$ times for
possible colour reassignments. In each such attempt, two colours $i$ and $j$
are randomly chosen and the relative distances $d$ of the corresponding
parton pairs $\langle i\bar{i}\rangle$ and $\langle j\bar{j}\rangle$ and the
swapped pairs $\langle i\bar{j}\rangle$ and $\langle j\bar{i}\rangle$
is calculated. For each pair $kl$, the distance in momentum space is given by
$d_{kl} = \log(1+(p_kp_l-m_km_l)/Q_0^2)$.
Based on this distance, a colour reassignment happens with a
probability given by
$P_{\mathrm{swap}}(i\leftrightarrow j) = R_c\{1-\exp[\eta_Q\,
    (d_{i\bar{i}}+d_{j\bar{j}}-d_{i\bar{j}}-d_{j\bar{i}})
    ]\}$.
In these equations, $Q_0$ is the infrared scale in the distance measure,
$R_c$ is the colour factor, and $\eta_Q$ is the weight of the distances
in the exponential. The (untuned) defaults are $Q_0\approx 1$ GeV,
$R_c\approx 1/9$, and $\eta_Q\approx 0.1$.

\subsubsection{Cluster hadronisation}
\label{sec:sherpa:had:cluster}

\Sherpa's default hadronisation model~\cite{Winter:2003tt,Chahal:2022rid}
is based on the twin concepts of local parton-hadron duality (LPHD)~\cite{Azimov:1984np}
and preconfinement~\cite{Amati:1979fg,Bassetto:1979vy,Marchesini:1980cr},
which postulate that the flow of quantum numbers, momenta, and energies at
the hadron level closely follows their counterparts at the parton level, and
that the transition from partons to hadrons proceeds through the formation
of colourless clusters with a {\em perturbatively} calculable mass spectrum.
The first realisation of the LPHD paradigm in the form of the Feynman--Field
independent fragmentation model~\cite{Field:1977fa} suffered from a range of
theoretical issues, among them lack of Lorentz invariance. These issues were
ultimately resolved by the concept of preconfinement, which
introduced the intermediate step of a non-perturbative splitting of gluons
into quark--\-anti-quark pairs~\cite{Hoyer:1979ta,Ali:1979em} and the
subsequent formation of colour-neutral clusters and their decay into
hadrons~\cite{Field:1982dg}.  The first cluster fragmentation model embedded
in a widely used event generator, \Herwig~\cite{Corcella:2000bw,Bewick:2023tfi},
was introduced shortly afterwards~\cite{Marchesini:1983bm,Webber:1983if}
and is continuously improved~\cite{Bellm:2015jjp,Masouminia:2023zhb,Hoang:2024zwl}.

\Sherpa's cluster model~\cite{Winter:2003tt,Chahal:2022rid}
differs from the \Herwig model in multiple ways.
Firstly, \Sherpa does not introduce non-perturbative gluon masses but rather
keeps the gluons massless, and allows not only light up and down quarks,
but also strange quarks and diquarks as decay products of their forced splitting
at the onset of hadronisation. This results in the presence of baryonic
clusters throughout the hadronisation process.  Secondly, the fission of
relatively heavy clusters into two lighter ones is not parametrised by
selecting masses of the latter in (typically) isotropic decays; instead,
\Sherpa distributes light-cone momentum fractions of the new clusters with
respect to the constituents of the decaying cluster according to
``fragmentation functions'', and applies a Gaussian transverse
momentum distribution to the decay kinematics. Finally, there are also differences in
the treatment of binary cluster decays into primary hadrons, including
kinematics and the way the hadron species are selected. Overall, despite the footing
of both models in the same underlying physics assumptions, these differences
result in a manifestly different hadronisation model with different sets of critical
parameters that need to be tuned to data.

\subsubsection[Interface to \Pythia]{Alternative hadronisation model via interface to Pythia 8}
\label{sec:sherpa:had:lund}

An alternative approach to hadronisation is provided by the string picture
of QCD~\cite{Artru:1974hr,Bowler:1981sb,Andersson:1983ia} which builds on
the observation of a potential between colour charges that increases
linearly with their position-space distance\footnote{
Approaches to combine the respective benefits of string and cluster
hadronisation~\cite{Gottschalk:1983fm,Gottschalk:1986bv} have not been
followed up in the past decades, arguably because they have not been provided
in the form of a widely used event generator.}.
The potential is represented by one-dimensional strings, which carry a finite
energy density per unit length.  As the strings are ``stretched'' with the
partons moving away from each other, their stored energy allows the dynamic
creation of quark--\-anti-quark pairs, akin to the Schwinger mechanism in
QED~\cite{Sauter:1931zz,Schwinger:1951nm}, essentially breaking the string into smaller, lighter fragments.  Successive refinements, including
symmetrising the string fragmentation function whilst respecting
Lorentz-invariance and causality~\cite{Andersson:1983jt}, extending the model
beyond the simple case of strings spanned by quark--\-anti-quark pairs and
including the effect of gluons~\cite{Andersson:1979ij,Sjostrand:1984ic}, and
the modelling of baryon production~\cite{Andersson:1981ce}, have contributed
to establishing the Lund model~\cite{Sjostrand:1982fn,Andersson:1997xwk} as
probably the most phenomenologically successful hadronisation model.

\Sherpa provides an interface to the Lund model implemented in
\Pythia8~\cite{Sjostrand:2006za,Bierlich:2022pfr}.
See App.~\ref{app:install} for instructions to enable it.
The support for both cluster hadronisation and string fragmentation available
in \Sherpa allows for direct comparisons of hadronisation models and their
observable effects, due to the identical treatment of the perturbative
phase of the event.

\subsection{Hadron decays}
\label{sec:sherpa:decays}

The \Sherpa framework contains a built-in module handling hadron
and tau-lepton decays~\cite{Siegert:2006xx,Laubrich:2006aa}.
It contains decay tables with branching ratios for approximately 2500
decay channels, many of which have their kinematics modelled according
to a matrix element with corresponding form factors.
In particular, decays of the tau lepton and heavy mesons have form-factor
models similar to dedicated codes like \Tauola~\cite{Jadach:1993hs}
and \EvtGen~\cite{Lange:2001uf}.

Several additional features are implemented: spin correlations can be
enabled to account for the correlation of the helicity of an unstable
particle between the production and decay matrix elements.
Neutral meson mixing can be described, including advanced features
like CP violation in the decay, in the mixing, and in the interference
between them~\cite{Campbell:2022qmc}.
Decay kinematics are adjusted to account for the finite-width
Breit--Wigner line shape of the decaying particle.
QED radiation can be simulated from all charged particles involved
in the hadron decay cascade within the formalism and implementation
described in Sec.~\ref{sec:sherpa:yfs}.
Aliases can be defined and used for a fine-tuned correlated steering
of open decay channels.
For hadrons with incomplete exclusive decay tables, these
can be completed by using the decays of their partonic content, corrected
for higher-order QCD effects using the parton shower of
Sec.\ \ref{sec:sherpa:showers}, and
subsequent hadronisation to yield a description of the missing decay channels.

\subsection{Event generation results and variations}\label{sec:sherpa:io}

After all stages of the simulation of a collider event
have been completed, the event exists as an internal representation
of flavours, momenta, weights and weight components.
We describe in this section how this information can be further
processed and stored in standardised output formats.

\subsubsection{Weighted vs.\ unweighted event generation}
\label{sec:sherpa:io:weights}

\Sherpa events are produced as tuples $\{\Phi_i,w_i,n_{\text{trial},i}\}$,
wherein the phase-space point $\Phi_i$ encodes the flavours and momenta
of all involved particles constituting this event.
All such events initially have a probabilistic Monte Carlo weight $w_i$
associated with them, which must be taken into account when calculating
expectation values for observables,
\begin{equation}
  \langle O\rangle
  =
    \frac{1}{N_{\text{trial}}} \cdot
    \sum\limits_{i=1}^n w_i\left(\Phi_i\right) O\left(\Phi_i\right) \,,
    \quad
    \text{with}\quad N_{\text{trial}}
    =\sum\limits_{i=1}^n n_{\text{trial},i}\,.
\end{equation}
Therein, $O\left(\Phi_i\right)$ is the value of the observable under
consideration, and $n_{\text{trial},i}$ is the number of trials needed to
successfully generate event $\Phi_i$, and $N_{\text{trial}}$ is
their total sum.
In essence, $N_\text{trial}$ keeps track of all attempts that
resulted in a weight $w_i=0$ during any stage of event generation
to retain the correct sample normalisation without the need to
write out events that will not contribute to any observable.

In applications with expensive post-processing steps of the event sample,
for example a full detector simulation, or if storage is a concern,
it is favourable to minimise the number of events in the sample
without reducing its statistical power.
This is achieved by an unweighting step, which accepts or rejects
events in accordance with their probabilistic weight.
The resulting sample consists of the accepted events only,
and their event weights are all normalised to a constant weight
while $n_\text{trials}$ book-keeps the rejected events.
Exceptions from such a uniformly-weighted sample exist for events
with particularly large weights or if a non-uniform bias is applied
to the event generation.
For further details, we refer the user to the full user manual
distributed with the \Sherpa code.

Applying the unweighting step as described above is the default behaviour of \Sherpa.
Since the fraction of events that survive the unweighting is typically very small,
deferring computations that do not affect the weight until after accepting an event
gives rise to major speed-ups of the overall event generation.
In~\cite{Bothmann:2022thx}, we have introduced a pilot-run strategy
leading to an overall reduction in computing time by about a factor of forty
for typical simulation setups used by the LHC collaborations.

\subsubsection{On-the-fly uncertainty estimates}
\label{sec:otf}

During event generation, \Sherpa can calculate a variety of alternative event weights
for various physical and algorithmic variations.
For each alternative weight, its fraction of the nominal weight
encodes the probability of the event to happen for that variation.
This multi-weight handling removes the need
to produce dedicated event samples for each variation
separately, for example to estimate an uncertainty on the nominal prediction.
Furthermore, downstream processing steps
such as physics analyses or detector-response simulations 
only need to process a single event sample and still retain all variations
by simply propagating through the alternative event weights.
The additional spread in weights which occurs when using this method, and
the resulting reduced statistical power of the event-weight sample,
is usually far outweighed by the benefits of the method, 
i.e.\ strongly reduced computing and storage needs.
Similar techniques
are used in the \Vincia~\cite{Giele:2011cb}, \Pythia~\cite{Mrenna:2016sih} and \Herwig~\cite{Bellm:2016voq} event generators.
For \Pythia, an extension of the approach to hadronisation models has recently been presented~\cite{Bierlich:2023fmh}.

For QCD uncertainty estimates,
\Sherpa supports factorisation and renormalisation scale variations,
as well as PDF variations~\cite{Bothmann:2016nao}.
To give some application examples, this allows one to quantify 7-point scale variation
uncertainties, study the spread of predictions due to different PDF fits by different
fitter groups, and/or derive the PDF uncertainty for each PDF set individually.
The strong coupling value for $\alpha_s(m_\mathrm{Z}^2)$ is usually
taken from the PDF set (and can thus be varied by selecting PDF sets with different
inputs for $\alpha_s(m_\mathrm{Z}^2)$),
but it can also be varied independently of the PDF.
By default, all such QCD parameter variations are applied to both the
hard process and the parton shower simultaneously.
Nonetheless, \Sherpa will additionally report variation results for
the hard process only.

The approximate EW corrections discussed in Sec.~\ref{sec:sherpa:ew}
are usually also given as alternative event weights, allowing the user
to compare the effect of including the corrections with the baseline
prediction~\cite{Bothmann:2021led}.
The same applies to the components of the heavy-flavour matching
discussed in Sec.~\ref{sec:sherpa:mm:fusing}.
Finally, on the algorithmic side, as of \Sherpa 3 one can vary the merging
parameter, $Q_\text{cut}$, on the fly when generating multijet-merged event
samples, see Sec.~\ref{sec:sherpa:mm:mets}.
This allows the user to study the effect of this formally higher-order variation,
e.g.\ to confirm that its impact is small compared to other uncertainties
in the phase space relevant for the analysis at hand.

Ultimately, \Sherpa reports the total cross section not only for the
nominal scale and input parameter choices, but also for each requested
variation thereof.
These alternative sample cross sections are passed to the \HepMC~\cite{Dobbs:2001ck,Buckley:2019xhk} event output
or directly to the \Rivet analysis framework~\cite{Bierlich:2019rhm,Bierlich:2024vqo}
via \Sherpa's internal interface. \Sherpa follows
the naming conventions for event-weight variations
specified in~\cite{Bothmann:2022pwf}. 

\subsubsection{Storing and analysing events}

As already mentioned, \Sherpa provides interfaces to the \HepMC and
\Rivet libraries, which can be used to facilitate the analysis of its output.
Generally, \HepMC serves as a common event-record format, allowing \Sherpa
to export its generated events in a standardised manner and ensuring
compatibility with likewise standard-compatible analysis tools and frameworks.
\HepMC itself supports a variety of structured formats for storing
the Monte Carlo event record to disk.
\Sherpa supports \HepMC version 3 onwards \cite{Buckley:2019xhk}.

\Rivet~\cite{Buckley:2010ar,Bierlich:2019rhm,Bierlich:2024vqo}
is a common analysis toolkit for the validation
of Monte Carlo event generators using experimental data.
It uses the Monte Carlo events in the \HepMC format as an input, either
reading the event record from file (independently of which generator
produced the events), or passed
programmatically as an object when a dedicated interface is in place.
\Sherpa supports both options, where its dedicated interface
is supported from \Rivet version~3 onwards.
Starting with \Rivet version~4~\cite{Bierlich:2024vqo},
\Sherpa supports the serialisation of the \Rivet output,
allowing for efficient data reductions in memory as part of MPI-collective
communications in high-performance applications.
This avoids the need for \emph{a posteriori} merging of histogram
files entirely.
Note that both \HepMC~3 and \Rivet~3, as well as later versions, have
native multi-weight support, which makes it very
straightforward to plot uncertainty bands via \Rivet when using
\Sherpa multi-weight event samples as an input.

\Rivet can also be used to fill cross-section interpolation grids from \Sherpa's fixed-order calculations.
This can be achieved using a \Rivet plugin called \MCgrid~\cite{DelDebbio:2013kxa,Bothmann:2015dba} that
projects individual events on differential observables and produces corresponding interpolation grids
in the \APPLgrid~\cite{Carli:2010rw} or \fastNLO~\cite{Britzger:2012bs} format. These grids can be
used for the fast and flexible evaluation of scale, $\alpha_s$, and PDF variations in leading- and
next-to-leading-order QCD calculations. The required event information is provided by \Sherpa via
auxiliary event weights.

For instructions on enabling
\Sherpa's \HepMC and \Rivet interfaces and its MPI support, see App.~\ref{app:install}.

\section{Development pipeline}
\label{sec:sherpa:pipeline}

In this section we briefly describe software and physics projects that have been
developed in the \Sherpa framework, but are not yet publicly
released with \Sherpa 3.0 or 3.1. They represent feature candidates to be included in near-future versions of the package.

\subsection{\texorpdfstring{\Caesar}{Caesar} resummation with \texorpdfstring{\Sherpa}{Sherpa}}
\label{sec:sherpa:pipeline:caesar}

The \Sherpa framework can be used to perform semi-analytic
QCD resummation calculations in the \Caesar
formalism~\cite{Banfi:2003je,Banfi:2004yd}, allowing for the
all-orders inclusion of leading and next-to-leading logarithms
in the observable value for suitable variables. The original
implementation of the \Caesar plugin to \Sherpa was presented
in~\cite{Gerwick:2014gya}. It utilises the event generation
framework, with \Sherpa facilitating
all the process management, providing access to matrix-element
generators, performing phase-space integration, and
providing event-analysis functionality. In the context of
matching the resummation to fixed-order calculations, aiming for
NLO+NLL' accuracy, the \Sherpa implementation of the Catani--Seymour
dipole subtraction and the interfaces to the one-loop providers
\Recola\ and \OpenLoops\ are employed. To correct the resummed
predictions for non-perturbative corrections from the underlying
event and hadronisation, multi-differential transfer matrices,
derived from corresponding \Sherpa simulations, can be employed,
capturing the kinematical migration of parton-level to particle-level
events~\cite{Marzani:2019evv,Reichelt:2021svh,Chien:2024uax}.

The \Caesar implementation of \Sherpa has been used to derive
resummed predictions for soft-drop thrust~\cite{Marzani:2019evv} in the context
of extractions of the strong coupling constant~\cite{dEnterria:2022hzv}
and multijet resolution scales~\cite{Baberuxki:2019ifp} in
electron--positron annihilation, as well as NLO+NLL' accurate
predictions for soft-drop groomed hadronic event shapes~\cite{Baron:2020xoi},
and jet angularities in proton--proton collisions at the
LHC~\cite{Caletti:2021oor, Caletti:2021ysv, Reichelt:2021svh}
and RHIC~\cite{Chien:2024uax}. Recently, it was applied
to plain and groomed event shapes in neutral-current deep inelastic
scattering~\cite{Knobbe:2023ehi,H1:2024aze,H1:2024pvu,Knobbe:2024rci}, as well
as event-shape observables in hadronic Higgs-boson decays at a future
lepton collider~\cite{Gehrmann-DeRidder:2024avt}.

\subsection{Precision resummation with SCET in \texorpdfstring{\Sherpa}{Sherpa}}
\label{sec:sherpa:pipeline:nkll}
In addition to the \Caesar resummation calculations described above,
a number of targeted highest-accuracy resummation
calculations were performed in the \SCET formalism~\cite{Bauer:2000yr,
  Bauer:2001yt} using the \Sherpa framework.
These calculations make use of a purpose-built resummation
routine which is interfaced to \Sherpa.
In this setup, \Sherpa supplies the exact fixed-order matrix elements,
using the interfaces to one-loop providers
(see Sec.~\ref{sec:sherpa:me:nlo}) and the
inbuilt Catani--Seymour dipole subtraction, and carries out the
matching to the resummation away from the infrared limits.
\Sherpa also handles the phase-space integration.

This has allowed the calculation of a number of selected observables
at high fixed-order and resummed precision.
For example,
the double-differential \qT--$\Delta\phi$ spectrum in both charged-
and neutral-current Drell--Yan production, as well as the ratio thereof, were
calculated at \NNNLLpNNLO accuracy, including the
non-negligible top-mass-dependent singlet contributions~\cite{Ju:2021lah}.
Further, these developments have been used to calculate various
observables in $t\bar{t}$ production.
Away from the $t\bar{t}$ production threshold, the projected transverse
momentum distributions were calculated at approximate \NNLLpNNLO
accuracy~\cite{Ju:2022wia}. In the entire $t\bar{t}$ production
region, including the threshold region, the \qT and $\Delta\phi$ spectra
were calculated at \NNLLNNLO~\cite{Ju:2024xhd}.
The \qT spectrum of Higgs production in gluon fusion was
calculated at subleading power (up to \NNLP) at \NLO in
\cite{Ferrera:2023vsw}.
This high-precision resummation interface will
be provided in future versions.

\subsection{High-performance and heterogeneous computing}
\label{sec:sherpa:pipeline:hpc}
\Sherpa is one of the workhorses of the modern experimental simulation
toolchain, in particular for the LHC experiments. While the code provides
enhanced physics modelling capabilities based on high-multiplicity multijet merged simulations,
the required matrix element calculations often strain the experimental computing
budgets~\cite{HSFPhysicsEventGeneratorWG:2020gxw,Valassi:2020ueh,
  HSFPhysicsEventGeneratorWG:2021xti,ATLAS:2021yza}. To reduce the computing
footprint and still facilitate cutting-edge physics simulations, \Sherpa has undergone
extensive performance improvements
in the past years~\cite{Bothmann:2022thx,Danziger:2021xvr,Campbell:2021vlt}.
These have been described in Secs.~\ref{sec:sherpa:is:pdfs},
\ref{sec:sherpa:me} and \ref{sec:sherpa:io:weights},
and have also been backported to the \Sherpa v2.2 series,
resulting in significant event generation speed gains
for the traditional compute model of running \Sherpa
on a single CPU core.

The current trend towards very large (exascale) HPC clusters,
and towards an increasing reliance on off-loading computations
to GPU-like accelerator hardware, brings about new challenges.
Large HPC clusters usually rely on parallel file systems (e.g.\ Lustre).
To take full advantage of such a file system,
one also needs to parallelise input/output operations.
On the other hand, using a GPU usually requires
reorganisation of the data in memory,
copying data to and from the GPU,
and compiling the compute kernel code
for the given GPU architecture,
thus requiring extensive changes to existing codebases like \Sherpa.

To improve performance on parallel file systems,
future versions of \Sherpa will include the \texttt{LHEH5} interface described
in~\cite{Hoche:2019flt,Bothmann:2023ozs}. When \Sherpa is run in parallel mode via MPI,
the \texttt{LHEH5} technology enables efficient parallel I/O operations across many nodes.
This technology also provides the means to store parton-level events
at leading and next-to-leading order, which
can be used for multijet merged simulations within \Sherpa or \Pythia.
This provides new options to cross-check simulations with different
parton showers or hadronisation modules, in order to derive
systematic uncertainty estimates.
In addition, \Sherpa now employs more efficient computing strategies,
including the recycling of particularly intensive parts of
the simulation, where possible.

The \texttt{LHEH5} interface will also ease the use of GPU resources, as it allows \Sherpa to
read in parton-level events from different matrix element generators, such as the GPU-enabled
simulation program \textsc{Pepper}~\cite{Bothmann:2021nch,Bothmann:2022itv,Bothmann:2023siu,Bothmann:2023gew}.
The excellent MPI performance of \Sherpa is further enhanced by the introduction of
a \Rivet~4~\cite{Bierlich:2024vqo} and \textsc{Yoda}~2~\cite{Buckley:2023xqh} interface, allowing
for an efficient in-memory merging of results across MPI nodes (see Sec.~\ref{sec:sherpa:io}).

\subsection{Machine learning for phase-space sampling and event unweighting}
\label{sec:sherpa:pipeline:ML}
Within the \Sherpa framework, several applications of modern machine learning techniques are
being explored with the aim of further improving the generator performance. The focus is currently
on the hard event component; in particular, on achieving the efficient sampling of high-dimensional
phase spaces. This involves the generation of momentum configurations distributed according to
the desired squared matrix element, and the efficient unweighting of events.

We have pioneered the development of novel sampling algorithms which work by
remapping the integration variables using trainable Normalising Flows
\cite{Bothmann:2020ywa,Gao:2020zvv,Bothmann:2023siu}. The task of optimising the performance
of the process-specific multi-channel importance sampler is traditionally accomplished
with the \textsc{Vegas} algorithm~\cite{Lepage:1980dq}. However, \textsc{Vegas}
assumes a factorisable target distribution, which is typically not found in
multi-particle transition matrix elements and the phase-space parametrisations which are employed,
limiting the potential for further optimisations. Normalising-flow maps are more
flexible and facilitate a better optimisation of the sampling distribution to the true target
functions. As a result, the statistical variance of cross-section predictions can be significantly reduced for many processes.

In general, event unweighting (the generation of events with
unit weight) is a rather inefficient process, especially
for high-multiplicity processes, due to the large spread in
event weights.
Unweighting efficiencies for these processes often fall below
the permille level~\cite{Hoche:2019flt}. At the same time, the required matrix element
evaluations are computationally costly. To improve the efficiency of this process, we have developed a novel
two-stage unweighting algorithm. It relies on a fast neural network surrogate for the
event weight in an initial unweighting phase, followed by a second rejection sampling
against the true event weight~\cite{Danziger:2021eeg}. The resulting
event sample is unbiased and follows the desired distribution, though statistically
somewhat diluted due to the possible appearance of overweights. However, the
effective gain factors turn out to be significant for complex final states. To improve the
algorithm's performance, physics knowledge about the target function can be incorporated into
the surrogate construction. To this end, we have studied network architectures that
reflect the dipole factorisation property of QCD real-emission matrix elements
\cite{Maitre:2021uaa}, resulting in a performance boost for the new unweighting
algorithm~\cite{Janssen:2023ahv}.

\subsection{Neutrino physics interface}
\label{sec:sherpa:pipeline:neutrino}
The current and next-generation neutrino experiments
are entering a precision era, in which the dominant uncertainty will shift from statistical
in nature to systematic~\cite{DUNE:2020fgq,DUNE:2020ypp,Hyper-Kamiokande:2018ofw}, enabling
a multitude of analyses that will probe physics beyond the Standard Model.
In particular, there has been a push to expand the searches for BSM physics at
accelerator neutrino experiments, such as the SBN program~\cite{MicroBooNE:2015bmn} and DUNE~\cite{DUNE:2020fgq}.
\Sherpa has proven to be a versatile tool for the corresponding event simulations.
To expedite the inclusion of novel models into the experimental pipeline, an interface to the
UFO module (see Sec.~\ref{sec:sherpa:me:ufo} for details) has been developed to return
only the leptonic current involved~\cite{Isaacson:2021xty}.

DUNE is expected to measure an unprecedented number of tau neutrino events in the far detector~\cite{DUNE:2020ypp}.
In the past, neutrino generators have simply assumed that the produced tau lepton is purely
left-handed. This has been shown to be a poor assumption~\cite{Hernandez:2022nmp}. To facilitate the needed
precision for DUNE, \Sherpa now provides an interface to enable neutrino generators to appropriately
include spin-correlations for tau decays~\cite{Isaacson:2023gwp}.
Further, within the neutrino community, only \Pythia is currently used
for estimating the hadronisation of particles
in the DIS region~\cite{GENIE:2021npt,Andreopoulos:2009rq,Buss:2011mx,Hayato:2021heg,Hayato:2009zz,Golan:2012rfa}. Extending the above interface to allow for neutrino generators to use \Sherpa
for hadronisation will enable a more robust estimate of the uncertainties in this energy region.

\section{Conclusions}
\label{sec:conclusions}
In this paper, we have described the new major release of the general-purpose
Monte Carlo event generator \Sherpa, a numerical simulation program
designed specifically to cope with the high centre-of-mass
energies at CERN's Large Hadron Collider and the associated
physics challenges.
Over the last few years, \Sherpa has been extended
to the simulation of a wider range of physics processes, such as polarised cross sections,
photoproduction and diffractive jet production.
The physics capabilities of the generator have been further enhanced through
improved models for soft physics and a universal
framework for NLO calculations in the complete Standard
Model. In addition to these and the other developments described here,
\Sherpa also provides a platform for various
other precision physics simulations, such as a generic
NLL resummation framework and a neutrino event generator.
Together with a number of technical improvements, the above
developments are released publicly and supported as
\Sherpa version 3, which will form the basis for further
refinement of the physics models in the eras of the High-Luminosity
LHC and the EIC, and for the preparation of
other potential future collider experiments.

\subsection*{Acknowledgements}
EB, MK and SS acknowledge financial support from BMBF
(projects 05H21MGCAB, 05D23MG1 and 05H24MGA) and funding by
the Deutsche Forschungsgemeinschaft (DFG,
German Research Foundation) - projects 456104544 and 510810461.
This research was supported by the Fermi National Accelerator
Laboratory (Fermilab), a U.S.\ Department of Energy, Office of
Science, HEP User Facility managed by Fermi Research Alliance,
LLC (FRA), acting under Contract No. DE--AC02--07CH11359.
AP acknowledges financial support from grant No. 2019/34/E/ST2/004571012
of the National Science Centre (NCN), Poland and also by the Priority Research Area
Digiworld under the program Excellence Initiative — Research University
at the Jagiellonian University in Krakow.
DR acknowledges funding by the European Union under the HORIZON program
in Marie Sk\l{}odowska-Curie project No.\ 101153541.
MS is funded by the Royal Society through a University Research Fellowship
(URF\textbackslash{}R1\textbackslash{}180549, URF\textbackslash{}R\textbackslash{}231031) and an Enhancement Award
(RGF\textbackslash{}EA\textbackslash{}181033,
 CEC19\textbackslash{}100349, and RF\textbackslash{}ERE\textbackslash{}210397)
as well as the STFC (ST/X003167/1 and ST/X000745/1).
LF is supported by Leverhulme Grant LIP-2021-014.
LF and PM acknowledge support from an STFC studentship under grant ST/P001246/1.
PM is supported by the Swiss National Science Foundation (SNF)
under contract 200020-204200 and acknowledges support from the STFC under grant agreement ST/P006744/1.
The work of JI was supported by the U.S.\ Department of Energy, Office of Science, Office of Advanced Scientific Computing Research, Scientific Discovery through Advanced Computing (SciDAC-5) program, grant “NeuCol”.
CG acknowledges funding via the SWIFT-HEP project (grant number ST/V002627/1).

\appendix
\section{Installing \texorpdfstring{\Sherpa}{Sherpa}}
\label{app:install}

\Sherpa is distributed as a tarred and gzipped file named
\texttt{sherpa-{<VERSION>}.tar.gz} available from the \textit{Downloads} section of the
project's webpage\footnote{\url{https://sherpa-team.gitlab.io}}.
The file can be unpacked in the current working directory with the shell command
\begin{verbatim}
   $ tar -xzf sherpa-<VERSION>.tar.gz
\end{verbatim}

Alternatively, \Sherpa can be accessed via Git, through

\begin{verbatim}
   $ git clone --single-branch -b rel-<VERSION> https://gitlab.com/sherpa-team/sherpa.git
\end{verbatim}

In either case, to guarantee successful installation, the following tools should be
available on the system:
\begin{itemize}
\item a recent C/C++ compiler toolchain,
\item a recent version of CMake to configure a build directory,
\item and Make or Ninja to build and install \Sherpa.
\end{itemize}
A Fortran compiler is recommended. For the use of \UFO models,
an installation of Python version 3.5 or later is required. 
Installations of the \LHAPDF and \texttt{libzip} libraries
are also recommended, but it is possible to let \Sherpa install its own copies
of both libraries, as will be discussed below.

Compilation and installation proceed through the following standard CMake workflow:
\begin{verbatim}
   $ cd sherpa-<VERSION>/
   $ cmake -S . -B <builddir> -DCMAKE_INSTALL_PREFIX=<installdir> [+ config options]
   $ cmake --build <builddir> [+ build options, e.g. -j 8]
   $ cmake --install <builddir>
\end{verbatim}
where \texttt{<builddir>} has to be replaced with the (temporary) directory in which intermediate files are stored for the build process,
and \texttt{<installdir>} with the installation directory into which the build products are installed. The structure of the program within \texttt{<installdir>} is as follows (if the installation procedure is not further customised):
\begin{itemize}
\item \texttt{<installdir>/bin}:
the main \texttt{Sherpa} executable and additional auxiliary executables and scripts,
\item \texttt{<installdir>/include}:
headers that define the API to use \Sherpa as an external framework from third party tools,
and which are used when \Sherpa writes out process libraries that must be compiled by the user,
\item \texttt{<installdir>/lib}:
basic library files,
\item \texttt{<installdir>/share}:
  PDF data files, tau lepton and hadron decay data, example run cards,
      command line auto-completion files and other auxiliary files.
\end{itemize}

\Sherpa can be interfaced with various external packages.
To enable this, the user has to add the corresponding options
to the \texttt{cmake} configuration command:
\begin{verbatim}
   $ cmake -S . -B <builddir> [...] -DSHERPA_ENABLE_<PACKAGENAME>=ON
\end{verbatim}
where \texttt{<PACKAGENAME>} is replaced by the external package name,
e.g.\ \texttt{RIVET}, \texttt{LHAPDF}, \texttt{HEPMC3} etc.
If the external package is not installed in a standard location,
the user might need to specify the installation directory of the package as follows:
\begin{verbatim}
   $ cmake -S . -B <builddir> [...] -D<PACKAGENAME>_DIR=<package_installdir>
\end{verbatim}
In Tab.~\ref{tab:config}, we list the configuration options
to enable interfaces to external packages and optional features
which are referred to in this article.
However, it is not a complete list of all available interfaces and options.
For this, we refer the reader to the manual distributed with the actual
code release and can also be found on the \Sherpa download webpage.
Alternatively, a complete list of possible configuration options
can be listed by running \texttt{cmake -LA <builddir>} or \texttt{ccmake <builddir>}.

\begin{table}
\begin{tabular}{llll}\toprule
   Option/package name & Enable option/interface & Specify package location & References \\
   \midrule
   MPI & \texttt{-DSHERPA\_ENABLE\_MPI=ON} & \texttt{-DMPI\_DIR=<path>} & --- \\
   \HepMC~3.0.0 or later & \texttt{-DSHERPA\_ENABLE\_HEPMC3=ON} & \texttt{-DHEPMC3\_DIR=<path>} & \cite{Dobbs:2001ck,Buckley:2019xhk} \\
   \LHAPDF & \texttt{-DSHERPA\_ENABLE\_LHAPDF=ON} & \texttt{-DLHAPDF\_DIR=<path>} & \cite{Buckley:2014ana} \\
   \MCFM & \texttt{-DSHERPA\_ENABLE\_MCFM=ON} & \texttt{-DMCFM\_DIR=<path>} & \cite{Campbell:2021vlt} \\
   \OpenLoops & \texttt{-DSHERPA\_ENABLE\_OPENLOOPS=ON} & \texttt{-DOPENLOOPS\_DIR=<path>} & \cite{Cascioli:2011va,Buccioni:2019sur} \\
   \Pythia8.220 or later & \texttt{-DSHERPA\_ENABLE\_PYTHIA8=ON} & \texttt{-DPYTHIA8\_DIR=<path>} & \cite{Bierlich:2022pfr} \\
   \Recola & \texttt{-DSHERPA\_ENABLE\_RECOLA=ON} & \texttt{-DRECOLA\_DIR=<path>} & \cite{Actis:2012qn,Denner:2017wsf,Biedermann:2017yoi} \\
   \Rivet~3.0.0 or later & \texttt{-DSHERPA\_ENABLE\_RIVET=ON} & \texttt{-DRIVET\_DIR=<path>} & \cite{Bierlich:2019rhm,Bierlich:2024vqo} \\
   EW Sudakovs & \texttt{-DSHERPA\_ENABLE\_EWSUD=ON} & --- & \cite{Bothmann:2020sxm,Bothmann:2021led} \\
   \UFO & \texttt{-DSHERPA\_ENABLE\_UFO=ON} & --- & \cite{Degrande:2011ua,Darme:2023jdn} \\
   \bottomrule
\end{tabular}
   \caption{\label{tab:config}
   Configuration options to enable some of \Sherpa's optional features
   and interfaces to external packages.
   A package location needs to be specified only if an external package is installed in a non-standard location,
   or to enforce the usage of a specific installation of the package.
   The \UFO and EW Sudakovs options do not rely on external packages
   and therefore have no associated package location option.}
\end{table}

\section{Input cards}
\label{app:input}

When \Sherpa is run without any arguments,
it scans for a configuration file called \texttt{Sherpa.yaml}
in the current working directory.
Such a configuration file is also called a runcard.
It usually defines the process(es) for which events should be generated,
the collider setup,
and other physics and technical settings the user wishes to customise.
An example for the production of an electron--neutrino pair with up to four additional jets
is given in Listing~\ref{listing:runcard}.

\begin{listing}
\begin{minted}[bgcolor=listingsbg]{yaml}
# set up beams for LHC run 2
BEAMS: 2212
BEAM_ENERGIES: 6500

TAGS:
  NJETS: 4

# request events for pp -> W[ev] with up to four additional final-state jets
PROCESSES:
- 93 93 -> 11 -12 93{$(NJETS)}:
    Order: {QCD: 0, EW: 2}
    CKKW: 20
    # use NLO accuracy for the lowest three multiplicities (2->2, 2->3 and 2->4)
    2->2-4:
      NLO_Mode: MC@NLO
      NLO_Order: {QCD: 1, EW: 0}
      Loop_Generator: OpenLoops
\end{minted}
\caption{An example \Sherpa runcard for generating $pp \to e^- \bar{\nu}_e + \text{jets}$ events.}
\label{listing:runcard}
\end{listing}

The settings in the runcard are given in YAML syntax~\cite{YAML2009}.
In the example, we set up symmetric proton beams corresponding to
a collision centre-of-mass energy of \SI{13}{\TeV},
by specifying the single-valued settings
\texttt{BEAMS: 2212} and \texttt{BEAM\_ENERGIES: 6500}.
Next, we specify a tag using the \texttt{TAGS} setting.
We call the tag \texttt{NJETS} and set it to the value \texttt{4}.
Any occurrence of \texttt{\$(NJETS)} in the runcard will now be replaced with that value.
Finally, we add a process using \texttt{PROCESSES}.
This setting also takes a list, since \Sherpa can generate events for more than one process.
Here, we only add one process, using \texttt{93 93 -> 11 -12 93\{\$(NJETS)\}},
which translates to $pp \to e^- \bar{\nu}_e + \text{up to four jets}$.
The process specification itself, \texttt{93 93 -> 11 -12 93\{\$(NJETS)\}},
takes various subsettings,
e.g.\ the orders in the strong and electroweak couplings using the \texttt{Order} subsetting.
For all parameter settings not specified explicitly in the runcard (here, for example, which parton shower or hadronisation model is to be used), their default values are assumed. Accordingly, the
given runcard would generate hadron-level events based on \Sherpa's Catani--Seymour dipole
shower, see \SecRef{sec:sherpa:showers:css}, and its cluster hadronisation model, described
in~\SecRef{sec:sherpa:had:cluster}.

All settings can be specified on the command line, too,
using the same syntax, for example:
\begin{verbatim}
   $ Sherpa 'EVENTS: 1M'
\end{verbatim}
This would set the number of events to one million,
taking precedence over any \Sherpa defaults or runcard settings.
Some settings have associated command-line arguments,
e.g.\ the following command is an equivalent way to request the generation of one million events:
\begin{verbatim}
   $ Sherpa -e 1M
\end{verbatim}
Other arguments on the command line are interpreted as paths to runcards to be used,
e.g.
\begin{verbatim}
   $ Sherpa -e 1M path/to/runcard.yaml
\end{verbatim}
will read in settings from \texttt{path/to/runcard.yaml},
and proceed to generate one million events for the process(es) defined therein.

Because it is easy to introduce a typo in a setting name,
or to use the wrong capitalisation (all setting names and values are case-sensitive),
\Sherpa will report a short summary of any unused settings in the output produced at the end of a run.
Additionally, details of all settings, used or unused, and the associated defaults and custom values,
are written to the \texttt{Settings\_Report} directory
within the current working directory,
which can be understood as a manifest of the run configuration.

For more details on the syntax, a complete documentation of all user settings,
available command-line arguments and the settings report,
we refer the reader to the manual available on 
the \Sherpa webpage or distributed with the code.

\section{Details on beam remnant handling}
\label{sec:sherpa:had:beamremnant_app}
In \Sherpa, the breakup of incoming hadrons and the formation of the beam
remnants is modelled after all multiple-parton interactions and associated
parton showering steps have terminated and a list of shower initiators can
be extracted from the incoming hadron. The physics model is the following:
\begin{enumerate}
\item Flavour compensation: assuming an incoming hadron to consist of a
      valence quark--diquark pair (the diquark is the carrier
      of the baryon number), and that di-quarks cannot act as shower
      initiators, the flavours of the shower initiators have to be compensated.
      One of them may be a valence quark -- shower initiators are assigned as
      valence quarks with a probability obtained from the PDFs at the lowest
      scale.  All other ``net'' quark flavours among the shower initiators
      are compensated with a corresponding anti-flavour spectator.
\item Colour compensation: as the overall hadron must form a singlet, \Sherpa
      assumes a colour-ordered list of partons of the type $q-g-g-\dots-(qq)$,
      where $q$ and $(qq)$ denote the valence quark and diquark.  The model
      also assumes that flavour--anti-flavour pairs, either formed by
      the shower initiators of independent MPI scatters or by compensating
      individual flavours with a corresponding anti-flavour spectator,
      emerge from a gluon and therefore will be in a relative colour-octet
      state.
\item Longitudinal momenta: The longitudinal momenta of the shower initiators
      are already fixed, and those for the spectators in the beam breakup
      are selected according to the PDFs.  The longitudinal momentum for the
      valence diquark is given by the residual at the end of the process.
\item Transverse momentum: \Sherpa models the finite ``intrinsic'' transverse
      momentum $k_\perp$ inside the hadrons, akin to Fermi motion by
      assuming a Gaussian distribution, cut-off at large values through a
      polynomial,
      \begin{equation}
      \mathcal{P}(k_\perp)\propto
      \exp(-(k_\perp-k_{\perp,0})^2/\sigma^2) (k_{\perp, \text{max}}-k_\perp)^\eta\,.
      \end{equation}
      The parameters of this distribution depend on the hadron forming the
      beam and scale with the centre-of-mass energy of the collision.
      \Sherpa allows different parameter values for parton-shower initiators
      and the spectator.
\item Overall momentum conservation:  To ensure overall momentum
      conservation after assigning individual intrinsic transverse momenta
      for the partons, \Sherpa allows two recoil strategies, which
      slightly modify both transverse and longitudinal momenta.  In a
      ``democratic approach'' the overall excess transverse momentum is
      compensated by subtracting it from the partons in proportion to their
      longitudinal momenta.  Alternatively, \Sherpa compensates in a similar
      fashion, the recoil of the shower initiators with the spectators and
      vice versa.  This leaves the question of overall momentum
      conservation within the beam break-up, as in most cases the combined
      invariant mass of partons coming from a hadron differs from its mass.
      In hadron--hadron collisions this is achieved by shuffling momenta
      between the two hadrons, while in collisions involving only one
      incident hadron, this compensation happens between the initial state
      and strongly-interacting final states.
\end{enumerate}

\bibliographystyle{bib/amsunsrt_mod}
\bibliography{bib/journal}

\begin{thebibliography}{100}

\bibitem{Buckley:2011ms}
A.~Buckley et~al., \emph{{General-purpose event generators for LHC physics}},
  Phys. Rept. \textbf{504} (2011),
  \href{http://inspirebeta.net/record/884202}{145--233},
  [\href{http://arXiv.org/pdf/1101.2599}{{\texttt{arXiv:1101.2599}}} [hep-ph]].
  \relax
 \relax
\bibitem{Campbell:2022qmc}
J.~M. Campbell et~al., \emph{{Event generators for high-energy physics
  experiments}}, SciPost Phys. \textbf{16} (2024), no.~5,
  \href{http://www.slac.stanford.edu/spires/find/hep/www?eprint=2203.11110}{130},
   [\href{http://arXiv.org/pdf/2203.11110}{{\texttt{arXiv:2203.11110}}}
  [hep-ph]]. \relax
 \relax
\bibitem{Bellm:2015jjp}
J.~Bellm et~al., \emph{{Herwig 7.0/Herwig++ 3.0 release note}}, Eur. Phys. J.
  \textbf{C76} (2016), no.~4,
  \href{http://www.slac.stanford.edu/spires/find/hep/www?eprint=1512.01178}{196},
   [\href{http://arXiv.org/pdf/1512.01178}{{\texttt{arXiv:1512.01178}}}
  [hep-ph]]. \relax
 \relax
\bibitem{Bewick:2023tfi}
\href{http://www.slac.stanford.edu/spires/find/hep/www?eprint=2312.05175}{G.~Bewick
  et~al.}, \emph{{Herwig 7.3 Release Note}},
  \href{http://arXiv.org/pdf/2312.05175}{{\texttt{arXiv:2312.05175}}} [hep-ph].
  \relax
 \relax
\bibitem{Sjostrand:2014zea}
T.~Sj{\"o}strand, S.~Ask, J.~R. Christiansen, R.~Corke, N.~Desai, P.~Ilten,
  S.~Mrenna, S.~Prestel, C.~O. Rasmussen and P.~Z. Skands, \emph{{An
  Introduction to PYTHIA 8.2}}, Comput. Phys. Commun. \textbf{191} (2015),
  \href{http://www.slac.stanford.edu/spires/find/hep/www?eprint=1410.3012}{159--177},
   [\href{http://arXiv.org/pdf/1410.3012}{{\texttt{arXiv:1410.3012}}}
  [hep-ph]]. \relax
 \relax
\bibitem{Bierlich:2022pfr}
C.~Bierlich et~al., \emph{{A comprehensive guide to the physics and usage of
  PYTHIA 8.3}}, SciPost Phys. Codeb. \textbf{2022} (2022),
  \href{http://www.slac.stanford.edu/spires/find/hep/www?eprint=2203.11601}{8},
   [\href{http://arXiv.org/pdf/2203.11601}{{\texttt{arXiv:2203.11601}}}
  [hep-ph]]. \relax
 \relax
\bibitem{EuropeanStrategyforParticlePhysicsPreparatoryGroup:2019qin}
\href{http://www.slac.stanford.edu/spires/find/hep/www?eprint=1910.11775}{R.~K.
  Ellis et~al.}, \emph{{Physics Briefing Book}: {Input for the European
  Strategy for Particle Physics Update 2020}},
  \href{http://arXiv.org/pdf/1910.11775}{{\texttt{arXiv:1910.11775}}} [hep-ex].
  \relax
 \relax
\bibitem{Butler:2023glv}
J.~N. Butler et~al., \emph{{Report of the 2021 U.S. Community Study on the
  Future of Particle Physics (Snowmass 2021)}}. \relax
 \relax
\bibitem{Gleisberg:2003xi}
T.~Gleisberg, S.~H{\"o}che, F.~Krauss, A.~Sch{\"a}licke, S.~Schumann and
  J.~Winter, \emph{{\Sherpa 1.$\alpha$, a proof-of-concept version}}, JHEP
  \textbf{02} (2004), \href{http://inspirehep.net/search?irn=5730570}{056},
  [\href{http://arXiv.org/pdf/hep-ph/0311263}{{\texttt{hep-ph/0311263}}}].
  \relax
 \relax
\bibitem{Gleisberg:2008ta}
T.~Gleisberg, S.~H{\"o}che, F.~Krauss, M.~Sch\"{o}nherr, S.~Schumann,
  F.~Siegert and J.~Winter, \emph{{Event generation with \Sherpa 1.1}}, JHEP
  \textbf{02} (2009), \href{http://inspirebeta.net/record/803708}{007},
  [\href{http://arXiv.org/pdf/0811.4622}{{\texttt{arXiv:0811.4622}}} [hep-ph]].
  \relax
 \relax
\bibitem{Bothmann:2019yzt}
E.~Bothmann et~al., \emph{{Event Generation with Sherpa 2.2}}, SciPost Phys.
  \textbf{7} (2019), no.~3,
  \href{http://www.slac.stanford.edu/spires/find/hep/www?eprint=1905.09127}{034},
   [\href{http://arXiv.org/pdf/1905.09127}{{\texttt{arXiv:1905.09127}}}
  [hep-ph]]. \relax
 \relax
\bibitem{vonWeizsacker:1934sx}
C.~von Weizs{\"a}cker, \emph{{Radiation emitted in collisions of very fast
  electrons}}, Z.Phys. \textbf{88} (1934),
  \href{http://www.slac.stanford.edu/spires/find/hep/www?j=ZPhys,88,612}{612--625}.
  \relax
 \relax
\bibitem{Williams:1934ad}
E.~Williams, \emph{{Nature of the high-energy particles of penetrating
  radiation and status of ionization and radiation formulae}}, Phys.Rev.
  \textbf{45} (1934),
  \href{http://www.slac.stanford.edu/spires/find/hep/www?j=PhysRev,45,729}{729--730}.
  \relax
 \relax
\bibitem{Budnev:1974de}
V.~M. Budnev, I.~F. Ginzburg, G.~V. Meledin and V.~G. Serbo, \emph{{The two
  photon particle production mechanism. Physical problems. Applications.
  Equivalent photon approximation}}, Phys. Rept. \textbf{15} (1974),
  \href{http://inspirehep.net/search?j=PRPLC,15,181}{181--281}. \relax
 \relax
\bibitem{Frixione:1993yw}
S.~Frixione, M.~L. Mangano, P.~Nason and G.~Ridolfi, \emph{{Improving the
  Weizsacker-Williams approximation in electron - proton collisions}}, Phys.
  Lett. B \textbf{319} (1993),
  \href{http://www.slac.stanford.edu/spires/find/hep/www?eprint=hep-ph/9310350}{339--345},
   [\href{http://arXiv.org/pdf/hep-ph/9310350}{{\texttt{hep-ph/9310350}}}].
  \relax
 \relax
\bibitem{Collins:1997sr}
J.~C. Collins, \emph{{Proof of factorization for diffractive hard scattering}},
  Phys. Rev. D \textbf{57} (1998),
  \href{http://www.slac.stanford.edu/spires/find/hep/www?eprint=hep-ph/9709499}{3051--3056},
   [\href{http://arXiv.org/pdf/hep-ph/9709499}{{\texttt{hep-ph/9709499}}}],
  [Erratum: Phys.Rev.D 61, 019902 (2000)]. \relax
 \relax
\bibitem{Ingelman:1984ns}
G.~Ingelman and P.~E. Schlein, \emph{{Jet Structure in High Mass Diffractive
  Scattering}}, Phys. Lett. B \textbf{152} (1985),
  \href{http://www.slac.stanford.edu/spires/find/hep/www?j=Phys%20Lett%20B,152,256}{256--260}.
  \relax
 \relax
\bibitem{Newman:2013ada}
P.~Newman and M.~Wing, \emph{{The Hadronic Final State at HERA}}, Rev. Mod.
  Phys. \textbf{86} (2014), no.~3,
  \href{http://www.slac.stanford.edu/spires/find/hep/www?eprint=1308.3368}{1037},
   [\href{http://arXiv.org/pdf/1308.3368}{{\texttt{arXiv:1308.3368}}}
  [hep-ex]]. \relax
 \relax
\bibitem{Buckley:2014ana}
A.~Buckley, J.~Ferrando, S.~Lloyd, K.~Nordstr{\"o}m, B.~Page, M.~R{\"u}fenacht,
  M.~Sch{\"o}nherr and G.~Watt, \emph{{LHAPDF6: parton density access in the
  LHC precision era}}, Eur. Phys. J. \textbf{C75} (2015),
  \href{http://www.slac.stanford.edu/spires/find/hep/www?eprint=1412.7420}{132},
   [\href{http://arXiv.org/pdf/1412.7420}{{\texttt{arXiv:1412.7420}}}
  [hep-ph]]. \relax
 \relax
\bibitem{PDF4LHCWorkingGroup:2022cjn}
R.~D. Ball et~al., PDF4LHC Working Group collaboration, \emph{{The PDF4LHC21
  combination of global PDF fits for the LHC Run III}}, J. Phys. G \textbf{49}
  (2022), no.~8,
  \href{http://www.slac.stanford.edu/spires/find/hep/www?eprint=2203.05506}{080501},
   [\href{http://arXiv.org/pdf/2203.05506}{{\texttt{arXiv:2203.05506}}}
  [hep-ph]]. \relax
 \relax
\bibitem{NNPDF:2017mvq}
R.~D. Ball et~al., NNPDF collaboration, \emph{{Parton distributions from
  high-precision collider data}}, Eur. Phys. J. C \textbf{77} (2017), no.~10,
  \href{http://www.slac.stanford.edu/spires/find/hep/www?eprint=1706.00428}{663},
   [\href{http://arXiv.org/pdf/1706.00428}{{\texttt{arXiv:1706.00428}}}
  [hep-ph]]. \relax
 \relax
\bibitem{Dulat:2015mca}
S.~Dulat, T.-J. Hou, J.~Gao, M.~Guzzi, J.~Huston, P.~Nadolsky, J.~Pumplin,
  C.~Schmidt, D.~Stump and C.~P. Yuan, \emph{{New parton distribution functions
  from a global analysis of quantum chromodynamics}}, Phys. Rev. \textbf{D93}
  (2016), no.~3,
  \href{http://www.slac.stanford.edu/spires/find/hep/www?eprint=1506.07443}{033006},
   [\href{http://arXiv.org/pdf/1506.07443}{{\texttt{arXiv:1506.07443}}}
  [hep-ph]]. \relax
 \relax
\bibitem{Bothmann:2022thx}
E.~Bothmann, A.~Buckley, I.~A. Christidi, C.~G\"utschow, S.~H\"oche, M.~Knobbe,
  T.~Martin and M.~Sch\"onherr, \emph{{Accelerating LHC event generation with
  simplified pilot runs and fast PDFs}}, Eur. Phys. J. C \textbf{82} (2022),
  no.~12,
  \href{http://www.slac.stanford.edu/spires/find/hep/www?eprint=2209.00843}{1128},
   [\href{http://arXiv.org/pdf/2209.00843}{{\texttt{arXiv:2209.00843}}}
  [hep-ph]]. \relax
 \relax
\bibitem{Gluck:1991ee}
M.~Gl{\"u}ck, E.~Reya and A.~Vogt, \emph{{Parton structure of the photon beyond
  the leading order}}, Phys. Rev. \textbf{D45} (1992),
  \href{http://inspirehep.net/search?j=PHRVA,D45,3986}{3986--3994}. \relax
 \relax
\bibitem{Gluck:1991jc}
M.~Gl{\"u}ck, E.~Reya and A.~Vogt, \emph{{Photonic parton distributions}},
  Phys. Rev. \textbf{D46} (1992),
  \href{http://inspirehep.net/search?j=PHRVA,D46,1973}{1973--1979}. \relax
 \relax
\bibitem{Gluck:1999ub}
M.~Gluck, E.~Reya and I.~Schienbein, \emph{{Radiatively generated parton
  distributions of real and virtual photons}}, Phys. Rev. D \textbf{60} (1999),
  \href{http://www.slac.stanford.edu/spires/find/hep/www?eprint=hep-ph/9903337}{054019},
   [\href{http://arXiv.org/pdf/hep-ph/9903337}{{\texttt{hep-ph/9903337}}}],
  [Erratum: Phys.Rev.D 62, 019902 (2000)]. \relax
 \relax
\bibitem{Slominski:2005bw}
W.~Slominski, H.~Abramowicz and A.~Levy, \emph{{NLO photon parton
  parametrization using ee and ep data}}, Eur. Phys. J. C \textbf{45} (2006),
  \href{http://www.slac.stanford.edu/spires/find/hep/www?eprint=hep-ph/0504003}{633--641},
   [\href{http://arXiv.org/pdf/hep-ph/0504003}{{\texttt{hep-ph/0504003}}}].
  \relax
 \relax
\bibitem{Cornet:2002iy}
F.~Cornet, P.~Jankowski, M.~Krawczyk and A.~Lorca, \emph{{A New five flavor LO
  analysis and parametrization of parton distributions in the real photon}},
  Phys. Rev. D \textbf{68} (2003),
  \href{http://www.slac.stanford.edu/spires/find/hep/www?eprint=hep-ph/0212160}{014010},
   [\href{http://arXiv.org/pdf/hep-ph/0212160}{{\texttt{hep-ph/0212160}}}].
  \relax
 \relax
\bibitem{Cornet:2004nb}
F.~Cornet, P.~Jankowski and M.~Krawczyk, \emph{{A New 5 flavor NLO analysis and
  parametrizations of parton distributions of the real photon}}, Phys. Rev. D
  \textbf{70} (2004),
  \href{http://www.slac.stanford.edu/spires/find/hep/www?eprint=hep-ph/0404063}{093004},
   [\href{http://arXiv.org/pdf/hep-ph/0404063}{{\texttt{hep-ph/0404063}}}].
  \relax
 \relax
\bibitem{Schuler:1995fk}
G.~A. Schuler and T.~Sj{\"o}strand, \emph{{Low and high mass components of the
  photon distribution functions}}, Z. Phys. C \textbf{68} (1995),
  \href{http://www.slac.stanford.edu/spires/find/hep/www?eprint=hep-ph/9503384}{607--624},
   [\href{http://arXiv.org/pdf/hep-ph/9503384}{{\texttt{hep-ph/9503384}}}].
  \relax
 \relax
\bibitem{Schuler:1996fc}
G.~A. Schuler and T.~Sj{\"o}strand, \emph{{Parton distributions of the virtual
  photon}}, Phys.Lett. \textbf{B376} (1996),
  \href{http://www.slac.stanford.edu/spires/find/hep/www?eprint=hep-ph/9601282}{193--200},
   [\href{http://arXiv.org/pdf/hep-ph/9601282}{{\texttt{arXiv:hep-ph/9601282}}}
  [hep-ph]]. \relax
 \relax
\bibitem{H1:2006zyl}
A.~Aktas et~al., H1 collaboration, \emph{{Measurement and QCD analysis of the
  diffractive deep-inelastic scattering cross-section at HERA}}, Eur. Phys. J.
  C \textbf{48} (2006),
  \href{http://www.slac.stanford.edu/spires/find/hep/www?eprint=hep-ex/0606004}{715--748},
   [\href{http://arXiv.org/pdf/hep-ex/0606004}{{\texttt{hep-ex/0606004}}}].
  \relax
 \relax
\bibitem{Kuraev:1985hb}
E.~A. Kuraev and V.~S. Fadin, \emph{{On Radiative Corrections to e+ e- Single
  Photon Annihilation at High-Energy}}, Sov. J. Nucl. Phys. \textbf{41} (1985),
  \href{http://www.slac.stanford.edu/spires/find/hep/www?j=Sov%20J%20Nucl%20Phys,41,466}{466--472}.
  \relax
 \relax
\bibitem{Bardin:1993bh}
D.~Y. Bardin, M.~S. Bilenky, A.~Olchevski and T.~Riemann, \emph{{Off-shell W
  pair production in e+ e- annihilation: Initial state radiation}}, Phys. Lett.
  B \textbf{308} (1993),
  \href{http://www.slac.stanford.edu/spires/find/hep/www?eprint=hep-ph/9507277}{403--410},
   [\href{http://arXiv.org/pdf/hep-ph/9507277}{{\texttt{hep-ph/9507277}}}],
  [Erratum: Phys.Lett.B 357, 725--726 (1995)]. \relax
 \relax
\bibitem{Beenakker:1994vn}
W.~Beenakker and A.~Denner, \emph{{Standard model predictions for $W$ pair
  production in electron - positron collisions}}, Int. J. Mod. Phys. A
  \textbf{9} (1994),
  \href{http://www.slac.stanford.edu/spires/find/hep/www?j=Int%20J%20Mod%20Phys%20A,9,4837}{4837--4920}.
  \relax
 \relax
\bibitem{Montagna:1994qu}
G.~Montagna, O.~Nicrosini, G.~Passarino and F.~Piccinini, \emph{{Semianalytical
  and Monte Carlo results for the production of four fermions in e+ e-
  collisions}}, Phys. Lett. B \textbf{348} (1995),
  \href{http://www.slac.stanford.edu/spires/find/hep/www?eprint=hep-ph/9411332}{178--184},
   [\href{http://arXiv.org/pdf/hep-ph/9411332}{{\texttt{hep-ph/9411332}}}].
  \relax
 \relax
\bibitem{Berends:1994pv}
F.~A. Berends, R.~Pittau and R.~Kleiss, \emph{{All electroweak four-fermion
  processes in electron-positron collisions}}, Nucl. Phys. \textbf{B424}
  (1994), \href{http://inspirehep.net/search?p=hep-ph/9404313}{308},
  [\href{http://arXiv.org/pdf/hep-ph/9404313}{{\texttt{hep-ph/9404313}}}].
  \relax
 \relax
\bibitem{Krauss:2022ajk}
F.~Krauss, A.~Price and M.~Sch\"onherr, \emph{{YFS Resummation for Future
  Lepton-Lepton Colliders in SHERPA}}, SciPost Phys. \textbf{13} (2022), no.~2,
  \href{http://www.slac.stanford.edu/spires/find/hep/www?eprint=2203.10948}{026},
   [\href{http://arXiv.org/pdf/2203.10948}{{\texttt{arXiv:2203.10948}}}
  [hep-ph]]. \relax
 \relax
\bibitem{Krauss:2001iv}
F.~Krauss, R.~Kuhn and G.~Soff, \emph{{AMEGIC++ 1.0: A Matrix Element Generator
  In C++}}, JHEP \textbf{02} (2002),
  \href{http://inspirehep.net/search?p=hep-ph/0109036}{044},
  [\href{http://arXiv.org/pdf/hep-ph/0109036}{{\texttt{hep-ph/0109036}}}].
  \relax
 \relax
\bibitem{Gleisberg:2008fv}
T.~Gleisberg and S.~H{\"o}che, \emph{{Comix, a new matrix element generator}},
  JHEP \textbf{12} (2008), \href{http://inspirehep.net/record/793879}{039},
  [\href{http://arXiv.org/pdf/0808.3674}{{\texttt{arXiv:0808.3674}}} [hep-ph]].
  \relax
 \relax
\bibitem{Ellis:1975ap}
J.~R. Ellis, M.~K. Gaillard and D.~V. Nanopoulos, \emph{{A Phenomenological
  Profile of the Higgs Boson}}, Nucl.Phys. \textbf{B106} (1976),
  \href{http://www.slac.stanford.edu/spires/find/hep/www?j=NuclPhys,B106,292}{292}.
  \relax
 \relax
\bibitem{Wilczek:1977zn}
F.~Wilczek, \emph{{Decays of Heavy Vector Mesons Into Higgs Particles}},
  Phys.Rev.Lett. \textbf{39} (1977),
  \href{http://www.slac.stanford.edu/spires/find/hep/www?j=PhysRevLett,39,1304}{1304}.
  \relax
 \relax
\bibitem{Shifman:1979eb}
M.~A. Shifman, A.~Vainshtein, M.~Voloshin and V.~I. Zakharov, \emph{{Low-Energy
  Theorems for Higgs Boson Couplings to Photons}}, Sov.J.Nucl.Phys. \textbf{30}
  (1979),
  \href{http://www.slac.stanford.edu/spires/find/hep/www?j=SovJNuclPhys,30,711}{711--716}.
  \relax
 \relax
\bibitem{Ellis:1979jy}
J.~R. Ellis, M.~Gaillard, D.~V. Nanopoulos and C.~T. Sachrajda, \emph{{Is the
  Mass of the Higgs Boson About 10-GeV?}}, Phys.Lett. \textbf{B83} (1979),
  \href{http://www.slac.stanford.edu/spires/find/hep/www?j=PhysLett,B83,339}{339}.
  \relax
 \relax
\bibitem{Degrande:2011ua}
C.~Degrande, C.~Duhr, B.~Fuks, D.~Grellscheid, O.~Mattelaer and T.~Reiter,
  \emph{{UFO - The Universal FeynRules Output}}, Comput.Phys.Commun.
  \textbf{183} (2012),
  \href{http://www.slac.stanford.edu/spires/find/hep/www?eprint=1108.2040}{1201--1214},
   [\href{http://arXiv.org/pdf/1108.2040}{{\texttt{arXiv:1108.2040}}}
  [hep-ph]]. \relax
 \relax
\bibitem{Darme:2023jdn}
L.~Darm\'e et~al., \emph{{UFO 2.0: the \textquoteleft{}Universal Feynman
  Output\textquoteright{} format}}, Eur. Phys. J. C \textbf{83} (2023), no.~7,
  \href{http://www.slac.stanford.edu/spires/find/hep/www?eprint=2304.09883}{631},
   [\href{http://arXiv.org/pdf/2304.09883}{{\texttt{arXiv:2304.09883}}}
  [hep-ph]]. \relax
 \relax
\bibitem{Byckling:1969sx}
E.~Byckling and K.~Kajantie, \emph{{N-particle phase space in terms of
  invariant momentum transfers}}, Nucl. Phys. \textbf{B9} (1969),
  \href{http://inspirehep.net/search?j=NUPHA,B9,568}{568--576}. \relax
 \relax
\bibitem{Kleiss:1994qy}
R.~Kleiss and R.~Pittau, \emph{{Weight optimization in multichannel Monte
  Carlo}}, Comput. Phys. Commun. \textbf{83} (1994),
  \href{http://www.slac.stanford.edu/spires/find/hep/www?eprint=hep-ph/9405257}{141--146},
   [\href{http://arXiv.org/pdf/hep-ph/9405257}{{\texttt{arXiv:hep-ph/9405257}}}
  [hep-ph]]. \relax
 \relax
\bibitem{Berends:1994xn}
F.~A. Berends, R.~Pittau and R.~Kleiss, \emph{{Excalibur: A Monte Carlo program
  to evaluate all four fermion processes at LEP-200 and beyond}}, Comput. Phys.
  Commun. \textbf{85} (1995),
  \href{http://www.slac.stanford.edu/spires/find/hep/www?eprint=hep-ph/9409326}{437--452},
   [\href{http://arXiv.org/pdf/hep-ph/9409326}{{\texttt{hep-ph/9409326}}}].
  \relax
 \relax
\bibitem{Lepage:1977sw}
G.~P. Lepage, \emph{{A New Algorithm for Adaptive Multidimensional
  Integration}}, J. Comput. Phys. \textbf{27} (1978),
  \href{http://inspirehep.net/search?j=JCTPA,27,192}{192}. \relax
 \relax
\bibitem{Ohl:1998jn}
T.~Ohl, \emph{{Vegas revisited: Adaptive Monte Carlo integration beyond
  factorization}}, Comput. Phys. Commun. \textbf{120} (1999),
  \href{http://inspirehep.net/search?p=hep-ph/9806432}{13--19},
  [\href{http://arXiv.org/pdf/hep-ph/9806432}{{\texttt{hep-ph/9806432}}}].
  \relax
 \relax
\bibitem{Cascioli:2013gfa}
F.~Cascioli, S.~H{\"o}che, F.~Krauss, P.~Maierh{\"o}fer, S.~Pozzorini and
  F.~Siegert, \emph{{Precise Higgs-background predictions: merging NLO QCD and
  squared quark-loop corrections to four-lepton + 0,1 jet production}}, JHEP
  \textbf{01} (2014),
  \href{http://www.slac.stanford.edu/spires/find/hep/www?eprint=1309.0500}{046},
   [\href{http://arXiv.org/pdf/1309.0500}{{\texttt{arXiv:1309.0500}}}
  [hep-ph]]. \relax
 \relax
\bibitem{Goncalves:2015mfa}
D.~Goncalves, F.~Krauss, S.~Kuttimalai and P.~Maierh{\"o}fer,
  \emph{{Higgs-Strahlung: Merging the NLO Drell-Yan and Loop-Induced 0+1 jet
  Multiplicities}}, Phys. Rev. \textbf{D92} (2015), no.~7,
  \href{http://www.slac.stanford.edu/spires/find/hep/www?eprint=1509.01597}{073006},
   [\href{http://arXiv.org/pdf/1509.01597}{{\texttt{arXiv:1509.01597}}}
  [hep-ph]]. \relax
 \relax
\bibitem{Jones:2017giv}
S.~Jones and S.~Kuttimalai, \emph{{Parton Shower and NLO-Matching uncertainties
  in Higgs Boson Pair Production}}, JHEP \textbf{02} (2018),
  \href{http://www.slac.stanford.edu/spires/find/hep/www?eprint=1711.03319}{176},
   [\href{http://arXiv.org/pdf/1711.03319}{{\texttt{arXiv:1711.03319}}}
  [hep-ph]]. \relax
 \relax
\bibitem{Bothmann:2021led}
E.~Bothmann, D.~Napoletano, M.~Sch\"onherr, S.~Schumann and S.~L. Villani,
  \emph{{Higher-order EW corrections in ZZ and ZZj production at the LHC}},
  JHEP \textbf{06} (2022),
  \href{http://www.slac.stanford.edu/spires/find/hep/www?eprint=2111.13453}{064},
   [\href{http://arXiv.org/pdf/2111.13453}{{\texttt{arXiv:2111.13453}}}
  [hep-ph]]. \relax
 \relax
\bibitem{Frixione:1995ms}
S.~Frixione, Z.~Kunszt and A.~Signer, \emph{{Three-jet cross-sections to
  next-to-leading order}}, Nucl. Phys. \textbf{B467} (1996),
  \href{http://inspirehep.net/search?p=hep-ph/9512328}{399--442},
  [\href{http://arXiv.org/pdf/hep-ph/9512328}{{\texttt{hep-ph/9512328}}}].
  \relax
 \relax
\bibitem{Catani:1996vz}
S.~Catani and M.~H. Seymour, \emph{{A general algorithm for calculating jet
  cross sections in NLO QCD}}, Nucl. Phys. \textbf{B485} (1997),
  \href{http://inspirehep.net/search?p=hep-ph/9605323}{291--419},
  [\href{http://arXiv.org/pdf/hep-ph/9605323}{{\texttt{hep-ph/9605323}}}].
  \relax
 \relax
\bibitem{Catani:2002hc}
S.~Catani, S.~Dittmaier, M.~H. Seymour and Z.~Trocsanyi, \emph{{The dipole
  formalism for next-to-leading order QCD calculations with massive partons}},
  Nucl. Phys. \textbf{B627} (2002),
  \href{http://inspirehep.net/search?p=hep-ph/0201036}{189--265},
  [\href{http://arXiv.org/pdf/hep-ph/0201036}{{\texttt{hep-ph/0201036}}}].
  \relax
 \relax
\bibitem{Gleisberg:2007md}
T.~Gleisberg and F.~Krauss, \emph{{Automating dipole subtraction for QCD NLO
  calculations}}, Eur. Phys. J. \textbf{C53} (2008),
  \href{http://inspirehep.net/search?p=arXiv:0709.2881}{501--523},
  [\href{http://arXiv.org/pdf/0709.2881}{{\texttt{arXiv:0709.2881}}} [hep-ph]].
  \relax
 \relax
\bibitem{Schonherr:2017qcj}
M.~Sch{\"o}nherr, \emph{{An automated subtraction of NLO EW infrared
  divergences}}, Eur. Phys. J. \textbf{C78} (2018), no.~2,
  \href{http://www.slac.stanford.edu/spires/find/hep/www?eprint=1712.07975}{119},
   [\href{http://arXiv.org/pdf/1712.07975}{{\texttt{arXiv:1712.07975}}}
  [hep-ph]]. \relax
 \relax
\bibitem{Kallweit:2017khh}
S.~Kallweit, J.~M. Lindert, S.~Pozzorini and M.~Sch{\"o}nherr, \emph{{NLO
  QCD+EW predictions for $2\ell2\nu$ diboson signatures at the LHC}}, JHEP
  \textbf{11} (2017),
  \href{http://www.slac.stanford.edu/spires/find/hep/www?eprint=1705.00598}{120},
   [\href{http://arXiv.org/pdf/1705.00598}{{\texttt{arXiv:1705.00598}}}
  [hep-ph]]. \relax
 \relax
\bibitem{Chiesa:2017gqx}
M.~Chiesa, N.~Greiner, M.~Sch{\"o}nherr and F.~Tramontano, \emph{{Electroweak
  corrections to diphoton plus jets}}, JHEP \textbf{10} (2017),
  \href{http://www.slac.stanford.edu/spires/find/hep/www?eprint=1706.09022}{181},
   [\href{http://arXiv.org/pdf/1706.09022}{{\texttt{arXiv:1706.09022}}}
  [hep-ph]]. \relax
 \relax
\bibitem{Greiner:2017mft}
N.~Greiner and M.~Sch{\"o}nherr, \emph{{NLO QCD+EW corrections to diphoton
  production in association with a vector boson}}, JHEP \textbf{01} (2018),
  \href{http://www.slac.stanford.edu/spires/find/hep/www?eprint=1710.11514}{079},
   [\href{http://arXiv.org/pdf/1710.11514}{{\texttt{arXiv:1710.11514}}}
  [hep-ph]]. \relax
 \relax
\bibitem{Reyer:2019obz}
M.~Reyer, M.~Sch{\"o}nherr and S.~Schumann, \emph{{Full NLO corrections to
  3-jet production and $R_{32}$ at the LHC}}, Eur. Phys. J. \textbf{C79}
  (2019), no.~4,
  \href{http://www.slac.stanford.edu/spires/find/hep/www?eprint=1902.01763}{321},
   [\href{http://arXiv.org/pdf/1902.01763}{{\texttt{arXiv:1902.01763}}}
  [hep-ph]]. \relax
 \relax
\bibitem{Cascioli:2011va}
F.~Cascioli, P.~Maierh{\"o}fer and S.~Pozzorini, \emph{{Scattering Amplitudes
  with Open Loops}}, Phys.Rev.Lett. \textbf{108} (2012),
  \href{http://inspirehep.net/record/946998}{111601},
  [\href{http://arXiv.org/pdf/1111.5206}{{\texttt{arXiv:1111.5206}}} [hep-ph]].
  \relax
 \relax
\bibitem{Buccioni:2019sur}
F.~Buccioni, J.-N. Lang, J.~M. Lindert, P.~Maierh{\"o}fer, S.~Pozzorini,
  H.~Zhang and M.~F. Zoller, \emph{{OpenLoops 2}}, Eur. Phys. J. C \textbf{79}
  (2019), no.~10,
  \href{http://www.slac.stanford.edu/spires/find/hep/www?eprint=1907.13071}{866},
   [\href{http://arXiv.org/pdf/1907.13071}{{\texttt{arXiv:1907.13071}}}
  [hep-ph]]. \relax
 \relax
\bibitem{Actis:2012qn}
S.~Actis, A.~Denner, L.~Hofer, A.~Scharf and S.~Uccirati, \emph{{Recursive
  generation of one-loop amplitudes in the Standard Model}}, JHEP \textbf{1304}
  (2013),
  \href{http://www.slac.stanford.edu/spires/find/hep/www?eprint=1211.6316}{037},
   [\href{http://arXiv.org/pdf/1211.6316}{{\texttt{arXiv:1211.6316}}}
  [hep-ph]]. \relax
 \relax
\bibitem{Denner:2017wsf}
A.~Denner, J.-N. Lang and S.~Uccirati, \emph{{Recola2: REcursive Computation of
  One-Loop Amplitudes 2}}, Comput. Phys. Commun. \textbf{224} (2018),
  \href{http://www.slac.stanford.edu/spires/find/hep/www?eprint=1711.07388}{346--361},
   [\href{http://arXiv.org/pdf/1711.07388}{{\texttt{arXiv:1711.07388}}}
  [hep-ph]]. \relax
 \relax
\bibitem{Biedermann:2017yoi}
B.~Biedermann, S.~Br{\"a}uer, A.~Denner, M.~Pellen, S.~Schumann and J.~M.
  Thompson, \emph{{Automation of NLO QCD and EW corrections with Sherpa and
  Recola}}, Eur. Phys. J. \textbf{C77} (2017),
  \href{http://www.slac.stanford.edu/spires/find/hep/www?eprint=1704.05783}{492},
   [\href{http://arXiv.org/pdf/1704.05783}{{\texttt{arXiv:1704.05783}}}
  [hep-ph]]. \relax
 \relax
\bibitem{Hirschi:2011pa}
V.~Hirschi, R.~Frederix, S.~Frixione, M.~V. Garzelli, F.~Maltoni and R.~Pittau,
  \emph{{Automation of one-loop QCD corrections}}, JHEP \textbf{05} (2011),
  \href{http://inspirebeta.net/record/891365}{044},
  [\href{http://arXiv.org/pdf/1103.0621}{{\texttt{arXiv:1103.0621}}} [hep-ph]].
  \relax
 \relax
\bibitem{Campbell:2021vlt}
J.~M. Campbell, S.~H\"oche and C.~T. Preuss, \emph{{Accelerating LHC
  phenomenology with analytic one-loop amplitudes: A C++ interface to MCFM}},
  Eur. Phys. J. C \textbf{81} (2021), no.~12,
  \href{http://www.slac.stanford.edu/spires/find/hep/www?eprint=2107.04472}{1117},
   [\href{http://arXiv.org/pdf/2107.04472}{{\texttt{arXiv:2107.04472}}}
  [hep-ph]]. \relax
 \relax
\bibitem{Hoche:2014uhw}
S.~H\"oche, Y.~Li and S.~Prestel, \emph{{Drell-Yan lepton pair production at
  NNLO QCD with parton showers}}, Phys. Rev. D \textbf{91} (2015), no.~7,
  \href{http://www.slac.stanford.edu/spires/find/hep/www?eprint=1405.3607}{074015},
   [\href{http://arXiv.org/pdf/1405.3607}{{\texttt{arXiv:1405.3607}}}
  [hep-ph]]. \relax
 \relax
\bibitem{Hoche:2014dla}
S.~H{\"o}che, Y.~Li and S.~Prestel, \emph{{Higgs-boson production through gluon
  fusion at NNLO QCD with parton showers}}, Phys.Rev. \textbf{D90} (2014),
  \href{http://www.slac.stanford.edu/spires/find/hep/www?eprint=1407.3773}{054011},
   [\href{http://arXiv.org/pdf/1407.3773}{{\texttt{arXiv:1407.3773}}}
  [hep-ph]]. \relax
 \relax
\bibitem{Hoche:2018gti}
S.~H{\"o}che, S.~Kuttimalai and Y.~Li, \emph{{Hadronic Final States in DIS at
  NNLO QCD with Parton Showers}}, Phys. Rev. \textbf{D98} (2018), no.~11,
  \href{http://www.slac.stanford.edu/spires/find/hep/www?eprint=1809.04192}{114013},
   [\href{http://arXiv.org/pdf/1809.04192}{{\texttt{arXiv:1809.04192}}}
  [hep-ph]]. \relax
 \relax
\bibitem{Catani:2007vq}
S.~Catani and M.~Grazzini, \emph{{An NNLO subtraction formalism in hadron
  collisions and its application to Higgs boson production at the LHC}},
  Phys.Rev.Lett. \textbf{98} (2007),
  \href{http://www.slac.stanford.edu/spires/find/hep/www?eprint=hep-ph/0703012}{222002},
   [\href{http://arXiv.org/pdf/hep-ph/0703012}{{\texttt{arXiv:hep-ph/0703012}}}
  [hep-ph]]. \relax
 \relax
\bibitem{Catani:2009sm}
S.~Catani, L.~Cieri, G.~Ferrera, D.~de~Florian and M.~Grazzini, \emph{{Vector
  boson production at hadron colliders: a fully exclusive QCD calculation at
  NNLO}}, Phys.Rev.Lett. \textbf{103} (2009),
  \href{http://www.slac.stanford.edu/spires/find/hep/www?eprint=0903.2120}{082001},
   [\href{http://arXiv.org/pdf/0903.2120}{{\texttt{arXiv:0903.2120}}}
  [hep-ph]]. \relax
 \relax
\bibitem{Cacciari:2015jma}
M.~Cacciari, F.~A. Dreyer, A.~Karlberg, G.~P. Salam and G.~Zanderighi,
  \emph{{Fully Differential Vector-Boson-Fusion Higgs Production at
  Next-to-Next-to-Leading Order}}, Phys. Rev. Lett. \textbf{115} (2015), no.~8,
  \href{http://www.slac.stanford.edu/spires/find/hep/www?eprint=1506.02660}{082002},
   [\href{http://arXiv.org/pdf/1506.02660}{{\texttt{arXiv:1506.02660}}}
  [hep-ph]], [Erratum: Phys. Rev. Lett.120,no.13,139901(2018)]. \relax
 \relax
\bibitem{Hoche:2014kca}
S.~H{\"o}che, S.~Kuttimalai, S.~Schumann and F.~Siegert, \emph{{Beyond Standard
  Model calculations with Sherpa}}, Eur. Phys. J. \textbf{C75} (2015), no.~3,
  \href{http://inspirehep.net/record/1335162}{135},
  [\href{http://arXiv.org/pdf/1412.6478}{{\texttt{arXiv:1412.6478}}} [hep-ph]].
  \relax
 \relax
\bibitem{Collins:1987cp}
J.~C. Collins, \emph{{Spin correlations in Monte Carlo event generators}},
  Nucl.Phys. \textbf{B304} (1988),
  \href{http://inspirehep.net/record/249338}{794}. \relax
 \relax
\bibitem{Knowles:1987cu}
I.~Knowles, \emph{{Angular Correlations in {QCD}}}, Nucl.Phys. \textbf{B304}
  (1988),
  \href{http://www.slac.stanford.edu/spires/find/hep/www?j=NuclPhys,B304,767}{767}.
  \relax
 \relax
\bibitem{Knowles:1988vs}
I.~Knowles, \emph{{Spin Correlations in Parton - Parton Scattering}},
  Nucl.Phys. \textbf{B310} (1988),
  \href{http://inspirehep.net/record/261839}{571}. \relax
 \relax
\bibitem{Richardson:2001df}
P.~Richardson, \emph{{Spin correlations in Monte Carlo simulations}}, JHEP
  \textbf{11} (2001),
  \href{http://inspirehep.net/search?p=hep-ph/0110108}{029},
  [\href{http://arXiv.org/pdf/hep-ph/0110108}{{\texttt{hep-ph/0110108}}}].
  \relax
 \relax
\bibitem{ILC:2013jhg}
\href{http://www.slac.stanford.edu/spires/find/hep/www?eprint=1306.6352}{}ILC
  collaboration, \emph{{The International Linear Collider Technical Design
  Report - Volume 2: Physics}},
  \href{http://arXiv.org/pdf/1306.6352}{{\texttt{arXiv:1306.6352}}} [hep-ph].
  \relax
 \relax
\bibitem{CLICdp:2018cto}
\href{http://www.slac.stanford.edu/spires/find/hep/www?eprint=1812.06018}{T.~K.
  Charles et~al.}, CLICdp, CLIC collaboration, \emph{{The Compact Linear
  Collider (CLIC) - 2018 Summary Report}},
  \href{http://arXiv.org/pdf/1812.06018}{{\texttt{arXiv:1812.06018}}}
  [physics.acc-ph]. \relax
 \relax
\bibitem{Moortgat-Pick:2005jsx}
G.~Moortgat-Pick et~al., \emph{{The Role of polarized positrons and electrons
  in revealing fundamental interactions at the linear collider}}, Phys. Rept.
  \textbf{460} (2008),
  \href{http://www.slac.stanford.edu/spires/find/hep/www?eprint=hep-ph/0507011}{131--243},
   [\href{http://arXiv.org/pdf/hep-ph/0507011}{{\texttt{hep-ph/0507011}}}].
  \relax
 \relax
\bibitem{Kilian:2007gr}
W.~Kilian, T.~Ohl and J.~Reuter, \emph{{WHIZARD: Simulating Multi-Particle
  Processes at LHC and ILC}}, Eur. Phys. J. \textbf{C71} (2007),
  \href{http://inspirehep.net/record/759495}{1742},
  [\href{http://arXiv.org/pdf/0708.4233}{{\texttt{arXiv:0708.4233}}} [hep-ph]].
  \relax
 \relax
\bibitem{Alwall:2014hca}
J.~Alwall, R.~Frederix, S.~Frixione, V.~Hirschi, F.~Maltoni, O.~Mattelaer,
  H.-S. Shao, T.~Stelzer, P.~Torrielli and M.~Zaro, \emph{{The automated
  computation of tree-level and next-to-leading order differential cross
  sections, and their matching to parton shower simulations}}, JHEP \textbf{07}
  (2014),
  \href{http://www.slac.stanford.edu/spires/find/hep/www?eprint=1405.0301}{079},
   [\href{http://arXiv.org/pdf/1405.0301}{{\texttt{arXiv:1405.0301}}}
  [hep-ph]]. \relax
 \relax
\bibitem{Hoppe:2023uux}
M.~Hoppe, M.~Sch\"onherr and F.~Siegert, \emph{{Polarised cross sections for
  vector boson production with Sherpa}}, JHEP \textbf{04} (2024),
  \href{http://www.slac.stanford.edu/spires/find/hep/www?eprint=2310.14803}{001},
   [\href{http://arXiv.org/pdf/2310.14803}{{\texttt{arXiv:2310.14803}}}
  [hep-ph]]. \relax
 \relax
\bibitem{Bothmann:2022pwf}
E.~Bothmann et~al., \emph{{A standard convention for particle-level Monte Carlo
  event-variation weights}}, SciPost Phys. Core \textbf{6} (2023),
  \href{http://www.slac.stanford.edu/spires/find/hep/www?eprint=2203.08230}{007},
   [\href{http://arXiv.org/pdf/2203.08230}{{\texttt{arXiv:2203.08230}}}
  [hep-ph]]. \relax
 \relax
\bibitem{Denner:2020eck}
A.~Denner and G.~Pelliccioli, \emph{{NLO QCD predictions for doubly-polarized
  WZ production at the LHC}}, Phys. Lett. B \textbf{814} (2021),
  \href{http://www.slac.stanford.edu/spires/find/hep/www?eprint=2010.07149}{136107},
   [\href{http://arXiv.org/pdf/2010.07149}{{\texttt{arXiv:2010.07149}}}
  [hep-ph]]. \relax
 \relax
\bibitem{Kuzmin:1985mm}
V.~A. Kuzmin, V.~A. Rubakov and M.~E. Shaposhnikov, \emph{{On the Anomalous
  Electroweak Baryon Number Nonconservation in the Early Universe}}, Phys.
  Lett. B \textbf{155} (1985),
  \href{http://www.slac.stanford.edu/spires/find/hep/www?j=Phys%20Lett%20B,155,36}{36}.
  \relax
 \relax
\bibitem{Fukugita:1986hr}
M.~Fukugita and T.~Yanagida, \emph{{Baryogenesis Without Grand Unification}},
  Phys. Lett. B \textbf{174} (1986),
  \href{http://www.slac.stanford.edu/spires/find/hep/www?j=Phys%20Lett%20B,174,45}{45--47}.
  \relax
 \relax
\bibitem{tHooft:1976rip}
G.~'t~Hooft, \emph{{Symmetry Breaking Through Bell-Jackiw Anomalies}}, Phys.
  Rev. Lett. \textbf{37} (1976),
  \href{http://www.slac.stanford.edu/spires/find/hep/www?j=Phys%20Rev%20Lett,37,8}{8--11}.
  \relax
 \relax
\bibitem{Khoze:2019jta}
V.~V. Khoze, F.~Krauss and M.~Schott, \emph{{Large Effects from Small QCD
  Instantons: Making Soft Bombs at Hadron Colliders}}, JHEP \textbf{04} (2020),
  \href{http://www.slac.stanford.edu/spires/find/hep/www?eprint=1911.09726}{201},
   [\href{http://arXiv.org/pdf/1911.09726}{{\texttt{arXiv:1911.09726}}}
  [hep-ph]]. \relax
 \relax
\bibitem{Christensen:2009jx}
N.~D. Christensen, P.~de~Aquino, C.~Degrande, C.~Duhr, B.~Fuks, M.~Herquet,
  F.~Maltoni and S.~Schumann, \emph{{A comprehensive approach to new physics
  simulations}}, Eur. Phys. J. \textbf{C71} (2011),
  \href{http://inspirebeta.net/record/823106}{1541},
  [\href{http://arXiv.org/pdf/0906.2474}{{\texttt{arXiv:0906.2474}}} [hep-ph]].
  \relax
 \relax
\bibitem{Martin:1997ns}
S.~P. Martin, \emph{{A Supersymmetry primer}}, Adv. Ser. Direct. High Energy
  Phys. \textbf{18} (1998),
  \href{http://www.slac.stanford.edu/spires/find/hep/www?eprint=hep-ph/9709356}{1--98},
   [\href{http://arXiv.org/pdf/hep-ph/9709356}{{\texttt{hep-ph/9709356}}}].
  \relax
 \relax
\bibitem{Skands:2003cj}
P.~Skands et~al., \emph{{SUSY Les Houches accord: interfacing SUSY spectrum
  calculators, decay packages, and event generators}}, JHEP \textbf{07} (2004),
  \href{http://inspirehep.net/search?p=hep-ph/0311123}{036},
  [\href{http://arXiv.org/pdf/hep-ph/0311123}{{\texttt{hep-ph/0311123}}}].
  \relax
 \relax
\bibitem{Hagiwara:1986vm}
K.~Hagiwara, R.~D. Peccei, D.~Zeppenfeld and K.~Hikasa, \emph{{Probing the weak
  boson sector in $e^+e^-\to W^+W^-$}}, Nucl. Phys. \textbf{B282} (1987),
  \href{http://inspirehep.net/search?j=NUPHA,B282,253}{253}. \relax
 \relax
\bibitem{Belyaev:1998ih}
A.~Belyaev, O.~J. Eboli, M.~Gonzalez-Garcia, J.~Mizukoshi, S.~Novaes and
  I.~Zacharov, \emph{{Strongly interacting vector bosons at the CERN LHC:
  Quartic anomalous couplings}}, Phys.Rev. \textbf{D59} (1999),
  \href{http://www.slac.stanford.edu/spires/find/hep/www?eprint=hep-ph/9805229}{015022},
   [\href{http://arXiv.org/pdf/hep-ph/9805229}{{\texttt{arXiv:hep-ph/9805229}}}
  [hep-ph]]. \relax
 \relax
\bibitem{Eboli:2000ad}
O.~J. Eboli, M.~Gonzalez-Garcia, S.~Lietti and S.~Novaes, \emph{{Anomalous
  quartic gauge boson couplings at hadron colliders}}, Phys.Rev. \textbf{D63}
  (2001),
  \href{http://www.slac.stanford.edu/spires/find/hep/www?eprint=hep-ph/0009262}{075008},
   [\href{http://arXiv.org/pdf/hep-ph/0009262}{{\texttt{arXiv:hep-ph/0009262}}}
  [hep-ph]]. \relax
 \relax
\bibitem{Eboli:2003nq}
O.~Eboli, M.~Gonzalez-Garcia and S.~Lietti, \emph{{Bosonic quartic couplings at
  CERN LHC}}, Phys.Rev. \textbf{D69} (2004),
  \href{http://www.slac.stanford.edu/spires/find/hep/www?eprint=hep-ph/0310141}{095005},
   [\href{http://arXiv.org/pdf/hep-ph/0310141}{{\texttt{arXiv:hep-ph/0310141}}}
  [hep-ph]]. \relax
 \relax
\bibitem{Eboli:2006wa}
O.~Eboli, M.~Gonzalez-Garcia and J.~Mizukoshi, \emph{{$p p\to j j e^\pm \mu^\pm
  \nu \nu$ and $j j e^\pm \mu^\mp \nu \nu$ at $O(\alpha_{em}^6)$ and
  $O(\alpha_{em}^4 \alpha_s^2)$ for the study of the quartic electroweak gauge
  boson vertex at CERN LHC}}, Phys.Rev. \textbf{D74} (2006),
  \href{http://www.slac.stanford.edu/spires/find/hep/www?eprint=hep-ph/0606118}{073005},
   [\href{http://arXiv.org/pdf/hep-ph/0606118}{{\texttt{hep-ph/0606118}}}].
  \relax
 \relax
\bibitem{Biekotter:2020flu}
A.~Biek\"otter, R.~Gomez-Ambrosio, P.~Gregg, F.~Krauss and M.~Sch\"onherr,
  \emph{{Constraining SMEFT operators with associated $h\gamma$ production in
  weak boson fusion}}, Phys. Lett. B \textbf{814} (2021),
  \href{http://www.slac.stanford.edu/spires/find/hep/www?eprint=2003.06379}{136079},
   [\href{http://arXiv.org/pdf/2003.06379}{{\texttt{arXiv:2003.06379}}}
  [hep-ph]]. \relax
 \relax
\bibitem{Biekotter:2021int}
A.~Biek\"otter, P.~Gregg, F.~Krauss and M.~Sch\"onherr, \emph{{Constraining CP
  violating operators in charged and neutral triple gauge couplings}}, Phys.
  Lett. B \textbf{817} (2021),
  \href{http://www.slac.stanford.edu/spires/find/hep/www?eprint=2102.01115}{136311},
   [\href{http://arXiv.org/pdf/2102.01115}{{\texttt{arXiv:2102.01115}}}
  [hep-ph]]. \relax
 \relax
\bibitem{Banerjee:2024eyo}
\href{http://www.slac.stanford.edu/spires/find/hep/www?eprint=2406.15640}{S.~Banerjee,
  D.~Reichelt and M.~Spannowsky}, \emph{{Electroweak Corrections and EFT
  Operators in $W^+W^-$ production at the LHC}},
  \href{http://arXiv.org/pdf/2406.15640}{{\texttt{arXiv:2406.15640}}} [hep-ph].
  \relax
 \relax
\bibitem{Christensen:2013aua}
N.~D. Christensen, P.~de~Aquino, N.~Deutschmann, C.~Duhr, B.~Fuks,
  C.~Garcia-Cely, O.~Mattelaer, K.~Mawatari, B.~Oexl and Y.~Takaesu,
  \emph{{Simulating spin-$ \frac{3}{2}$ particles at colliders}}, Eur. Phys. J.
  C \textbf{73} (2013), no.~10,
  \href{http://www.slac.stanford.edu/spires/find/hep/www?eprint=1308.1668}{2580},
   [\href{http://arXiv.org/pdf/1308.1668}{{\texttt{arXiv:1308.1668}}}
  [hep-ph]]. \relax
 \relax
\bibitem{Alwall:2014bza}
J.~Alwall, C.~Duhr, B.~Fuks, O.~Mattelaer, D.~G. \"Ozt\"urk and C.-H. Shen,
  \emph{{Computing decay rates for new physics theories with FeynRules and
  MadGraph 5\_aMC@NLO}}, Comput. Phys. Commun. \textbf{197} (2015),
  \href{http://www.slac.stanford.edu/spires/find/hep/www?eprint=1402.1178}{312--323},
   [\href{http://arXiv.org/pdf/1402.1178}{{\texttt{arXiv:1402.1178}}}
  [hep-ph]]. \relax
 \relax
\bibitem{Aoude:2022aro}
R.~Aoude, F.~Maltoni, O.~Mattelaer, C.~Severi and E.~Vryonidou,
  \emph{{Renormalisation group effects on SMEFT interpretations of LHC data}},
  JHEP \textbf{09} (2023),
  \href{http://www.slac.stanford.edu/spires/find/hep/www?eprint=2212.05067}{191},
   [\href{http://arXiv.org/pdf/2212.05067}{{\texttt{arXiv:2212.05067}}}
  [hep-ph]]. \relax
 \relax
\bibitem{Webber:1983if}
B.~R. Webber, \emph{{A QCD model for jet fragmentation including soft gluon
  interference}}, Nucl. Phys. \textbf{B238} (1984),
  \href{http://inspirehep.net/search?j=NUPHA,B238,492}{492}. \relax
 \relax
\bibitem{Bengtsson:1986gz}
M.~Bengtsson, T.~Sj{\"o}strand and M.~van Zijl, \emph{{Initial State Radiation
  Effects on $W$ and Jet Production}}, Z. Phys. \textbf{C32} (1986),
  \href{http://inspirehep.net/search?j=ZEPYA,C32,67}{67}. \relax
 \relax
\bibitem{Bengtsson:1986et}
M.~Bengtsson and T.~Sj{\"o}strand, \emph{{A comparative study of coherent and
  non-coherent parton shower evolution}}, Nucl. Phys. \textbf{B289} (1987),
  \href{http://inspirehep.net/search?j=NUPHA,B289,810}{810}. \relax
 \relax
\bibitem{Marchesini:1987cf}
G.~Marchesini and B.~R. Webber, \emph{{Monte Carlo Simulation of General Hard
  Processes with Coherent QCD Radiation}}, Nucl. Phys. \textbf{B310} (1988),
  \href{http://inspirehep.net/search?j=NUPHA,B310,461}{461}. \relax
 \relax
\bibitem{Gustafson:1987rq}
G.~Gustafson and U.~Pettersson, \emph{{Dipole formulation of QCD cascades}},
  Nucl. Phys. \textbf{B306} (1988),
  \href{http://inspirehep.net/search?j=NUPHA,B306,746}{746}. \relax
 \relax
\bibitem{Andersson:1989ki}
B.~Andersson, G.~Gustafson and L.~L{\"o}nnblad, \emph{{Gluon splitting in the
  color dipole cascades}}, Nucl. Phys. \textbf{B339} (1990),
  \href{http://inspirehep.net/search?j=NUPHA,B339,393}{393--406}. \relax
 \relax
\bibitem{Lonnblad:1992tz}
L.~L{\"o}nnblad, \emph{{Ariadne version 4: A program for simulation of QCD
  cascades implementing the colour dipole model}}, Comput. Phys. Commun.
  \textbf{71} (1992),
  \href{http://inspirehep.net/search?j=CPHCB,71,15}{15--31}. \relax
 \relax
\bibitem{Schumann:2007mg}
S.~Schumann and F.~Krauss, \emph{{A parton shower algorithm based on
  Catani-Seymour dipole factorisation}}, JHEP \textbf{03} (2008),
  \href{http://inspirehep.net/search?p=arXiv:0709.1027}{038},
  [\href{http://arXiv.org/pdf/0709.1027}{{\texttt{arXiv:0709.1027}}} [hep-ph]].
  \relax
 \relax
\bibitem{Nagy:2006kb}
\href{http://inspirehep.net/search?p=hep-ph/0601021}{Z.~Nagy and D.~E. Soper},
  \emph{{A new parton shower algorithm: Shower evolution, matching at leading
  and next-to-leading order level}},
  \href{http://arXiv.org/pdf/hep-ph/0601021}{{\texttt{hep-ph/0601021}}}. \relax
 \relax
\bibitem{Hoeche:2014lxa}
S.~H{\"o}che, F.~Krauss and M.~Sch{\"o}nherr, \emph{{Uncertainties in MEPS@NLO
  calculations of h+jets}}, Phys. Rev. \textbf{D90} (2014), no.~1,
  \href{http://www.slac.stanford.edu/spires/find/hep/www?eprint=1401.7971}{014012},
   [\href{http://arXiv.org/pdf/1401.7971}{{\texttt{arXiv:1401.7971}}}
  [hep-ph]]. \relax
 \relax
\bibitem{Krauss:2016orf}
F.~Krauss, D.~Napoletano and S.~Schumann, \emph{{Simulating $b$-associated
  production of $Z$ and Higgs bosons with the SHERPA event generator}}, Phys.
  Rev. \textbf{D95} (2017), no.~3,
  \href{http://www.slac.stanford.edu/spires/find/hep/www?eprint=1612.04640}{036012},
   [\href{http://arXiv.org/pdf/1612.04640}{{\texttt{arXiv:1612.04640}}}
  [hep-ph]]. \relax
 \relax
\bibitem{Krauss:2017wmx}
F.~Krauss and D.~Napoletano, \emph{{Towards a fully massive five-flavor
  scheme}}, Phys. Rev. \textbf{D98} (2018), no.~9,
  \href{http://www.slac.stanford.edu/spires/find/hep/www?eprint=1712.06832}{096002},
   [\href{http://arXiv.org/pdf/1712.06832}{{\texttt{arXiv:1712.06832}}}
  [hep-ph]]. \relax
 \relax
\bibitem{Hoeche:2009xc}
S.~H{\"o}che, S.~Schumann and F.~Siegert, \emph{{Hard photon production and
  matrix-element parton-shower merging}}, Phys. Rev. \textbf{D81} (2010),
  \href{http://inspirehep.net/search?p=arXiv:0912.3501}{034026},
  [\href{http://arXiv.org/pdf/0912.3501}{{\texttt{arXiv:0912.3501}}} [hep-ph]].
  \relax
 \relax
\bibitem{Hoche:2015sya}
S.~H{\"o}che and S.~Prestel, \emph{{The midpoint between dipole and parton
  showers}}, Eur. Phys. J. \textbf{C75} (2015), no.~9,
  \href{http://www.slac.stanford.edu/spires/find/hep/www?eprint=1506.05057}{461},
   [\href{http://arXiv.org/pdf/1506.05057}{{\texttt{arXiv:1506.05057}}}
  [hep-ph]]. \relax
 \relax
\bibitem{Hoche:2017iem}
S.~H{\"o}che and S.~Prestel, \emph{{Triple collinear emissions in parton
  showers}}, Phys. Rev. \textbf{D96} (2017), no.~7,
  \href{http://www.slac.stanford.edu/spires/find/hep/www?eprint=1705.00742}{074017},
   [\href{http://arXiv.org/pdf/1705.00742}{{\texttt{arXiv:1705.00742}}}
  [hep-ph]]. \relax
 \relax
\bibitem{Dulat:2018vuy}
F.~Dulat, S.~H{\"o}che and S.~Prestel, \emph{{Leading-Color Fully Differential
  Two-Loop Soft Corrections to QCD Dipole Showers}}, Phys. Rev. \textbf{D98}
  (2018), no.~7,
  \href{http://www.slac.stanford.edu/spires/find/hep/www?eprint=1805.03757}{074013},
   [\href{http://arXiv.org/pdf/1805.03757}{{\texttt{arXiv:1805.03757}}}
  [hep-ph]]. \relax
 \relax
\bibitem{Gellersen:2021eci}
L.~Gellersen, S.~H\"oche and S.~Prestel, \emph{{Disentangling soft and
  collinear effects in QCD parton showers}}, Phys. Rev. D \textbf{105} (2022),
  no.~11,
  \href{http://www.slac.stanford.edu/spires/find/hep/www?eprint=2110.05964}{114012},
   [\href{http://arXiv.org/pdf/2110.05964}{{\texttt{arXiv:2110.05964}}}
  [hep-ph]]. \relax
 \relax
\bibitem{Dasgupta:2018nvj}
M.~Dasgupta, F.~A. Dreyer, K.~Hamilton, P.~F. Monni and G.~P. Salam,
  \emph{{Logarithmic accuracy of parton showers: a fixed-order study}}, JHEP
  \textbf{09} (2018),
  \href{http://www.slac.stanford.edu/spires/find/hep/www?eprint=1805.09327}{033},
   [\href{http://arXiv.org/pdf/1805.09327}{{\texttt{arXiv:1805.09327}}}
  [hep-ph]]. \relax
 \relax
\bibitem{Herren:2022jej}
F.~Herren, S.~H\"oche, F.~Krauss, D.~Reichelt and M.~Sch\"onherr, \emph{{A new
  approach to color-coherent parton evolution}}, JHEP \textbf{10} (2023),
  \href{http://www.slac.stanford.edu/spires/find/hep/www?eprint=2208.06057}{091},
   [\href{http://arXiv.org/pdf/2208.06057}{{\texttt{arXiv:2208.06057}}}
  [hep-ph]]. \relax
 \relax
\bibitem{Dasgupta:2020fwr}
M.~Dasgupta, F.~A. Dreyer, K.~Hamilton, P.~F. Monni, G.~P. Salam and G.~Soyez,
  \emph{{Parton showers beyond leading logarithmic accuracy}}, Phys. Rev. Lett.
  \textbf{125} (2020), no.~5,
  \href{http://www.slac.stanford.edu/spires/find/hep/www?eprint=2002.11114}{052002},
   [\href{http://arXiv.org/pdf/2002.11114}{{\texttt{arXiv:2002.11114}}}
  [hep-ph]]. \relax
 \relax
\bibitem{Assi:2023rbu}
B.~Assi and S.~H\"oche, \emph{{New approach to QCD final-state evolution in
  processes with massive partons}}, Phys. Rev. D \textbf{109} (2024), no.~11,
  \href{http://www.slac.stanford.edu/spires/find/hep/www?eprint=2307.00728}{114008},
   [\href{http://arXiv.org/pdf/2307.00728}{{\texttt{arXiv:2307.00728}}}
  [hep-ph]]. \relax
 \relax
\bibitem{Hoche:2024dee}
\href{http://www.slac.stanford.edu/spires/find/hep/www?eprint=2404.14360}{S.~H\"oche,
  F.~Krauss and D.~Reichelt}, \emph{{The Alaric parton shower for hadron
  colliders}},
  \href{http://arXiv.org/pdf/2404.14360}{{\texttt{arXiv:2404.14360}}} [hep-ph].
  \relax
 \relax
\bibitem{Hoeche:2017jsi}
S.~H{\"o}che, D.~Reichelt and F.~Siegert, \emph{{Momentum conservation and
  unitarity in parton showers and NLL resummation}}, JHEP \textbf{01} (2018),
  \href{http://www.slac.stanford.edu/spires/find/hep/www?eprint=1711.03497}{118},
   [\href{http://arXiv.org/pdf/1711.03497}{{\texttt{arXiv:1711.03497}}}
  [hep-ph]]. \relax
 \relax
\bibitem{ATLAS:2012nnf}
G.~Aad et~al., ATLAS collaboration, \emph{{ATLAS Measurements of the Properties
  of Jets for Boosted Particle Searches}}, Phys. Rev. D \textbf{86} (2012),
  \href{http://www.slac.stanford.edu/spires/find/hep/www?eprint=1206.5369}{072006},
   [\href{http://arXiv.org/pdf/1206.5369}{{\texttt{arXiv:1206.5369}}}
  [hep-ex]]. \relax
 \relax
\bibitem{Frixione:2002ik}
S.~Frixione and B.~R. Webber, \emph{{Matching NLO QCD computations and parton
  shower simulations}}, JHEP \textbf{06} (2002),
  \href{http://inspirehep.net/search?p=hep-ph/0204244}{029},
  [\href{http://arXiv.org/pdf/hep-ph/0204244}{{\texttt{hep-ph/0204244}}}].
  \relax
 \relax
\bibitem{Nason:2004rx}
P.~Nason, \emph{{A new method for combining NLO QCD with shower Monte Carlo
  algorithms}}, JHEP \textbf{11} (2004),
  \href{http://inspirebeta.net/record/659055}{040},
  [\href{http://arXiv.org/pdf/hep-ph/0409146}{{\texttt{hep-ph/0409146}}}].
  \relax
 \relax
\bibitem{Frixione:2007vw}
S.~Frixione, P.~Nason and C.~Oleari, \emph{{Matching NLO QCD computations with
  parton shower simulations: the POWHEG method}}, JHEP \textbf{11} (2007),
  \href{http://inspirehep.net/search?p=arXiv:0709.2092}{070},
  [\href{http://arXiv.org/pdf/0709.2092}{{\texttt{arXiv:0709.2092}}} [hep-ph]].
  \relax
 \relax
\bibitem{Jadach:2015mza}
S.~Jadach, W.~P\l{}aczek, S.~Sapeta, A.~Si\'odmok and M.~Skrzypek,
  \emph{{Matching NLO QCD with parton shower in Monte Carlo scheme
  \textemdash{} the KrkNLO method}}, JHEP \textbf{10} (2015),
  \href{http://www.slac.stanford.edu/spires/find/hep/www?eprint=1503.06849}{052},
   [\href{http://arXiv.org/pdf/1503.06849}{{\texttt{arXiv:1503.06849}}}
  [hep-ph]]. \relax
 \relax
\bibitem{Lonnblad:2012ix}
L.~L{\"o}nnblad and S.~Prestel, \emph{{Merging Multi-leg NLO Matrix Elements
  with Parton Showers}}, JHEP \textbf{03} (2013),
  \href{http://inspirehep.net/record/1205021}{166},
  [\href{http://arXiv.org/pdf/1211.7278}{{\texttt{arXiv:1211.7278}}} [hep-ph]].
  \relax
 \relax
\bibitem{Nason:2021xke}
P.~Nason and G.~P. Salam, \emph{{Multiplicative-accumulative matching of NLO
  calculations with parton showers}}, JHEP \textbf{01} (2022),
  \href{http://www.slac.stanford.edu/spires/find/hep/www?eprint=2111.03553}{067},
   [\href{http://arXiv.org/pdf/2111.03553}{{\texttt{arXiv:2111.03553}}}
  [hep-ph]]. \relax
 \relax
\bibitem{Hoeche:2011fd}
S.~H{\"o}che, F.~Krauss, M.~Sch{\"o}nherr and F.~Siegert, \emph{{A critical
  appraisal of NLO+PS matching methods}}, JHEP \textbf{09} (2012),
  \href{http://inspirehep.net/record/944643}{049},
  [\href{http://arXiv.org/pdf/1111.1220}{{\texttt{arXiv:1111.1220}}} [hep-ph]].
  \relax
 \relax
\bibitem{Hoeche:2012fm}
S.~H{\"o}che and M.~Sch{\"o}nherr, \emph{{Uncertainties in next-to-leading
  order plus parton shower matched simulations of inclusive jet and dijet
  production}}, Phys.Rev. \textbf{D86} (2012),
  \href{http://inspirehep.net/record/1127523}{094042},
  [\href{http://arXiv.org/pdf/1208.2815}{{\texttt{arXiv:1208.2815}}} [hep-ph]].
  \relax
 \relax
\bibitem{Hoeche:2013mua}
S.~H{\"o}che, J.~Huang, G.~Luisoni, M.~Sch{\"o}nherr and J.~Winter, \emph{{Zero
  and one jet combined NLO analysis of the top quark forward-backward
  asymmetry}}, Phys.Rev. \textbf{D88} (2013),
  \href{http://inspirehep.net/record/1238288}{014040},
  [\href{http://arXiv.org/pdf/1306.2703}{{\texttt{arXiv:1306.2703}}} [hep-ph]].
  \relax
 \relax
\bibitem{Bellm:2019yyh}
J.~Bellm et~al., \emph{{Jet Cross Sections at the LHC and the Quest for Higher
  Precision}}, Eur. Phys. J. C \textbf{80} (2020), no.~2,
  \href{http://www.slac.stanford.edu/spires/find/hep/www?eprint=1903.12563}{93},
   [\href{http://arXiv.org/pdf/1903.12563}{{\texttt{arXiv:1903.12563}}}
  [hep-ph]]. \relax
 \relax
\bibitem{Buckley:2021gfw}
A.~Buckley et~al., \emph{{A comparative study of Higgs boson production from
  vector-boson fusion}}, JHEP \textbf{11} (2021),
  \href{http://www.slac.stanford.edu/spires/find/hep/www?eprint=2105.11399}{108},
   [\href{http://arXiv.org/pdf/2105.11399}{{\texttt{arXiv:2105.11399}}}
  [hep-ph]]. \relax
 \relax
\bibitem{Hoche:2012tae}
S.~H\"oche, F.~Krauss, M.~Sch\"onherr and F.~Siegert, \emph{{W+ $n$-Jet
  predictions at the Large Hadron Collider at next-to-leading order matched
  with a parton shower}}, Phys. Rev. Lett. \textbf{110} (2013), no.~5,
  \href{http://www.slac.stanford.edu/spires/find/hep/www?eprint=1201.5882}{052001},
   [\href{http://arXiv.org/pdf/1201.5882}{{\texttt{arXiv:1201.5882}}}
  [hep-ph]]. \relax
 \relax
\bibitem{Andersen:2016qtm}
J.~R. Andersen et~al., \emph{{Les Houches 2015: Physics at TeV Colliders
  Standard Model Working Group Report}}, {9th Les Houches Workshop on Physics
  at TeV Colliders}, 5 2016. \relax
 \relax
\bibitem{ATLAS:2018hxb}
M.~Aaboud et~al., ATLAS collaboration, \emph{{Measurements of Higgs boson
  properties in the diphoton decay channel with 36 fb$^{-1}$ of $pp$ collision
  data at $\sqrt{s} = 13$ TeV with the ATLAS detector}}, Phys. Rev. D
  \textbf{98} (2018),
  \href{http://www.slac.stanford.edu/spires/find/hep/www?eprint=1802.04146}{052005},
   [\href{http://arXiv.org/pdf/1802.04146}{{\texttt{arXiv:1802.04146}}}
  [hep-ex]]. \relax
 \relax
\bibitem{Danziger:2021xvr}
\href{http://www.slac.stanford.edu/spires/find/hep/www?eprint=2110.15211}{K.~Danziger,
  S.~H\"oche and F.~Siegert}, \emph{{Reducing negative weights in Monte Carlo
  event generation with Sherpa}},
  \href{http://arXiv.org/pdf/2110.15211}{{\texttt{arXiv:2110.15211}}} [hep-ph].
  \relax
 \relax
\bibitem{L3:2004ehh}
P.~Achard et~al., L3 collaboration, \emph{{Inclusive jet production in
  two-photon collisions at LEP}}, Phys. Lett. B \textbf{602} (2004),
  \href{http://www.slac.stanford.edu/spires/find/hep/www?eprint=hep-ex/0410012}{157--166},
   [\href{http://arXiv.org/pdf/hep-ex/0410012}{{\texttt{hep-ex/0410012}}}].
  \relax
 \relax
\bibitem{OPAL:1996iet}
K.~Ackerstaff et~al., OPAL collaboration, \emph{{Inclusive jet production in
  photon-photon collisions at $\sqrt{s}$ = 130-GeV and 136-GeV}}, Z. Phys. C
  \textbf{73} (1997),
  \href{http://www.slac.stanford.edu/spires/find/hep/www?j=Z%20Phys%20C,73,433}{433--442}.
  \relax
 \relax
\bibitem{OPAL:2007jeb}
G.~Abbiendi et~al., OPAL collaboration, \emph{{Inclusive Jet Production in
  Photon-Photon Collisions at s(ee)**(1/2) from 189 to 209-GeV}}, Phys. Lett. B
  \textbf{658} (2008),
  \href{http://www.slac.stanford.edu/spires/find/hep/www?eprint=0706.4382}{185--192},
   [\href{http://arXiv.org/pdf/0706.4382}{{\texttt{arXiv:0706.4382}}}
  [hep-ex]]. \relax
 \relax
\bibitem{ZEUS:1997fwn}
J.~Breitweg et~al., ZEUS collaboration, \emph{{Dijet cross-sections in
  photoproduction at HERA}}, Eur. Phys. J. C \textbf{1} (1998),
  \href{http://www.slac.stanford.edu/spires/find/hep/www?eprint=hep-ex/9710018}{109--122},
   [\href{http://arXiv.org/pdf/hep-ex/9710018}{{\texttt{hep-ex/9710018}}}].
  \relax
 \relax
\bibitem{ZEUS:2012pcn}
H.~Abramowicz et~al., ZEUS collaboration, \emph{{Inclusive-jet photoproduction
  at HERA and determination of alphas}}, Nucl. Phys. B \textbf{864} (2012),
  \href{http://www.slac.stanford.edu/spires/find/hep/www?eprint=1205.6153}{1--37},
   [\href{http://arXiv.org/pdf/1205.6153}{{\texttt{arXiv:1205.6153}}}
  [hep-ex]]. \relax
 \relax
\bibitem{H1:2006rre}
A.~Aktas et~al., H1 collaboration, \emph{{Photoproduction of dijets with high
  transverse momenta at HERA}}, Phys. Lett. B \textbf{639} (2006),
  \href{http://www.slac.stanford.edu/spires/find/hep/www?eprint=hep-ex/0603014}{21--31},
   [\href{http://arXiv.org/pdf/hep-ex/0603014}{{\texttt{hep-ex/0603014}}}].
  \relax
 \relax
\bibitem{ATLAS:2017fur}
M.~Aaboud et~al., ATLAS collaboration, \emph{{Evidence for light-by-light
  scattering in heavy-ion collisions with the ATLAS detector at the LHC}},
  Nature Phys. \textbf{13} (2017), no.~9,
  \href{http://www.slac.stanford.edu/spires/find/hep/www?eprint=1702.01625}{852--858},
   [\href{http://arXiv.org/pdf/1702.01625}{{\texttt{arXiv:1702.01625}}}
  [hep-ex]]. \relax
 \relax
\bibitem{STAR:2004bzo}
J.~Adams et~al., STAR collaboration, \emph{{Production of e+ e- pairs
  accompanied by nuclear dissociation in ultra-peripheral heavy ion
  collision}}, Phys. Rev. C \textbf{70} (2004),
  \href{http://www.slac.stanford.edu/spires/find/hep/www?eprint=nucl-ex/0404012}{031902},
   [\href{http://arXiv.org/pdf/nucl-ex/0404012}{{\texttt{nucl-ex/0404012}}}].
  \relax
 \relax
\bibitem{STAR:2019wlg}
J.~Adam et~al., STAR collaboration, \emph{{Measurement of $e^+e^-$ Momentum and
  Angular Distributions from Linearly Polarized Photon Collisions}}, Phys. Rev.
  Lett. \textbf{127} (2021), no.~5,
  \href{http://www.slac.stanford.edu/spires/find/hep/www?eprint=1910.12400}{052302},
   [\href{http://arXiv.org/pdf/1910.12400}{{\texttt{arXiv:1910.12400}}}
  [nucl-ex]]. \relax
 \relax
\bibitem{STAR:2009giy}
B.~I. Abelev et~al., STAR collaboration, \emph{{Observation of pi+ pi- pi+ pi-
  Photoproduction in Ultra-Peripheral Heavy Ion Collisions at STAR}}, Phys.
  Rev. C \textbf{81} (2010),
  \href{http://www.slac.stanford.edu/spires/find/hep/www?eprint=0912.0604}{044901},
   [\href{http://arXiv.org/pdf/0912.0604}{{\texttt{arXiv:0912.0604}}}
  [nucl-ex]]. \relax
 \relax
\bibitem{Schuler:1996en}
G.~A. Schuler and T.~Sj{\"o}strand, \emph{{A Scenario for high-energy gamma
  gamma interactions}}, Z. Phys. C \textbf{73} (1997),
  \href{http://www.slac.stanford.edu/spires/find/hep/www?eprint=hep-ph/9605240}{677--688},
   [\href{http://arXiv.org/pdf/hep-ph/9605240}{{\texttt{hep-ph/9605240}}}].
  \relax
 \relax
\bibitem{Hoeche:2023gme}
S.~Hoeche, F.~Krauss and P.~Meinzinger, \emph{{Resolved photons in Sherpa}},
  Eur. Phys. J. C \textbf{84} (2024), no.~2,
  \href{http://www.slac.stanford.edu/spires/find/hep/www?eprint=2310.18674}{178},
   [\href{http://arXiv.org/pdf/2310.18674}{{\texttt{arXiv:2310.18674}}}
  [hep-ph]]. \relax
 \relax
\bibitem{Meinzinger:2023xuf}
P.~Meinzinger and F.~Krauss, \emph{{Hadron-level NLO predictions for QCD
  observables in photo-production at the Electron-Ion Collider}}, Phys. Rev. D
  \textbf{109} (2024), no.~3,
  \href{http://www.slac.stanford.edu/spires/find/hep/www?eprint=2311.14571}{034037},
   [\href{http://arXiv.org/pdf/2311.14571}{{\texttt{arXiv:2311.14571}}}
  [hep-ph]]. \relax
 \relax
\bibitem{H1:2015okx}
V.~Andreev et~al., H1 collaboration, \emph{{Diffractive Dijet Production with a
  Leading Proton in $ep$ Collisions at HERA}}, JHEP \textbf{05} (2015),
  \href{http://www.slac.stanford.edu/spires/find/hep/www?eprint=1502.01683}{056},
   [\href{http://arXiv.org/pdf/1502.01683}{{\texttt{arXiv:1502.01683}}}
  [hep-ex]]. \relax
 \relax
\bibitem{AbdulKhalek:2021gbh}
R.~Abdul~Khalek et~al., \emph{{Science Requirements and Detector Concepts for
  the Electron-Ion Collider}: {EIC Yellow Report}}, Nucl. Phys. A \textbf{1026}
  (2022),
  \href{http://www.slac.stanford.edu/spires/find/hep/www?eprint=2103.05419}{122447},
   [\href{http://arXiv.org/pdf/2103.05419}{{\texttt{arXiv:2103.05419}}}
  [physics.ins-det]]. \relax
 \relax
\bibitem{Krauss:2024dzj}
F.~Krauss and P.~Meinzinger, \emph{{Hard diffraction in SHERPA}}, Eur. Phys. J.
  C \textbf{84} (2024), no.~9,
  \href{http://www.slac.stanford.edu/spires/find/hep/www?eprint=2407.02133}{894},
   [\href{http://arXiv.org/pdf/2407.02133}{{\texttt{arXiv:2407.02133}}}
  [hep-ph]]. \relax
 \relax
\bibitem{Gehrmann:2012yg}
T.~Gehrmann, S.~H{\"o}che, F.~Krauss, M.~Sch{\"o}nherr and F.~Siegert,
  \emph{{NLO QCD matrix elements + parton showers in $e^+e^-\to$hadrons}}, JHEP
  \textbf{01} (2013),
  \href{http://www.slac.stanford.edu/spires/find/hep/www?eprint=1207.5031}{144},
   [\href{http://arXiv.org/pdf/1207.5031}{{\texttt{arXiv:1207.5031}}}
  [hep-ph]]. \relax
 \relax
\bibitem{Baberuxki:2019ifp}
N.~Baberuxki, C.~T. Preuss, D.~Reichelt and S.~Schumann, \emph{{Resummed
  predictions for jet-resolution scales in multijet production in
  e$^{+}$e$^{-}$ annihilation}}, JHEP \textbf{04} (2020),
  \href{http://www.slac.stanford.edu/spires/find/hep/www?eprint=1912.09396}{112},
   [\href{http://arXiv.org/pdf/1912.09396}{{\texttt{arXiv:1912.09396}}}
  [hep-ph]]. \relax
 \relax
\bibitem{Knobbe:2023ehi}
M.~Knobbe, D.~Reichelt and S.~Schumann, \emph{{(N)NLO+NLL\textquoteright{}
  accurate predictions for plain and groomed 1-jettiness in neutral current
  DIS}}, JHEP \textbf{09} (2023),
  \href{http://www.slac.stanford.edu/spires/find/hep/www?eprint=2306.17736}{194},
   [\href{http://arXiv.org/pdf/2306.17736}{{\texttt{arXiv:2306.17736}}}
  [hep-ph]]. \relax
 \relax
\bibitem{Baron:2020xoi}
J.~Baron, D.~Reichelt, S.~Schumann, N.~Schwanemann and V.~Theeuwes,
  \emph{{Soft-drop grooming for hadronic event shapes}}, JHEP \textbf{07}
  (2021),
  \href{http://www.slac.stanford.edu/spires/find/hep/www?eprint=2012.09574}{142},
   [\href{http://arXiv.org/pdf/2012.09574}{{\texttt{arXiv:2012.09574}}}
  [hep-ph]]. \relax
 \relax
\bibitem{Hoeche:2012yf}
S.~H{\"o}che, F.~Krauss, M.~Sch{\"o}nherr and F.~Siegert, \emph{{QCD matrix
  elements + parton showers: The NLO case}}, JHEP \textbf{04} (2013),
  \href{http://inspirehep.net/record/1123387}{027},
  [\href{http://arXiv.org/pdf/1207.5030}{{\texttt{arXiv:1207.5030}}} [hep-ph]].
  \relax
 \relax
\bibitem{ATLAS:2019iaa}
G.~Aad et~al., ATLAS collaboration, \emph{{Measurement of isolated-photon plus
  two-jet production in $pp$ collisions at $\sqrt s=13$ TeV with the ATLAS
  detector}}, JHEP \textbf{03} (2020),
  \href{http://www.slac.stanford.edu/spires/find/hep/www?eprint=1912.09866}{179},
   [\href{http://arXiv.org/pdf/1912.09866}{{\texttt{arXiv:1912.09866}}}
  [hep-ex]]. \relax
 \relax
\bibitem{Lindert:2022ejn}
J.~M. Lindert, S.~Pozzorini and M.~Sch\"onherr, \emph{{Precise predictions for
  V + 2 jet backgrounds in searches for invisible Higgs decays}}, JHEP
  \textbf{01} (2023),
  \href{http://www.slac.stanford.edu/spires/find/hep/www?eprint=2204.07652}{070},
   [\href{http://arXiv.org/pdf/2204.07652}{{\texttt{arXiv:2204.07652}}}
  [hep-ph]]. \relax
 \relax
\bibitem{Brauer:2020kfv}
S.~Br\"auer, A.~Denner, M.~Pellen, M.~Sch\"onherr and S.~Schumann,
  \emph{{Fixed-order and merged parton-shower predictions for WW and WWj
  production at the LHC including NLO QCD and EW corrections}}, JHEP
  \textbf{10} (2020),
  \href{http://www.slac.stanford.edu/spires/find/hep/www?eprint=2005.12128}{159},
   [\href{http://arXiv.org/pdf/2005.12128}{{\texttt{arXiv:2005.12128}}}
  [hep-ph]]. \relax
 \relax
\bibitem{Krause:2017nxq}
J.~Krause and F.~Siegert, \emph{{NLO QCD predictions for $Z+\gamma$ + jets
  production with Sherpa}}, Eur. Phys. J. \textbf{C78} (2018), no.~2,
  \href{http://www.slac.stanford.edu/spires/find/hep/www?eprint=1708.06283}{161},
   [\href{http://arXiv.org/pdf/1708.06283}{{\texttt{arXiv:1708.06283}}}
  [hep-ph]]. \relax
 \relax
\bibitem{Siegert:2016bre}
F.~Siegert, \emph{{A practical guide to event generation for prompt photon
  production with Sherpa}}, J. Phys. \textbf{G44} (2017), no.~4,
  \href{http://www.slac.stanford.edu/spires/find/hep/www?eprint=1611.07226}{044007},
   [\href{http://arXiv.org/pdf/1611.07226}{{\texttt{arXiv:1611.07226}}}
  [hep-ph]]. \relax
 \relax
\bibitem{ATLAS:2021mbt}
G.~Aad et~al., ATLAS collaboration, \emph{{Measurement of the production cross
  section of pairs of isolated photons in $pp$ collisions at 13 TeV with the
  ATLAS detector}}, JHEP \textbf{11} (2021),
  \href{http://www.slac.stanford.edu/spires/find/hep/www?eprint=2107.09330}{169},
   [\href{http://arXiv.org/pdf/2107.09330}{{\texttt{arXiv:2107.09330}}}
  [hep-ex]]. \relax
 \relax
\bibitem{Hoeche:2014rya}
S.~H{\"o}che, F.~Krauss, S.~Pozzorini, M.~Sch{\"o}nherr, J.~Thompson,
  S.~Pozzorini and K.~C. Zapp, \emph{{Triple vector boson production through
  Higgs-Strahlung with NLO multijet merging}}, Phys.Rev. \textbf{D89} (2014),
  \href{http://www.slac.stanford.edu/spires/find/hep/www?eprint=1403.7516}{093015},
   [\href{http://arXiv.org/pdf/1403.7516}{{\texttt{arXiv:1403.7516}}}
  [hep-ph]]. \relax
 \relax
\bibitem{Goncalves:2016bkl}
D.~Goncalves, F.~Krauss, S.~Kuttimalai and P.~Maierh\"ofer, \emph{{Boosting
  invisible searches via ZH : From the Higgs boson to dark matter simplified
  models}}, Phys. Rev. D \textbf{94} (2016), no.~5,
  \href{http://www.slac.stanford.edu/spires/find/hep/www?eprint=1605.08039}{053014},
   [\href{http://arXiv.org/pdf/1605.08039}{{\texttt{arXiv:1605.08039}}}
  [hep-ph]]. \relax
 \relax
\bibitem{Denner:2024ufg}
A.~Denner, M.~Pellen, M.~Sch\"onherr and S.~Schumann, \emph{{Tri-boson and WH
  production in the W$^{+}$W$^{+}$jj channel: predictions at full NLO accuracy
  and beyond}}, JHEP \textbf{08} (2024),
  \href{http://www.slac.stanford.edu/spires/find/hep/www?eprint=2406.11516}{043},
   [\href{http://arXiv.org/pdf/2406.11516}{{\texttt{arXiv:2406.11516}}}
  [hep-ph]]. \relax
 \relax
\bibitem{ATLAS:2023zkw}
G.~Aad et~al., ATLAS collaboration, \emph{{Observation of WZ\ensuremath{\gamma}
  Production in pp Collisions at s=13\,\,TeV with the ATLAS Detector}}, Phys.
  Rev. Lett. \textbf{132} (2024), no.~2,
  \href{http://www.slac.stanford.edu/spires/find/hep/www?eprint=2305.16994}{021802},
   [\href{http://arXiv.org/pdf/2305.16994}{{\texttt{arXiv:2305.16994}}}
  [hep-ex]]. \relax
 \relax
\bibitem{ATLAS:2022wmu}
G.~Aad et~al., ATLAS collaboration, \emph{{Measurement of $Z\gamma \gamma $
  production in pp collisions at $\sqrt{s}= 13$~TeV with the ATLAS detector}},
  Eur. Phys. J. C \textbf{83} (2023), no.~6,
  \href{http://www.slac.stanford.edu/spires/find/hep/www?eprint=2211.14171}{539},
   [\href{http://arXiv.org/pdf/2211.14171}{{\texttt{arXiv:2211.14171}}}
  [hep-ex]]. \relax
 \relax
\bibitem{ATLAS:2023avk}
G.~Aad et~al., ATLAS collaboration, \emph{{Observation of $W\gamma\gamma$
  triboson production in proton-proton collisions at $\sqrt{s} = 13$ TeV with
  the ATLAS detector}}, Phys. Lett. B \textbf{848} (2024),
  \href{http://www.slac.stanford.edu/spires/find/hep/www?eprint=2308.03041}{138400},
   [\href{http://arXiv.org/pdf/2308.03041}{{\texttt{arXiv:2308.03041}}}
  [hep-ex]]. \relax
 \relax
\bibitem{ATLAS:2021lsc}
ATLAS collaboration, \emph{{Modelling of isolated multi-photon production in
  Monte Carlo generators in ATLAS}}. \relax
 \relax
\bibitem{Bothmann:2017jfv}
E.~Bothmann, F.~Krauss and M.~Sch{\"o}nherr, \emph{{Single top-quark production
  with SHERPA}}, Eur. Phys. J. \textbf{C78} (2018), no.~3,
  \href{http://www.slac.stanford.edu/spires/find/hep/www?eprint=1711.02568}{220},
   [\href{http://arXiv.org/pdf/1711.02568}{{\texttt{arXiv:1711.02568}}}
  [hep-ph]]. \relax
 \relax
\bibitem{Hoeche:2014qda}
S.~H{\"o}che, F.~Krauss, P.~Maierh{\"o}fer, S.~Pozzorini, M.~Sch{\"o}nherr and
  F.~Siegert, \emph{{Next-to-leading order QCD predictions for top-quark pair
  production with up to two jets merged with a parton shower}}, Phys. Lett.
  \textbf{B748} (2015),
  \href{http://www.slac.stanford.edu/spires/find/hep/www?eprint=1402.6293}{74--78},
   [\href{http://arXiv.org/pdf/1402.6293}{{\texttt{arXiv:1402.6293}}}
  [hep-ph]]. \relax
 \relax
\bibitem{Gutschow:2018tuk}
C.~G{\"u}tschow, J.~M. Lindert and M.~Sch{\"o}nherr, \emph{{Multi-jet merged
  top-pair production including electroweak corrections}}, Eur. Phys. J.
  \textbf{C78} (2018), no.~4,
  \href{http://www.slac.stanford.edu/spires/find/hep/www?eprint=1803.00950}{317},
   [\href{http://arXiv.org/pdf/1803.00950}{{\texttt{arXiv:1803.00950}}}
  [hep-ph]]. \relax
 \relax
\bibitem{ATLAS:2020esn}
ATLAS collaboration, \emph{{Modelling of rare top quark processes at $\sqrt{s}$
  = 13 TeV in ATLAS}}. \relax
 \relax
\bibitem{Cascioli:2013era}
F.~Cascioli, P.~Maierh{\"o}fer, N.~Moretti, S.~Pozzorini and F.~Siegert,
  \emph{{NLO matching for $t\bar t b \bar b$ production with massive
  $b$-quarks}}, Phys. Lett. \textbf{B734} (2014),
  \href{http://www.slac.stanford.edu/spires/find/hep/www?eprint=1309.5912}{210--214},
   [\href{http://arXiv.org/pdf/1309.5912}{{\texttt{arXiv:1309.5912}}}
  [hep-ph]]. \relax
 \relax
\bibitem{Hoche:2019ncc}
S.~H\"oche, J.~Krause and F.~Siegert, \emph{{Multijet Merging in a Variable
  Flavor Number Scheme}}, Phys. Rev. D \textbf{100} (2019), no.~1,
  \href{http://www.slac.stanford.edu/spires/find/hep/www?eprint=1904.09382}{014011},
   [\href{http://arXiv.org/pdf/1904.09382}{{\texttt{arXiv:1904.09382}}}
  [hep-ph]]. \relax
 \relax
\bibitem{Ferencz:2024pay}
L.~Ferencz, S.~H\"oche, J.~Katzy and F.~Siegert, \emph{{$t\bar{t}b\bar{b}$ at
  NLO precision in a variable flavor number scheme}}, JHEP \textbf{07} (2024),
  \href{http://www.slac.stanford.edu/spires/find/hep/www?eprint=2402.15497}{026},
   [\href{http://arXiv.org/pdf/2402.15497}{{\texttt{arXiv:2402.15497}}}
  [hep-ph]]. \relax
 \relax
\bibitem{Hoche:2010kg}
S.~H\"oche, F.~Krauss, M.~Sch\"onherr and F.~Siegert, \emph{{NLO matrix
  elements and truncated showers}}, JHEP \textbf{08} (2011),
  \href{http://www.slac.stanford.edu/spires/find/hep/www?eprint=1009.1127}{123},
   [\href{http://arXiv.org/pdf/1009.1127}{{\texttt{arXiv:1009.1127}}}
  [hep-ph]]. \relax
 \relax
\bibitem{Catani:2001cc}
S.~Catani, F.~Krauss, R.~Kuhn and B.~R. Webber, \emph{{QCD matrix elements +
  parton showers}}, JHEP \textbf{11} (2001),
  \href{http://inspirehep.net/search?p=hep-ph/0109231}{063},
  [\href{http://arXiv.org/pdf/hep-ph/0109231}{{\texttt{hep-ph/0109231}}}].
  \relax
 \relax
\bibitem{Lonnblad:2001iq}
L.~L{\"o}nnblad, \emph{{Correcting the colour-dipole cascade model with fixed
  order matrix elements}}, JHEP \textbf{05} (2002),
  \href{http://inspirehep.net/search?p=hep-ph/0112284}{046},
  [\href{http://arXiv.org/pdf/hep-ph/0112284}{{\texttt{hep-ph/0112284}}}].
  \relax
 \relax
\bibitem{Krauss:2002up}
F.~Krauss, \emph{{Matrix elements and parton showers in hadronic
  interactions}}, JHEP \textbf{08} (2002),
  \href{http://inspirehep.net/search?p=hep-ph/0205283}{015},
  [\href{http://arXiv.org/pdf/hep-ph/0205283}{{\texttt{hep-ph/0205283}}}].
  \relax
 \relax
\bibitem{Alwall:2007fs}
J.~Alwall et~al., \emph{{Comparative study of various algorithms for the
  merging of parton showers and matrix elements in hadronic collisions}}, Eur.
  Phys. J. \textbf{C53} (2008),
  \href{http://inspirehep.net/record/753397}{473--500},
  [\href{http://arXiv.org/pdf/0706.2569}{{\texttt{arXiv:0706.2569}}} [hep-ph]].
  \relax
 \relax
\bibitem{Hoeche:2009rj}
S.~H{\"o}che, F.~Krauss, S.~Schumann and F.~Siegert, \emph{{QCD matrix elements
  and truncated showers}}, JHEP \textbf{05} (2009),
  \href{http://inspirehep.net/search?p=arXiv:0903.1219}{053},
  [\href{http://arXiv.org/pdf/0903.1219}{{\texttt{arXiv:0903.1219}}} [hep-ph]].
  \relax
 \relax
\bibitem{Lonnblad:2011xx}
L.~L{\"o}nnblad and S.~Prestel, \emph{{Matching Tree-Level Matrix Elements with
  Interleaved Showers}}, JHEP \textbf{03} (2012),
  \href{http://www.slac.stanford.edu/spires/find/hep/www?eprint=1109.4829}{019},
   [\href{http://arXiv.org/pdf/1109.4829}{{\texttt{arXiv:1109.4829}}}
  [hep-ph]]. \relax
 \relax
\bibitem{Frederix:2012ps}
R.~Frederix and S.~Frixione, \emph{{Merging meets matching in MC@NLO}}, JHEP
  \textbf{12} (2012), \href{http://inspirehep.net/record/1188307}{061},
  [\href{http://arXiv.org/pdf/1209.6215}{{\texttt{arXiv:1209.6215}}} [hep-ph]].
  \relax
 \relax
\bibitem{Platzer:2012bs}
S.~Pl{\"a}tzer, \emph{{Controlling inclusive cross sections in parton shower +
  matrix element merging}}, JHEP \textbf{08} (2013),
  \href{http://www.slac.stanford.edu/spires/find/hep/www?eprint=1211.5467}{114},
   [\href{http://arXiv.org/pdf/1211.5467}{{\texttt{arXiv:1211.5467}}}
  [hep-ph]]. \relax
 \relax
\bibitem{Hoche:2019flt}
S.~H\"oche, S.~Prestel and H.~Schulz, \emph{{Simulation of Vector Boson Plus
  Many Jet Final States at the High Luminosity LHC}}, Phys. Rev. D \textbf{100}
  (2019), no.~1,
  \href{http://www.slac.stanford.edu/spires/find/hep/www?eprint=1905.05120}{014024},
   [\href{http://arXiv.org/pdf/1905.05120}{{\texttt{arXiv:1905.05120}}}
  [hep-ph]]. \relax
 \relax
\bibitem{Bothmann:2023ozs}
E.~Bothmann, T.~Childers, C.~G\"utschow, S.~H\"oche, P.~Hovland, J.~Isaacson,
  M.~Knobbe and R.~Latham, \emph{{Efficient precision simulation of processes
  with many-jet final states at the LHC}}, Phys. Rev. D \textbf{109} (2024),
  no.~1,
  \href{http://www.slac.stanford.edu/spires/find/hep/www?eprint=2309.13154}{014013},
   [\href{http://arXiv.org/pdf/2309.13154}{{\texttt{arXiv:2309.13154}}}
  [hep-ph]]. \relax
 \relax
\bibitem{Hoche:2021mkv}
S.~H\"oche, S.~Mrenna, S.~Payne, C.~T. Preuss and P.~Skands, \emph{{A Study of
  QCD Radiation in VBF Higgs Production with Vincia and Pythia}}, SciPost Phys.
  \textbf{12} (2022), no.~1,
  \href{http://www.slac.stanford.edu/spires/find/hep/www?eprint=2106.10987}{010},
   [\href{http://arXiv.org/pdf/2106.10987}{{\texttt{arXiv:2106.10987}}}
  [hep-ph]]. \relax
 \relax
\bibitem{Carli:2009cg}
T.~Carli, T.~Gehrmann and S.~H{\"o}che, \emph{{Hadronic final states in
  deep-inelastic scattering with \Sherpa}}, Eur. Phys. J. \textbf{C67} (2010),
  \href{http://inspirehep.net/search?rawcmd=f+eprint+0912.3715}{73},
  [\href{http://arXiv.org/pdf/0912.3715}{{\texttt{arXiv:0912.3715}}} [hep-ph]].
  \relax
 \relax
\bibitem{Forte:2016sja}
S.~Forte, D.~Napoletano and M.~Ubiali, \emph{{Higgs production in bottom-quark
  fusion: matching beyond leading order}}, Phys. Lett. \textbf{B763} (2016),
  \href{http://www.slac.stanford.edu/spires/find/hep/www?eprint=1607.00389}{190--196},
   [\href{http://arXiv.org/pdf/1607.00389}{{\texttt{arXiv:1607.00389}}}
  [hep-ph]]. \relax
 \relax
\bibitem{Denner:2024yut}
A.~Denner and S.~Rode, \emph{{Automated resummation of electroweak Sudakov
  logarithms in diboson production at future colliders}}, Eur. Phys. J. C
  \textbf{84} (2024), no.~5,
  \href{http://www.slac.stanford.edu/spires/find/hep/www?eprint=2402.10503}{542},
   [\href{http://arXiv.org/pdf/2402.10503}{{\texttt{arXiv:2402.10503}}}
  [hep-ph]]. \relax
 \relax
\bibitem{Kallweit:2015dum}
S.~Kallweit, J.~M. Lindert, P.~Maierh{\"o}fer, S.~Pozzorini and
  M.~Sch{\"o}nherr, \emph{{NLO QCD+EW predictions for V + jets including
  off-shell vector-boson decays and multijet merging}}, JHEP \textbf{04}
  (2016),
  \href{http://www.slac.stanford.edu/spires/find/hep/www?eprint=1511.08692}{021},
   [\href{http://arXiv.org/pdf/1511.08692}{{\texttt{arXiv:1511.08692}}}
  [hep-ph]]. \relax
 \relax
\bibitem{Gutschow:2020cug}
C.~G\"utschow and M.~Sch\"onherr, \emph{{Four lepton production and the
  accuracy of QED FSR}}, Eur. Phys. J. C \textbf{81} (2021), no.~1,
  \href{http://www.slac.stanford.edu/spires/find/hep/www?eprint=2007.15360}{48},
   [\href{http://arXiv.org/pdf/2007.15360}{{\texttt{arXiv:2007.15360}}}
  [hep-ph]]. \relax
 \relax
\bibitem{Denner:2000jv}
A.~Denner and S.~Pozzorini, \emph{{One loop leading logarithms in electroweak
  radiative corrections. 1. Results}}, Eur.Phys.J. \textbf{C18} (2001),
  \href{http://www.slac.stanford.edu/spires/find/hep/www?eprint=hep-ph/0010201}{461--480},
   [\href{http://arXiv.org/pdf/hep-ph/0010201}{{\texttt{arXiv:hep-ph/0010201}}}
  [hep-ph]]. \relax
 \relax
\bibitem{Denner:2001gw}
A.~Denner and S.~Pozzorini, \emph{{One loop leading logarithms in electroweak
  radiative corrections. 2. Factorization of collinear singularities}},
  Eur.Phys.J. \textbf{C21} (2001),
  \href{http://www.slac.stanford.edu/spires/find/hep/www?eprint=hep-ph/0104127}{63--79},
   [\href{http://arXiv.org/pdf/hep-ph/0104127}{{\texttt{arXiv:hep-ph/0104127}}}
  [hep-ph]]. \relax
 \relax
\bibitem{Bothmann:2020sxm}
E.~Bothmann and D.~Napoletano, \emph{{Automated evaluation of electroweak
  Sudakov logarithms in Sherpa}}, Eur. Phys. J. C \textbf{80} (2020), no.~11,
  \href{http://www.slac.stanford.edu/spires/find/hep/www?eprint=2006.14635}{1024},
   [\href{http://arXiv.org/pdf/2006.14635}{{\texttt{arXiv:2006.14635}}}
  [hep-ph]]. \relax
 \relax
\bibitem{Pagani:2021vyk}
D.~Pagani and M.~Zaro, \emph{{One-loop electroweak Sudakov logarithms: a
  revisitation and automation}}, JHEP \textbf{02} (2022),
  \href{http://www.slac.stanford.edu/spires/find/hep/www?eprint=2110.03714}{161},
   [\href{http://arXiv.org/pdf/2110.03714}{{\texttt{arXiv:2110.03714}}}
  [hep-ph]]. \relax
 \relax
\bibitem{Pagani:2023wgc}
D.~Pagani, T.~Vitos and M.~Zaro, \emph{{Improving NLO QCD event generators with
  high-energy EW corrections}}, Eur. Phys. J. C \textbf{84} (2024), no.~5,
  \href{http://www.slac.stanford.edu/spires/find/hep/www?eprint=2309.00452}{514},
   [\href{http://arXiv.org/pdf/2309.00452}{{\texttt{arXiv:2309.00452}}}
  [hep-ph]]. \relax
 \relax
\bibitem{Lindert:2023fcu}
\href{http://www.slac.stanford.edu/spires/find/hep/www?eprint=2312.07927}{J.~M.
  Lindert and L.~Mai}, \emph{{Logarithmic EW corrections at one-loop}},
  \href{http://arXiv.org/pdf/2312.07927}{{\texttt{arXiv:2312.07927}}} [hep-ph].
  \relax
 \relax
\bibitem{Yennie:1961ad}
D.~R. Yennie, S.~C. Frautschi and H.~Suura, \emph{{The Infrared Divergence
  Phenomena and High-Energy Processes}}, Ann. Phys. \textbf{13} (1961),
  \href{http://inspirehep.net/search?j=APNYA,13,379}{379--452}. \relax
 \relax
\bibitem{Schonherr:2008av}
M.~Sch\"{o}nherr and F.~Krauss, \emph{Soft photon radiation in particle decays
  in \Sherpa{}}, JHEP \textbf{12} (2008),
  \href{http://inspirehep.net/search?p=arXiv:0810.5071}{018},
  [\href{http://arXiv.org/pdf/0810.5071}{{\texttt{arXiv:0810.5071}}} [hep-ph]].
  \relax
 \relax
\bibitem{Alioli:2016fum}
S.~Alioli et~al., \emph{{Precision studies of observables in $p p \rightarrow W
  \rightarrow l\nu _l$ and $ pp \rightarrow \gamma ,Z \rightarrow l^+ l^-$
  processes at the LHC}}, Eur. Phys. J. \textbf{C77} (2017), no.~5,
  \href{http://www.slac.stanford.edu/spires/find/hep/www?eprint=1606.02330}{280},
   [\href{http://arXiv.org/pdf/1606.02330}{{\texttt{arXiv:1606.02330}}}
  [hep-ph]]. \relax
 \relax
\bibitem{Bernlochner:2010fc}
\href{http://www.slac.stanford.edu/spires/find/hep/www?eprint=1010.5997}{F.~U.
  Bernlochner and M.~Sch{\"o}nherr}, \emph{{Comparing different ansatzes to
  describe electroweak radiative corrections to exclusive semileptonic B meson
  decays into (pseudo)scalar final state mesons using Monte-Carlo techniques}},
   \href{http://arXiv.org/pdf/1010.5997}{{\texttt{arXiv:1010.5997}}} [hep-ph].
  \relax
 \relax
\bibitem{Krauss:2018djz}
F.~Krauss, J.~M. Lindert, R.~Linten and M.~Sch{\"o}nherr, \emph{{Accurate
  simulation of W, Z and Higgs boson decays in Sherpa}}, Eur. Phys. J.
  \textbf{C79} (2019), no.~2,
  \href{http://www.slac.stanford.edu/spires/find/hep/www?eprint=1809.10650}{143},
   [\href{http://arXiv.org/pdf/1809.10650}{{\texttt{arXiv:1809.10650}}}
  [hep-ph]]. \relax
 \relax
\bibitem{Flower:2022iew}
L.~Flower and M.~Sch\"{o}nherr, \emph{{Photon splitting corrections to
  soft-photon resummation}}, JHEP \textbf{03} (2023),
  \href{http://www.slac.stanford.edu/spires/find/hep/www?eprint=2210.07007}{238},
   [\href{http://arXiv.org/pdf/2210.07007}{{\texttt{arXiv:2210.07007}}}
  [hep-ph]]. \relax
 \relax
\bibitem{Altarelli:1977zs}
G.~Altarelli and G.~Parisi, \emph{{Asymptotic freedom in parton language}},
  Nucl. Phys. \textbf{B126} (1977),
  \href{http://inspirehep.net/search?j=NUPHA,B126,298}{298--318}. \relax
 \relax
\bibitem{Gribov:1972ri}
V.~N. Gribov and L.~N. Lipatov, \emph{{Deep inelastic $e$-$p$ scattering in
  perturbation theory}}, Sov. J. Nucl. Phys. \textbf{15} (1972),
  \href{http://inspirehep.net/search?j=SJNCA,15,438}{438--450}. \relax
 \relax
\bibitem{Lipatov:1974qm}
L.~N. Lipatov, \emph{{The parton model and perturbation theory}}, Sov. J. Nucl.
  Phys. \textbf{20} (1975),
  \href{http://inspirehep.net/search?j=SJNCA,20,94}{94--102}. \relax
 \relax
\bibitem{Dokshitzer:1977sg}
Y.~L. Dokshitzer, \emph{{Calculation of the structure functions for deep
  inelastic scattering and $e^+e^-$ annihilation by perturbation theory in
  quantum chromodynamics}}, Sov. Phys. JETP \textbf{46} (1977),
  \href{http://inspirehep.net/search?j=SPHJA,46,641}{641--653}. \relax
 \relax
\bibitem{Skrzypek:1990qs}
M.~Skrzypek and S.~Jadach, \emph{{Exact and approximate solutions for the
  electron nonsinglet structure function in QED}}, Z. Phys. C \textbf{49}
  (1991),
  \href{http://www.slac.stanford.edu/spires/find/hep/www?j=Z%20Phys%20C,49,577}{577--584}.
  \relax
 \relax
\bibitem{Jadach:2001mp}
S.~Jadach, W.~P{\l}aczek, M.~Skrzypek, B.~F.~L. Ward and Z.~W{\c a}s,
  \emph{{The Monte Carlo program KoralW version 1.51 and the concurrent Monte
  Carlo KoralW\&YFSWW3 with all background graphs and first order corrections
  to W pair production}}, Comput. Phys. Commun. \textbf{140} (2001),
  \href{http://inspirehep.net/search?p=hep-ph/0104049}{475--512},
  [\href{http://arXiv.org/pdf/hep-ph/0104049}{{\texttt{hep-ph/0104049}}}].
  \relax
 \relax
\bibitem{Jadach:1999vf}
S.~Jadach, B.~F.~L. Ward and Z.~W{\c a}s, \emph{{The precision Monte Carlo
  event generator $\mathcal{KK}$ for two- fermion final states in $e^+ e^-$
  collisions}}, Comput. Phys. Commun. \textbf{130} (2000),
  \href{http://inspirehep.net/search?p=hep-ph/9912214}{260--325},
  [\href{http://arXiv.org/pdf/hep-ph/9912214}{{\texttt{hep-ph/9912214}}}].
  \relax
 \relax
\bibitem{Jadach:1991by}
S.~Jadach, E.~Richter-Was, B.~F.~L. Ward and Z.~Was, \emph{{Monte Carlo program
  BHLUMI-2.01 for Bhabha scattering at low angles with Yennie-Frautschi-Suura
  exponentiation}}, Comput. Phys. Commun. \textbf{70} (1992),
  \href{http://www.slac.stanford.edu/spires/find/hep/www?j=Comput%20Phys%20Commun,70,305}{305--344}.
  \relax
 \relax
\bibitem{Jadach:2022mbe}
S.~Jadach, B.~F.~L. Ward, Z.~Was, S.~A. Yost and A.~Siodmok,
  \emph{{Multi-photon Monte Carlo event generator KKMCee for lepton and quark
  pair production in lepton colliders}}, Comput. Phys. Commun. \textbf{283}
  (2023),
  \href{http://www.slac.stanford.edu/spires/find/hep/www?eprint=2204.11949}{108556},
   [\href{http://arXiv.org/pdf/2204.11949}{{\texttt{arXiv:2204.11949}}}
  [hep-ph]]. \relax
 \relax
\bibitem{ALEPH:2005ab}
S.~Schael et~al., ALEPH, DELPHI, L3, OPAL, SLD, LEP Electroweak Working Group,
  SLD Electroweak Group, SLD Heavy Flavour Group collaboration,
  \emph{{Precision electroweak measurements on the $Z$ resonance}}, Phys. Rept.
  \textbf{427} (2006),
  \href{http://www.slac.stanford.edu/spires/find/hep/www?eprint=hep-ex/0509008}{257--454},
   [\href{http://arXiv.org/pdf/hep-ex/0509008}{{\texttt{hep-ex/0509008}}}].
  \relax
 \relax
\bibitem{Sjostrand:1987su}
T.~Sj{\"o}strand and M.~van Zijl, \emph{{A multiple-interaction model for the
  event structure in hadron collisions}}, Phys. Rev. \textbf{D36} (1987),
  \href{http://inspirehep.net/search?j=PHRVA,D36,2019}{2019}. \relax
 \relax
\bibitem{Sjostrand:2004pf}
T.~Sj{\"o}strand and P.~Z. Skands, \emph{{Multiple interactions and the
  structure of beam remnants}}, JHEP \textbf{03} (2004),
  \href{http://inspirehep.net/search?p=hep-ph/0402078}{053},
  [\href{http://arXiv.org/pdf/hep-ph/0402078}{{\texttt{hep-ph/0402078}}}].
  \relax
 \relax
\bibitem{Sjostrand:2004ef}
T.~Sj{\"o}strand and P.~Z. Skands, \emph{{Transverse-momentum-ordered showers
  and interleaved multiple interactions}}, Eur. Phys. J. \textbf{C39} (2005),
  \href{http://inspirehep.net/search?p=hep-ph/0408302}{129--154},
  [\href{http://arXiv.org/pdf/hep-ph/0408302}{{\texttt{hep-ph/0408302}}}].
  \relax
 \relax
\bibitem{Corke:2009tk}
R.~Corke and T.~Sjostrand, \emph{{Multiparton Interactions and Rescattering}},
  JHEP \textbf{01} (2010),
  \href{http://www.slac.stanford.edu/spires/find/hep/www?eprint=0911.1909}{035},
   [\href{http://arXiv.org/pdf/0911.1909}{{\texttt{arXiv:0911.1909}}}
  [hep-ph]]. \relax
 \relax
\bibitem{Alekhin:2005dx}
A.~De~Roeck and H.~Jung (Eds.), \emph{{HERA and the LHC: A Workshop on the
  implications of HERA for LHC physics: Proceedings Part A}}, Geneva, CERN,
  CERN, 2005. \relax
 \relax
\bibitem{Kaidalov:2001iz}
A.~B. Kaidalov, V.~A. Khoze, A.~D. Martin and M.~G. Ryskin,
  \emph{{Probabilities of rapidity gaps in high-energy interactions}}, Eur.
  Phys. J. C \textbf{21} (2001),
  \href{http://www.slac.stanford.edu/spires/find/hep/www?eprint=hep-ph/0105145}{521--529},
   [\href{http://arXiv.org/pdf/hep-ph/0105145}{{\texttt{hep-ph/0105145}}}].
  \relax
 \relax
\bibitem{Kaidalov:2003xf}
A.~B. Kaidalov, V.~A. Khoze, A.~D. Martin and M.~G. Ryskin, \emph{{Unitarity
  effects in hard diffraction at HERA}}, Phys. Lett. B \textbf{567} (2003),
  \href{http://www.slac.stanford.edu/spires/find/hep/www?eprint=hep-ph/0306134}{61--68},
   [\href{http://arXiv.org/pdf/hep-ph/0306134}{{\texttt{hep-ph/0306134}}}].
  \relax
 \relax
\bibitem{Kaidalov:2009fp}
A.~B. Kaidalov, V.~A. Khoze, A.~D. Martin and M.~G. Ryskin,
  \emph{{Factorization breaking in diffractive dijet photoproduction at HERA}},
  Eur. Phys. J. C \textbf{66} (2010),
  \href{http://www.slac.stanford.edu/spires/find/hep/www?eprint=0911.3716}{373--376},
   [\href{http://arXiv.org/pdf/0911.3716}{{\texttt{arXiv:0911.3716}}}
  [hep-ph]]. \relax
 \relax
\bibitem{Dokshitzer:1991wu}
Y.~L. Dokshitzer, V.~A. Khoze, A.~H. Mueller and S.~I. Troyan, \emph{{Basics of
  perturbative QCD}}, Gif-sur-Yvette, France: Ed. Frontieres, 1991. \relax
 \relax
\bibitem{Skands:2007zg}
P.~Z. Skands and D.~Wicke, \emph{{Non-perturbative QCD effects and the top mass
  at the Tevatron}}, Eur. Phys. J. C \textbf{52} (2007),
  \href{http://www.slac.stanford.edu/spires/find/hep/www?eprint=hep-ph/0703081}{133--140},
   [\href{http://arXiv.org/pdf/hep-ph/0703081}{{\texttt{hep-ph/0703081}}}].
  \relax
 \relax
\bibitem{Sjostrand:1993hi}
T.~Sj{\"o}strand and V.~A. Khoze, \emph{{On Color rearrangement in hadronic W+
  W- events}}, Z. Phys. C \textbf{62} (1994),
  \href{http://www.slac.stanford.edu/spires/find/hep/www?eprint=hep-ph/9310242}{281--310},
   [\href{http://arXiv.org/pdf/hep-ph/9310242}{{\texttt{hep-ph/9310242}}}].
  \relax
 \relax
\bibitem{Khoze:1994fu}
V.~A. Khoze and T.~Sj{\"o}strand, \emph{{Color correlations and multiplicities
  in top events}}, Phys. Lett. B \textbf{328} (1994),
  \href{http://www.slac.stanford.edu/spires/find/hep/www?eprint=hep-ph/9403394}{466--476},
   [\href{http://arXiv.org/pdf/hep-ph/9403394}{{\texttt{hep-ph/9403394}}}].
  \relax
 \relax
\bibitem{Khoze:1999up}
V.~A. Khoze and T.~Sj{\"o}strand, \emph{{QCD interconnection studies at linear
  colliders}}, Eur. Phys. J. direct \textbf{2} (2000), no.~1,
  \href{http://www.slac.stanford.edu/spires/find/hep/www?eprint=hep-ph/9912297}{1},
   [\href{http://arXiv.org/pdf/hep-ph/9912297}{{\texttt{hep-ph/9912297}}}].
  \relax
 \relax
\bibitem{Lonnblad:2023stc}
L.~L\"onnblad and H.~Shah, \emph{{A spatially constrained QCD colour
  reconnection in $\textrm{pp}$, $\textrm{p}A$, and $AA$ collisions in the
  Pythia8/Angantyr model}}, Eur. Phys. J. C \textbf{83} (2023), no.~7,
  \href{http://www.slac.stanford.edu/spires/find/hep/www?eprint=2303.11747}{575},
   [\href{http://arXiv.org/pdf/2303.11747}{{\texttt{arXiv:2303.11747}}}
  [hep-ph]], [Erratum: Eur.Phys.J.C 83, 639 (2023)]. \relax
 \relax
\bibitem{Webber:1997iw}
B.~R. Webber, \emph{{Color reconnection and Bose-Einstein effects}}, J. Phys. G
  \textbf{24} (1998),
  \href{http://www.slac.stanford.edu/spires/find/hep/www?eprint=hep-ph/9708463}{287--296},
   [\href{http://arXiv.org/pdf/hep-ph/9708463}{{\texttt{hep-ph/9708463}}}].
  \relax
 \relax
\bibitem{Gieseke:2012ft}
S.~Gieseke, C.~Rohr and A.~Siodmok, \emph{{Colour reconnections in Herwig++}},
  Eur. Phys. J. C \textbf{72} (2012),
  \href{http://www.slac.stanford.edu/spires/find/hep/www?eprint=1206.0041}{2225},
   [\href{http://arXiv.org/pdf/1206.0041}{{\texttt{arXiv:1206.0041}}}
  [hep-ph]]. \relax
 \relax
\bibitem{Gieseke:2017clv}
S.~Gieseke, P.~Kirchgae\ss{}er and S.~Pl\"atzer, \emph{{Baryon production from
  cluster hadronisation}}, Eur. Phys. J. C \textbf{78} (2018), no.~2,
  \href{http://www.slac.stanford.edu/spires/find/hep/www?eprint=1710.10906}{99},
   [\href{http://arXiv.org/pdf/1710.10906}{{\texttt{arXiv:1710.10906}}}
  [hep-ph]]. \relax
 \relax
\bibitem{Gieseke:2018gff}
S.~Gieseke, P.~Kirchgae\ss{}er, S.~Pl\"atzer and A.~Siodmok, \emph{{Colour
  Reconnection from Soft Gluon Evolution}}, JHEP \textbf{11} (2018),
  \href{http://www.slac.stanford.edu/spires/find/hep/www?eprint=1808.06770}{149},
   [\href{http://arXiv.org/pdf/1808.06770}{{\texttt{arXiv:1808.06770}}}
  [hep-ph]]. \relax
 \relax
\bibitem{Bellm:2019wrh}
J.~Bellm, C.~B. Duncan, S.~Gieseke, M.~Myska and A.~Si\'odmok, \emph{{Spacetime
  colour reconnection in Herwig 7}}, Eur. Phys. J. C \textbf{79} (2019),
  no.~12,
  \href{http://www.slac.stanford.edu/spires/find/hep/www?eprint=1909.08850}{1003},
   [\href{http://arXiv.org/pdf/1909.08850}{{\texttt{arXiv:1909.08850}}}
  [hep-ph]]. \relax
 \relax
\bibitem{Platzer:2022jny}
S.~Pl\"atzer, \emph{{Colour evolution and infrared physics}}, JHEP \textbf{07}
  (2023),
  \href{http://www.slac.stanford.edu/spires/find/hep/www?eprint=2204.06956}{126},
   [\href{http://arXiv.org/pdf/2204.06956}{{\texttt{arXiv:2204.06956}}}
  [hep-ph]]. \relax
 \relax
\bibitem{Winter:2003tt}
J.-C. Winter, F.~Krauss and G.~Soff, \emph{{A modified cluster-hadronisation
  model}}, Eur. Phys. J. \textbf{C36} (2004),
  \href{http://inspirehep.net/search?p=hep-ph/0311085}{381--395},
  [\href{http://arXiv.org/pdf/hep-ph/0311085}{{\texttt{hep-ph/0311085}}}].
  \relax
 \relax
\bibitem{Chahal:2022rid}
G.~S. Chahal and F.~Krauss, \emph{{Cluster Hadronisation in Sherpa}}, SciPost
  Phys. \textbf{13} (2022), no.~2,
  \href{http://www.slac.stanford.edu/spires/find/hep/www?eprint=2203.11385}{019},
   [\href{http://arXiv.org/pdf/2203.11385}{{\texttt{arXiv:2203.11385}}}
  [hep-ph]]. \relax
 \relax
\bibitem{Azimov:1984np}
Y.~I. Azimov, Y.~L. Dokshitzer, V.~A. Khoze and S.~Troyan, \emph{{Similarity of
  Parton and Hadron Spectra in QCD Jets}}, Z.Phys. \textbf{C27} (1985),
  \href{http://www.slac.stanford.edu/spires/find/hep/www?j=ZPhys,C27,65}{65--72}.
  \relax
 \relax
\bibitem{Amati:1979fg}
D.~Amati and G.~Veneziano, \emph{{Preconfinement as a Property of Perturbative
  QCD}}, Phys.Lett. \textbf{B83} (1979),
  \href{http://www.slac.stanford.edu/spires/find/hep/www?j=PhysLett,B83,87}{87}.
  \relax
 \relax
\bibitem{Bassetto:1979vy}
A.~Bassetto, M.~Ciafaloni and G.~Marchesini, \emph{{Color Singlet Distributions
  and Mass Damping in Perturbative QCD}}, Phys.Lett. \textbf{B83} (1979),
  \href{http://www.slac.stanford.edu/spires/find/hep/www?j=PhysLett,B83,207}{207}.
  \relax
 \relax
\bibitem{Marchesini:1980cr}
G.~Marchesini, L.~Trentadue and G.~Veneziano, \emph{{Space-time Description of
  Color Screening via Jet Calculus Techniques}}, Nucl. Phys. B \textbf{181}
  (1981),
  \href{http://www.slac.stanford.edu/spires/find/hep/www?j=Nucl%20Phys%20B,181,335}{335--346}.
  \relax
 \relax
\bibitem{Field:1977fa}
R.~D. Field and R.~P. Feynman, \emph{{A parametrization of the properties of
  quark jets}}, Nucl. Phys. \textbf{B136} (1978),
  \href{http://inspirehep.net/search?j=NUPHA,B136,1}{1}. \relax
 \relax
\bibitem{Hoyer:1979ta}
P.~Hoyer, P.~Osland, H.~G. Sander, T.~F. Walsh and P.~M. Zerwas, \emph{{Quantum
  Chromodynamics and Jets in e+ e-}}, Nucl. Phys. B \textbf{161} (1979),
  \href{http://www.slac.stanford.edu/spires/find/hep/www?j=Nucl%20Phys%20B,161,349}{349--372}.
  \relax
 \relax
\bibitem{Ali:1979em}
A.~Ali, E.~Pietarinen, G.~Kramer and J.~Willrodt, \emph{{A QCD Analysis of the
  High-Energy e+ e- Data from PETRA}}, Phys. Lett. B \textbf{93} (1980),
  \href{http://www.slac.stanford.edu/spires/find/hep/www?j=Phys%20Lett%20B,93,155}{155--160}.
  \relax
 \relax
\bibitem{Field:1982dg}
R.~D. Field and S.~Wolfram, \emph{{A QCD model for $e^+e^-$ annihilation}},
  Nucl. Phys. \textbf{B213} (1983),
  \href{http://inspirehep.net/search?j=NUPHA,B213,65}{65}. \relax
 \relax
\bibitem{Corcella:2000bw}
G.~Corcella et~al., \emph{{HERWIG 6: an event generator for hadron emission
  reactions with interfering gluons (including supersymmetric processes)}},
  JHEP \textbf{01} (2001),
  \href{http://inspirehep.net/search?p=hep-ph/0011363}{010},
  [\href{http://arXiv.org/pdf/hep-ph/0011363}{{\texttt{hep-ph/0011363}}}].
  \relax
 \relax
\bibitem{Marchesini:1983bm}
G.~Marchesini and B.~R. Webber, \emph{{Simulation of QCD Jets Including Soft
  Gluon Interference}}, Nucl. Phys. \textbf{B238} (1984),
  \href{http://inspirehep.net/search?j=NUPHA,B238,1}{1}. \relax
 \relax
\bibitem{Masouminia:2023zhb}
M.~R. Masouminia and P.~Richardson, \emph{{Hadronization and decay of excited
  heavy hadrons in Herwig 7}}, JHEP \textbf{07} (2024),
  \href{http://www.slac.stanford.edu/spires/find/hep/www?eprint=2312.02757}{278},
   [\href{http://arXiv.org/pdf/2312.02757}{{\texttt{arXiv:2312.02757}}}
  [hep-ph]]. \relax
 \relax
\bibitem{Hoang:2024zwl}
\href{http://www.slac.stanford.edu/spires/find/hep/www?eprint=2404.09856}{A.~H.
  Hoang, O.~L. Jin, S.~Pl\"atzer and D.~Samitz}, \emph{{Matching Hadronization
  and Perturbative Evolution: The Cluster Model in Light of Infrared Shower
  Cutoff Dependence}},
  \href{http://arXiv.org/pdf/2404.09856}{{\texttt{arXiv:2404.09856}}} [hep-ph].
  \relax
 \relax
\bibitem{Artru:1974hr}
X.~Artru and G.~Mennessier, \emph{{String model and multiproduction}}, Nucl.
  Phys. \textbf{B70} (1974),
  \href{http://inspirehep.net/search?j=NUPHA,B70,93}{93--115}. \relax
 \relax
\bibitem{Bowler:1981sb}
M.~G. Bowler, \emph{{$e^+e^-$ production of heavy quarks in the string model}},
  Z. Phys. \textbf{C11} (1981),
  \href{http://inspirehep.net/search?j=ZEPYA,C11,169}{169}. \relax
 \relax
\bibitem{Andersson:1983ia}
B.~Andersson, G.~Gustafson, G.~Ingelman and T.~Sj{\"o}strand, \emph{{Parton
  Fragmentation and String Dynamics}}, Phys. Rept. \textbf{97} (1983),
  \href{http://www.slac.stanford.edu/spires/find/hep/www?j=Phys%20Rept,97,31}{31--145}.
  \relax
 \relax
\bibitem{Gottschalk:1983fm}
T.~D. Gottschalk, \emph{{An improved description of hadronization in the QCD
  cluster model for $e^+e^-$ annihilation}}, Nucl. Phys. \textbf{B239} (1984),
  \href{http://inspirehep.net/search?j=NUPHA,B239,349}{349}. \relax
 \relax
\bibitem{Gottschalk:1986bv}
T.~D. Gottschalk and D.~A. Morris, \emph{{A new model for hadronization and
  $e^+e^-$ annihilation}}, Nucl. Phys. \textbf{B288} (1987),
  \href{http://inspirehep.net/search?j=NUPHA,B288,729}{729}. \relax
 \relax
\bibitem{Sauter:1931zz}
F.~Sauter, \emph{{Uber das Verhalten eines Elektrons im homogenen elektrischen
  Feld nach der relativistischen Theorie Diracs}}, Z. Phys. \textbf{69} (1931),
  \href{http://www.slac.stanford.edu/spires/find/hep/www?j=Z%20Phys,69,742}{742--764}.
  \relax
 \relax
\bibitem{Schwinger:1951nm}
J.~S. Schwinger, \emph{{On gauge invariance and vacuum polarization}}, Phys.
  Rev. \textbf{82} (1951),
  \href{http://www.slac.stanford.edu/spires/find/hep/www?j=Phys%20Rev,82,664}{664--679}.
  \relax
 \relax
\bibitem{Andersson:1983jt}
B.~Andersson, G.~Gustafson and B.~S{\"o}derberg, \emph{{A general model for jet
  fragmentation}}, Z. Phys. \textbf{C20} (1983),
  \href{http://inspirehep.net/search?j=ZEPYA,C20,317}{317}. \relax
 \relax
\bibitem{Andersson:1979ij}
B.~Andersson and G.~Gustafson, \emph{{Semiclassical Models for Gluon Jets and
  Leptoproduction Based on the Massless Relativistic String}}, Z. Phys. C
  \textbf{3} (1980),
  \href{http://www.slac.stanford.edu/spires/find/hep/www?j=Z%20Phys%20C,3,223}{223}.
  \relax
 \relax
\bibitem{Sjostrand:1984ic}
T.~Sj{\"o}strand, \emph{{Jet Fragmentation of Nearby Partons}}, Nucl.Phys.
  \textbf{B248} (1984),
  \href{http://www.slac.stanford.edu/spires/find/hep/www?j=NuclPhys,B248,469}{469}.
  \relax
 \relax
\bibitem{Andersson:1981ce}
B.~Andersson, G.~Gustafson and T.~Sj{\"o}strand, \emph{{A Model for Baryon
  Production in Quark and Gluon Jets}}, Nucl. Phys. \textbf{B197} (1982),
  \href{http://www.slac.stanford.edu/spires/find/hep/www?j=Nucl%20Phys,B197,45}{45}.
  \relax
 \relax
\bibitem{Sjostrand:1982fn}
T.~Sj{\"o}strand, \emph{{The Lund Monte Carlo for Jet Fragmentation}}, Comput.
  Phys. Commun. \textbf{27} (1982),
  \href{http://inspirehep.net/search?j=CPHCB,27,243}{243}. \relax
 \relax
\bibitem{Andersson:1997xwk}
B.~Andersson, \emph{{The Lund Model}}, Cambridge Monographs on Particle
  Physics, Nuclear Physics and Cosmology, vol.~7, Cambridge University Press, 7
  2023. \relax
 \relax
\bibitem{Sjostrand:2006za}
T.~Sj{\"o}strand, S.~Mrenna and P.~Skands, \emph{{PYTHIA 6.4 physics and
  manual}}, JHEP \textbf{05} (2006),
  \href{http://inspirehep.net/search?p=hep-ph/0603175}{026},
  [\href{http://arXiv.org/pdf/hep-ph/0603175}{{\texttt{hep-ph/0603175}}}].
  \relax
 \relax
\bibitem{Siegert:2006xx}
F.~Siegert, \emph{{Simulation of hadron decays in \Sherpa}}, Diploma thesis.
  \relax
 \relax
\bibitem{Laubrich:2006aa}
T.~Laubrich, \emph{{Implementation of tau-lepton decays into the event
  generator Sherpa}}, Diploma Thesis. \relax
 \relax
\bibitem{Jadach:1993hs}
S.~Jadach, Z.~Was, R.~Decker and J.~H. K{\"u}hn, \emph{{The tau decay library
  TAUOLA: Version 2.4}}, Comput. Phys. Commun. \textbf{76} (1993),
  \href{http://www.slac.stanford.edu/spires/find/hep/www?j=Comput%20Phys%20Commun,76,361}{361--380}.
  \relax
 \relax
\bibitem{Lange:2001uf}
D.~J. Lange, \emph{{The EvtGen particle decay simulation package}}, Nucl.
  Instrum. Meth. \textbf{A462} (2001),
  \href{http://inspirehep.net/search?j=NUIMA,A462,152}{152--155}. \relax
 \relax
\bibitem{Giele:2011cb}
W.~T. Giele, D.~A. Kosower and P.~Z. Skands, \emph{{Higher-Order Corrections to
  Timelike Jets}}, Phys. Rev. \textbf{D84} (2011),
  \href{http://inspirehep.net/record/889142}{054003},
  [\href{http://arXiv.org/pdf/1102.2126}{{\texttt{arXiv:1102.2126}}} [hep-ph]].
  \relax
 \relax
\bibitem{Mrenna:2016sih}
S.~Mrenna and P.~Skands, \emph{{Automated Parton-Shower Variations in Pythia
  8}}, Phys. Rev. \textbf{D94} (2016), no.~7,
  \href{http://www.slac.stanford.edu/spires/find/hep/www?eprint=1605.08352}{074005},
   [\href{http://arXiv.org/pdf/1605.08352}{{\texttt{arXiv:1605.08352}}}
  [hep-ph]]. \relax
 \relax
\bibitem{Bellm:2016voq}
J.~Bellm, S.~Pl{\"a}tzer, P.~Richardson, A.~Si{\'o}dmok and S.~Webster,
  \emph{{Reweighting Parton Showers}}, Phys. Rev. \textbf{D94} (2016), no.~3,
  \href{http://www.slac.stanford.edu/spires/find/hep/www?eprint=1605.08256}{034028},
   [\href{http://arXiv.org/pdf/1605.08256}{{\texttt{arXiv:1605.08256}}}
  [hep-ph]]. \relax
 \relax
\bibitem{Bierlich:2023fmh}
C.~Bierlich, P.~Ilten, T.~Menzo, S.~Mrenna, M.~Szewc, M.~K. Wilkinson,
  A.~Youssef and J.~Zupan, \emph{{Reweighting Monte Carlo predictions and
  automated fragmentation variations in Pythia 8}}, SciPost Phys. \textbf{16}
  (2024), no.~5,
  \href{http://www.slac.stanford.edu/spires/find/hep/www?eprint=2308.13459}{134},
   [\href{http://arXiv.org/pdf/2308.13459}{{\texttt{arXiv:2308.13459}}}
  [hep-ph]]. \relax
 \relax
\bibitem{Bothmann:2016nao}
E.~Bothmann, M.~Sch{\"o}nherr and S.~Schumann, \emph{{Reweighting QCD
  matrix-element and parton-shower calculations}}, Eur. Phys. J. \textbf{C76}
  (2016), no.~11,
  \href{http://www.slac.stanford.edu/spires/find/hep/www?eprint=1606.08753}{590},
   [\href{http://arXiv.org/pdf/1606.08753}{{\texttt{arXiv:1606.08753}}}
  [hep-ph]]. \relax
 \relax
\bibitem{Dobbs:2001ck}
M.~Dobbs and J.~B. Hansen, \emph{{The HepMC C++ Monte Carlo event record for
  High Energy Physics}}, Comput. Phys. Commun. \textbf{134} (2001),
  \href{http://www.slac.stanford.edu/spires/find/hep/www?j=Comput%20Phys%20Commun,134,41}{41--46}.
  \relax
 \relax
\bibitem{Buckley:2019xhk}
A.~Buckley, P.~Ilten, D.~Konstantinov, L.~L\"onnblad, J.~Monk, W.~Pokorski,
  T.~Przedzinski and A.~Verbytskyi, \emph{{The HepMC3 event record library for
  Monte Carlo event generators}}, Comput. Phys. Commun. \textbf{260} (2021),
  \href{http://www.slac.stanford.edu/spires/find/hep/www?eprint=1912.08005}{107310},
   [\href{http://arXiv.org/pdf/1912.08005}{{\texttt{arXiv:1912.08005}}}
  [hep-ph]]. \relax
 \relax
\bibitem{Bierlich:2019rhm}
C.~Bierlich et~al., \emph{{Robust Independent Validation of Experiment and
  Theory: Rivet version 3}}, SciPost Phys. \textbf{8} (2020),
  \href{http://www.slac.stanford.edu/spires/find/hep/www?eprint=1912.05451}{026},
   [\href{http://arXiv.org/pdf/1912.05451}{{\texttt{arXiv:1912.05451}}}
  [hep-ph]]. \relax
 \relax
\bibitem{Bierlich:2024vqo}
\href{http://www.slac.stanford.edu/spires/find/hep/www?eprint=2404.15984}{C.~Bierlich,
  A.~Buckley, J.~Butterworth, C.~Gutschow, L.~Lonnblad, T.~Procter,
  P.~Richardson and Y.~Yeh}, \emph{{Robust Independent Validation of Experiment
  and Theory: Rivet version 4 release note}},
  \href{http://arXiv.org/pdf/2404.15984}{{\texttt{arXiv:2404.15984}}} [hep-ph].
  \relax
 \relax
\bibitem{Buckley:2010ar}
A.~Buckley, J.~Butterworth, L.~L{\"o}nnblad, D.~Grellscheid, H.~Hoeth et~al.,
  \emph{{Rivet user manual}}, Comput.Phys.Commun. \textbf{184} (2013),
  \href{http://www.slac.stanford.edu/spires/find/hep/www?eprint=1003.0694}{2803--2819},
   [\href{http://arXiv.org/pdf/1003.0694}{{\texttt{arXiv:1003.0694}}}
  [hep-ph]]. \relax
 \relax
\bibitem{DelDebbio:2013kxa}
L.~Del~Debbio, N.~P. Hartland and S.~Schumann, \emph{{MCgrid: projecting cross
  section calculations on grids}}, Comput. Phys. Commun. \textbf{185} (2014),
  \href{http://www.slac.stanford.edu/spires/find/hep/www?eprint=1312.4460}{2115--2126},
   [\href{http://arXiv.org/pdf/1312.4460}{{\texttt{arXiv:1312.4460}}}
  [hep-ph]]. \relax
 \relax
\bibitem{Bothmann:2015dba}
E.~Bothmann, N.~Hartland and S.~Schumann, \emph{{Introducing MCgrid 2.0:
  Projecting cross section calculations on grids}}, Comput. Phys. Commun.
  \textbf{196} (2015),
  \href{http://www.slac.stanford.edu/spires/find/hep/www?j=Comput%20Phys%20Commun,196,617}{617--618}.
  \relax
 \relax
\bibitem{Carli:2010rw}
T.~Carli et~al., \emph{{A posteriori inclusion of parton density functions in
  NLO QCD final-state calculations at hadron colliders: The APPLGRID Project}},
  Eur. Phys. J. \textbf{C} (2010),
  \href{http://inspirehep.net/search?p=arXiv:0911.2985}{503--524},
  [\href{http://arXiv.org/pdf/0911.2985}{{\texttt{arXiv:0911.2985}}} [hep-ph]].
  \relax
 \relax
\bibitem{Britzger:2012bs}
D.~Britzger, K.~Rabbertz, F.~Stober and M.~Wobisch, \emph{{New features in
  version 2 of the fastNLO project}}, {Proceedings, 20th International Workshop
  on Deep-Inelastic Scattering and Related Subjects (DIS 2012): Bonn, Germany,
  March 26-30, 2012}, 2012, pp.~217--221. \relax
 \relax
\bibitem{Banfi:2003je}
A.~Banfi, G.~P. Salam and G.~Zanderighi, \emph{{Generalized resummation of QCD
  final state observables}}, Phys.Lett. \textbf{B584} (2004),
  \href{http://www.slac.stanford.edu/spires/find/hep/www?eprint=hep-ph/0304148}{298--305},
   [\href{http://arXiv.org/pdf/hep-ph/0304148}{{\texttt{arXiv:hep-ph/0304148}}}
  [hep-ph]]. \relax
 \relax
\bibitem{Banfi:2004yd}
A.~Banfi, G.~P. Salam and G.~Zanderighi, \emph{{Principles of general
  final-state resummation and automated implementation}}, JHEP \textbf{03}
  (2005), \href{http://inspirehep.net/record/655163}{073},
  [\href{http://arXiv.org/pdf/hep-ph/0407286}{{\texttt{arXiv:hep-ph/0407286}}}
  [hep-ph]]. \relax
 \relax
\bibitem{Gerwick:2014gya}
E.~Gerwick, S.~H{\"o}che, S.~Marzani and S.~Schumann, \emph{{Soft evolution of
  multi-jet final states}}, JHEP \textbf{02} (2015),
  \href{http://www.slac.stanford.edu/spires/find/hep/www?eprint=1411.7325}{106},
   [\href{http://arXiv.org/pdf/1411.7325}{{\texttt{arXiv:1411.7325}}}
  [hep-ph]]. \relax
 \relax
\bibitem{Marzani:2019evv}
S.~Marzani, D.~Reichelt, S.~Schumann, G.~Soyez and V.~Theeuwes, \emph{{Fitting
  the Strong Coupling Constant with Soft-Drop Thrust}}, JHEP \textbf{11}
  (2019),
  \href{http://www.slac.stanford.edu/spires/find/hep/www?eprint=1906.10504}{179},
   [\href{http://arXiv.org/pdf/1906.10504}{{\texttt{arXiv:1906.10504}}}
  [hep-ph]]. \relax
 \relax
\bibitem{Reichelt:2021svh}
D.~Reichelt, S.~Caletti, O.~Fedkevych, S.~Marzani, S.~Schumann and G.~Soyez,
  \emph{{Phenomenology of jet angularities at the LHC}}, JHEP \textbf{03}
  (2022),
  \href{http://www.slac.stanford.edu/spires/find/hep/www?eprint=2112.09545}{131},
   [\href{http://arXiv.org/pdf/2112.09545}{{\texttt{arXiv:2112.09545}}}
  [hep-ph]]. \relax
 \relax
\bibitem{Chien:2024uax}
Y.-T. Chien, O.~Fedkevych, D.~Reichelt and S.~Schumann, \emph{{Jet angularities
  in dijet production in proton-proton and heavy-ion collisions at RHIC}}, JHEP
  \textbf{07} (2024),
  \href{http://www.slac.stanford.edu/spires/find/hep/www?eprint=2404.04168}{230},
   [\href{http://arXiv.org/pdf/2404.04168}{{\texttt{arXiv:2404.04168}}}
  [hep-ph]]. \relax
 \relax
\bibitem{dEnterria:2022hzv}
\href{http://www.slac.stanford.edu/spires/find/hep/www?eprint=2203.08271}{D.~d'Enterria
  et~al.}, \emph{{The strong coupling constant: State of the art and the decade
  ahead}},  \href{http://arXiv.org/pdf/2203.08271}{{\texttt{arXiv:2203.08271}}}
  [hep-ph]. \relax
 \relax
\bibitem{Caletti:2021oor}
S.~Caletti, O.~Fedkevych, S.~Marzani, D.~Reichelt, S.~Schumann, G.~Soyez and
  V.~Theeuwes, \emph{{Jet angularities in Z+jet production at the LHC}}, JHEP
  \textbf{07} (2021),
  \href{http://www.slac.stanford.edu/spires/find/hep/www?eprint=2104.06920}{076},
   [\href{http://arXiv.org/pdf/2104.06920}{{\texttt{arXiv:2104.06920}}}
  [hep-ph]]. \relax
 \relax
\bibitem{Caletti:2021ysv}
S.~Caletti, O.~Fedkevych, S.~Marzani and D.~Reichelt, \emph{{Tagging the
  initial-state gluon}}, Eur. Phys. J. C \textbf{81} (2021), no.~9,
  \href{http://www.slac.stanford.edu/spires/find/hep/www?eprint=2108.10024}{844},
   [\href{http://arXiv.org/pdf/2108.10024}{{\texttt{arXiv:2108.10024}}}
  [hep-ph]]. \relax
 \relax
\bibitem{H1:2024aze}
V.~Andreev et~al., H1 collaboration, \emph{{Measurement of the 1-jettiness
  event shape observable in deep-inelastic electron-proton scattering at
  HERA}}, Eur. Phys. J. C \textbf{84} (2024), no.~8,
  \href{http://www.slac.stanford.edu/spires/find/hep/www?eprint=2403.10109}{785},
   [\href{http://arXiv.org/pdf/2403.10109}{{\texttt{arXiv:2403.10109}}}
  [hep-ex]]. \relax
 \relax
\bibitem{H1:2024pvu}
V.~Andreev et~al., H1 collaboration, \emph{{Measurement of groomed event shape
  observables in deep-inelastic electron-proton scattering at HERA}}, Eur.
  Phys. J. C \textbf{84} (2024), no.~7,
  \href{http://www.slac.stanford.edu/spires/find/hep/www?eprint=2403.10134}{718},
   [\href{http://arXiv.org/pdf/2403.10134}{{\texttt{arXiv:2403.10134}}}
  [hep-ex]]. \relax
 \relax
\bibitem{Knobbe:2024rci}
\href{http://www.slac.stanford.edu/spires/find/hep/www?eprint=2407.02456}{M.~Knobbe,
  D.~Reichelt, S.~Schumann and L.~St\"ocker}, \emph{{Precision calculations for
  groomed event shapes at HERA}},
  \href{http://arXiv.org/pdf/2407.02456}{{\texttt{arXiv:2407.02456}}} [hep-ph].
  \relax
 \relax
\bibitem{Gehrmann-DeRidder:2024avt}
A.~Gehrmann-De~Ridder, C.~T. Preuss, D.~Reichelt and S.~Schumann,
  \emph{{NLO+NLL' accurate predictions for three-jet event shapes in hadronic
  Higgs decays}}, JHEP \textbf{07} (2024),
  \href{http://www.slac.stanford.edu/spires/find/hep/www?eprint=2403.06929}{160},
   [\href{http://arXiv.org/pdf/2403.06929}{{\texttt{arXiv:2403.06929}}}
  [hep-ph]]. \relax
 \relax
\bibitem{Bauer:2000yr}
C.~W. Bauer, S.~Fleming, D.~Pirjol and I.~W. Stewart, \emph{{An Effective field
  theory for collinear and soft gluons: Heavy to light decays}}, Phys.Rev.
  \textbf{D63} (2001),
  \href{http://www.slac.stanford.edu/spires/find/hep/www?eprint=hep-ph/0011336}{114020},
   [\href{http://arXiv.org/pdf/hep-ph/0011336}{{\texttt{arXiv:hep-ph/0011336}}}
  [hep-ph]]. \relax
 \relax
\bibitem{Bauer:2001yt}
C.~W. Bauer, D.~Pirjol and I.~W. Stewart, \emph{{Soft collinear factorization
  in effective field theory}}, Phys.Rev. \textbf{D65} (2002),
  \href{http://www.slac.stanford.edu/spires/find/hep/www?eprint=hep-ph/0109045}{054022},
   [\href{http://arXiv.org/pdf/hep-ph/0109045}{{\texttt{arXiv:hep-ph/0109045}}}
  [hep-ph]]. \relax
 \relax
\bibitem{Ju:2021lah}
W.-L. Ju and M.~Sch\"onherr, \emph{{The q$_{T}$ and
  \ensuremath{\Delta}\ensuremath{\phi} spectra in W and Z production at the LHC
  at N$^{3}$LL'+N$^{2}$LO}}, JHEP \textbf{10} (2021),
  \href{http://www.slac.stanford.edu/spires/find/hep/www?eprint=2106.11260}{088},
   [\href{http://arXiv.org/pdf/2106.11260}{{\texttt{arXiv:2106.11260}}}
  [hep-ph]]. \relax
 \relax
\bibitem{Ju:2022wia}
W.-L. Ju and M.~Sch\"onherr, \emph{{Projected transverse momentum resummation
  in top-antitop pair production at LHC}}, JHEP \textbf{02} (2023),
  \href{http://www.slac.stanford.edu/spires/find/hep/www?eprint=2210.09272}{075},
   [\href{http://arXiv.org/pdf/2210.09272}{{\texttt{arXiv:2210.09272}}}
  [hep-ph]]. \relax
 \relax
\bibitem{Ju:2024xhd}
\href{http://www.slac.stanford.edu/spires/find/hep/www?eprint=2407.03501}{W.-L.
  Ju and M.~Sch\"onherr}, \emph{{The $q_{\mathrm{T}}$ and
  $\Delta\phi_{t\bar{t}}$ spectra in top-antitop hadroproduction at NNLL+NNLO:
  the interplay of soft-collinear resummation and Coulomb singularities}},
  \href{http://arXiv.org/pdf/2407.03501}{{\texttt{arXiv:2407.03501}}} [hep-ph].
  \relax
 \relax
\bibitem{Ferrera:2023vsw}
G.~Ferrera, W.-L. Ju and M.~Sch\"onherr, \emph{{Zero-bin subtraction and the
  q$_{T}$ spectrum beyond leading power}}, JHEP \textbf{04} (2024),
  \href{http://www.slac.stanford.edu/spires/find/hep/www?eprint=2312.14911}{005},
   [\href{http://arXiv.org/pdf/2312.14911}{{\texttt{arXiv:2312.14911}}}
  [hep-ph]]. \relax
 \relax
\bibitem{HSFPhysicsEventGeneratorWG:2020gxw}
S.~Amoroso et~al., HSF Physics Event Generator WG collaboration,
  \emph{{Challenges in Monte Carlo Event Generator Software for High-Luminosity
  LHC}}, Comput. Softw. Big Sci. \textbf{5} (2021), no.~1,
  \href{http://www.slac.stanford.edu/spires/find/hep/www?eprint=2004.13687}{12},
   [\href{http://arXiv.org/pdf/2004.13687}{{\texttt{arXiv:2004.13687}}}
  [hep-ph]]. \relax
 \relax
\bibitem{Valassi:2020ueh}
\href{http://www.slac.stanford.edu/spires/find/hep/www?eprint=2004.13687}{S.~Amoroso
  et~al.}, HSF Physics Event Generator WG collaboration, \emph{{Challenges in
  Monte Carlo event generator software for High-Luminosity LHC}},
  \href{http://arXiv.org/pdf/2004.13687}{{\texttt{arXiv:2004.13687}}} [hep-ph].
  \relax
 \relax
\bibitem{HSFPhysicsEventGeneratorWG:2021xti}
\href{http://www.slac.stanford.edu/spires/find/hep/www?eprint=2109.14938}{E.~Yazgan
  et~al.}, HSF Physics Event Generator WG collaboration, \emph{{HL-LHC
  Computing Review Stage-2, Common Software Projects: Event Generators}},
  \href{http://arXiv.org/pdf/2109.14938}{{\texttt{arXiv:2109.14938}}} [hep-ph].
  \relax
 \relax
\bibitem{ATLAS:2021yza}
G.~Aad et~al., ATLAS collaboration, \emph{{Modelling and computational
  improvements to the simulation of single vector-boson plus jet processes for
  the ATLAS experiment}}, JHEP \textbf{08} (2022),
  \href{http://www.slac.stanford.edu/spires/find/hep/www?eprint=2112.09588}{089},
   [\href{http://arXiv.org/pdf/2112.09588}{{\texttt{arXiv:2112.09588}}}
  [hep-ex]]. \relax
 \relax
\bibitem{Bothmann:2021nch}
E.~Bothmann, W.~Giele, S.~H\"oche, J.~Isaacson and M.~Knobbe, \emph{{Many-gluon
  tree amplitudes on modern GPUs: A case study for novel event generators}},
  SciPost Phys. Codeb. \textbf{2022} (2022),
  \href{http://www.slac.stanford.edu/spires/find/hep/www?eprint=2106.06507}{3},
   [\href{http://arXiv.org/pdf/2106.06507}{{\texttt{arXiv:2106.06507}}}
  [hep-ph]]. \relax
 \relax
\bibitem{Bothmann:2022itv}
E.~Bothmann, J.~Isaacson, M.~Knobbe, S.~H\"oche and W.~Giele, \emph{{QCD tree
  amplitudes on modern GPUs: A case study for novel event generators}}, PoS
  \textbf{ICHEP2022} (2022),
  \href{http://www.slac.stanford.edu/spires/find/hep/www?j=PoS,ICHEP2022,222}{222}.
  \relax
 \relax
\bibitem{Bothmann:2023siu}
E.~Bothmann, T.~Childers, W.~Giele, F.~Herren, S.~H\"oche, J.~Isaacson,
  M.~Knobbe and R.~Wang, \emph{{Efficient phase-space generation for hadron
  collider event simulation}}, SciPost Phys. \textbf{15} (2023), no.~4,
  \href{http://www.slac.stanford.edu/spires/find/hep/www?eprint=2302.10449}{169},
   [\href{http://arXiv.org/pdf/2302.10449}{{\texttt{arXiv:2302.10449}}}
  [hep-ph]]. \relax
 \relax
\bibitem{Bothmann:2023gew}
E.~Bothmann, T.~Childers, W.~Giele, S.~H\"oche, J.~Isaacson and M.~Knobbe,
  \emph{{A Portable Parton-Level Event Generator for the High-Luminosity LHC}},
  SciPost Phys. \textbf{17} (2024),
  \href{http://www.slac.stanford.edu/spires/find/hep/www?eprint=2311.06198}{081},
   [\href{http://arXiv.org/pdf/2311.06198}{{\texttt{arXiv:2311.06198}}}
  [hep-ph]]. \relax
 \relax
\bibitem{Buckley:2023xqh}
\href{http://www.slac.stanford.edu/spires/find/hep/www?eprint=2312.15070}{A.~Buckley,
  L.~Corpe, M.~Filipovich, C.~Gutschow, N.~Rozinsky, S.~Thor, Y.~Yeh and
  J.~Yellen}, \emph{{Consistent, multidimensional differential histogramming
  and summary statistics with YODA 2}},
  \href{http://arXiv.org/pdf/2312.15070}{{\texttt{arXiv:2312.15070}}} [hep-ph].
  \relax
 \relax
\bibitem{Bothmann:2020ywa}
E.~Bothmann, T.~Jan\ss{}en, M.~Knobbe, T.~Schmale and S.~Schumann,
  \emph{{Exploring phase space with Neural Importance Sampling}}, SciPost Phys.
  \textbf{8} (2020), no.~4,
  \href{http://www.slac.stanford.edu/spires/find/hep/www?eprint=2001.05478}{069},
   [\href{http://arXiv.org/pdf/2001.05478}{{\texttt{arXiv:2001.05478}}}
  [hep-ph]]. \relax
 \relax
\bibitem{Gao:2020zvv}
C.~Gao, S.~H\"oche, J.~Isaacson, C.~Krause and H.~Schulz, \emph{{Event
  Generation with Normalizing Flows}}, Phys. Rev. D \textbf{101} (2020), no.~7,
  \href{http://www.slac.stanford.edu/spires/find/hep/www?eprint=2001.10028}{076002},
   [\href{http://arXiv.org/pdf/2001.10028}{{\texttt{arXiv:2001.10028}}}
  [hep-ph]]. \relax
 \relax
\bibitem{Lepage:1980dq}
\href{http://inspirehep.net/search?r=CLNS-80/447}{G.~P. Lepage}, \emph{{VEGAS -
  An Adaptive Multi-dimensional Integration Program}}, CLNS-80/447. \relax
 \relax
\bibitem{Danziger:2021eeg}
K.~Danziger, T.~Jan\ss{}en, S.~Schumann and F.~Siegert, \emph{{Accelerating
  Monte Carlo event generation -- rejection sampling using neural network
  event-weight estimates}}, SciPost Phys. \textbf{12} (2022),
  \href{http://www.slac.stanford.edu/spires/find/hep/www?eprint=2109.11964}{164},
   [\href{http://arXiv.org/pdf/2109.11964}{{\texttt{arXiv:2109.11964}}}
  [hep-ph]]. \relax
 \relax
\bibitem{Maitre:2021uaa}
D.~Ma\^\i{}tre and H.~Truong, \emph{{A factorisation-aware Matrix element
  emulator}}, JHEP \textbf{11} (2021),
  \href{http://www.slac.stanford.edu/spires/find/hep/www?eprint=2107.06625}{066},
   [\href{http://arXiv.org/pdf/2107.06625}{{\texttt{arXiv:2107.06625}}}
  [hep-ph]]. \relax
 \relax
\bibitem{Janssen:2023ahv}
T.~Jan\ss{}en, D.~Ma\^\i{}tre, S.~Schumann, F.~Siegert and H.~Truong,
  \emph{{Unweighting multijet event generation using factorisation-aware neural
  networks}}, SciPost Phys. \textbf{15} (2023), no.~3,
  \href{http://www.slac.stanford.edu/spires/find/hep/www?eprint=2301.13562}{107},
   [\href{http://arXiv.org/pdf/2301.13562}{{\texttt{arXiv:2301.13562}}}
  [hep-ph]]. \relax
 \relax
\bibitem{DUNE:2020fgq}
B.~Abi et~al., DUNE collaboration, \emph{{Prospects for beyond the Standard
  Model physics searches at the Deep Underground Neutrino Experiment}}, Eur.
  Phys. J. C \textbf{81} (2021), no.~4,
  \href{http://www.slac.stanford.edu/spires/find/hep/www?eprint=2008.12769}{322},
   [\href{http://arXiv.org/pdf/2008.12769}{{\texttt{arXiv:2008.12769}}}
  [hep-ex]]. \relax
 \relax
\bibitem{DUNE:2020ypp}
\href{http://www.slac.stanford.edu/spires/find/hep/www?eprint=2002.03005}{B.~Abi
  et~al.}, DUNE collaboration, \emph{{Deep Underground Neutrino Experiment
  (DUNE), Far Detector Technical Design Report, Volume II: DUNE Physics}},
  \href{http://arXiv.org/pdf/2002.03005}{{\texttt{arXiv:2002.03005}}} [hep-ex].
  \relax
 \relax
\bibitem{Hyper-Kamiokande:2018ofw}
\href{http://www.slac.stanford.edu/spires/find/hep/www?eprint=1805.04163}{K.~Abe
  et~al.}, Hyper-Kamiokande collaboration, \emph{{Hyper-Kamiokande Design
  Report}},
  \href{http://arXiv.org/pdf/1805.04163}{{\texttt{arXiv:1805.04163}}}
  [physics.ins-det]. \relax
 \relax
\bibitem{MicroBooNE:2015bmn}
\href{http://www.slac.stanford.edu/spires/find/hep/www?eprint=1503.01520}{R.~Acciarri
  et~al.}, MicroBooNE, LAr1-ND, ICARUS-WA104 collaboration, \emph{{A Proposal
  for a Three Detector Short-Baseline Neutrino Oscillation Program in the
  Fermilab Booster Neutrino Beam}},
  \href{http://arXiv.org/pdf/1503.01520}{{\texttt{arXiv:1503.01520}}}
  [physics.ins-det]. \relax
 \relax
\bibitem{Isaacson:2021xty}
J.~Isaacson, S.~H\"oche, D.~Lopez~Gutierrez and N.~Rocco, \emph{{Novel event
  generator for the automated simulation of neutrino scattering}}, Phys. Rev. D
  \textbf{105} (2022), no.~9,
  \href{http://www.slac.stanford.edu/spires/find/hep/www?eprint=2110.15319}{096006},
   [\href{http://arXiv.org/pdf/2110.15319}{{\texttt{arXiv:2110.15319}}}
  [hep-ph]]. \relax
 \relax
\bibitem{Hernandez:2022nmp}
E.~Hern\'andez, J.~Nieves, F.~S\'anchez and J.~E. Sobczyk, \emph{{Tau
  longitudinal and transverse polarizations from visible kinematics in
  (anti-)neutrino nucleus scattering}}, Phys. Lett. B \textbf{829} (2022),
  \href{http://www.slac.stanford.edu/spires/find/hep/www?eprint=2202.07539}{137046},
   [\href{http://arXiv.org/pdf/2202.07539}{{\texttt{arXiv:2202.07539}}}
  [hep-ph]]. \relax
 \relax
\bibitem{Isaacson:2023gwp}
J.~Isaacson, S.~H\"oche, F.~Siegert and S.~Wang, \emph{{Tau polarization and
  correlated decays in neutrino experiments}}, Phys. Rev. D \textbf{108}
  (2023), no.~9,
  \href{http://www.slac.stanford.edu/spires/find/hep/www?eprint=2303.08104}{093004},
   [\href{http://arXiv.org/pdf/2303.08104}{{\texttt{arXiv:2303.08104}}}
  [hep-ph]]. \relax
 \relax
\bibitem{GENIE:2021npt}
L.~Alvarez-Ruso et~al., GENIE collaboration, \emph{{Recent highlights from
  GENIE v3}}, Eur. Phys. J. ST \textbf{230} (2021), no.~24,
  \href{http://www.slac.stanford.edu/spires/find/hep/www?eprint=2106.09381}{4449--4467},
   [\href{http://arXiv.org/pdf/2106.09381}{{\texttt{arXiv:2106.09381}}}
  [hep-ph]]. \relax
 \relax
\bibitem{Andreopoulos:2009rq}
C.~Andreopoulos et~al., \emph{{The GENIE Neutrino Monte Carlo Generator}},
  Nucl. Instrum. Meth. A \textbf{614} (2010),
  \href{http://www.slac.stanford.edu/spires/find/hep/www?eprint=0905.2517}{87--104},
   [\href{http://arXiv.org/pdf/0905.2517}{{\texttt{arXiv:0905.2517}}}
  [hep-ph]]. \relax
 \relax
\bibitem{Buss:2011mx}
O.~Buss, T.~Gaitanos, K.~Gallmeister, H.~van Hees, M.~Kaskulov, O.~Lalakulich,
  A.~B. Larionov, T.~Leitner, J.~Weil and U.~Mosel,
  \emph{{Transport-theoretical Description of Nuclear Reactions}}, Phys. Rept.
  \textbf{512} (2012),
  \href{http://www.slac.stanford.edu/spires/find/hep/www?eprint=1106.1344}{1--124},
   [\href{http://arXiv.org/pdf/1106.1344}{{\texttt{arXiv:1106.1344}}}
  [hep-ph]]. \relax
 \relax
\bibitem{Hayato:2021heg}
Y.~Hayato and L.~Pickering, \emph{{The NEUT neutrino interaction simulation
  program library}}, Eur. Phys. J. ST \textbf{230} (2021), no.~24,
  \href{http://www.slac.stanford.edu/spires/find/hep/www?eprint=2106.15809}{4469--4481},
   [\href{http://arXiv.org/pdf/2106.15809}{{\texttt{arXiv:2106.15809}}}
  [hep-ph]]. \relax
 \relax
\bibitem{Hayato:2009zz}
Y.~Hayato, \emph{{A neutrino interaction simulation program library NEUT}},
  Acta Phys. Polon. B \textbf{40} (2009),
  \href{http://www.slac.stanford.edu/spires/find/hep/www?j=Acta%20Phys%20Polon%20B,40,2477}{2477--2489}.
  \relax
 \relax
\bibitem{Golan:2012rfa}
T.~Golan, J.~T. Sobczyk and J.~Zmuda, \emph{{NuWro: the Wroclaw Monte Carlo
  Generator of Neutrino Interactions}}, Nucl. Phys. B Proc. Suppl.
  \textbf{229-232} (2012),
  \href{http://www.slac.stanford.edu/spires/find/hep/www?j=Nucl%20Phys%20B%20Proc%20Suppl,229-232,499}{499--499}.
  \relax
 \relax
\bibitem{YAML2009}
O.~Ben-Kiki, C.~Evans and B.~Ingerson, \emph{{YAML} Ain't Markup Language
  ({YAML}) (tm) Version 1.2}, Tech. report, YAML.org, 10 2009. \relax
 \relax
\end{thebibliography}

\end{document}